\newif\ifdraft
\draftfalse
\ifdraft
\documentclass[manuscript]{aastex63}
\else
\documentclass[twocolumn]{aastex63}
\fi

\usepackage{amsmath,here,afterpage}
\usepackage{longtable}
\usepackage{multirow,lineno}
\usepackage{comment}

\ifdraft
\pagewiselinenumbers
\fi

\received{}
\revised{}
\accepted{}
\submitjournal{ApJ}

\tighten
\shorttitle{Physical Properties of Dense Gas Clumps and Probability of Star Formation}
\shortauthors{Tanaka et al.}
\tighten

\usepackage{float}
\input{defines.etex}

\ifdraft
\fi

\begin{document}

%%%%%%%%%%%%%%%%%%%%%%%%%%%%%%%%%%%%%%
%%\baselineskip=7mm %%!!!!!!!!!!!!!!!!!
%%%%%%%%%%%%%%%%%%%%%%%%%%%%%%%%%%%%%%

\title{HCN $J$=4--3, HNC $J$=1--0, $\mathrm{H^{13}CN}$ $J$=1--0, and $\mathrm{HC_3N}$ $J$=10--9 Maps of Galactic Center Region II: Physical Properties of Dense Gas Clumps and Probability of Star Formation}

\author{Kunihiko Tanaka}
\email{ktanaka@phys.keio.ac.jp}
\affil{Department of Physics, Faculty of Science and Technology, Keio University, 3-14-1 Hiyoshi, Yokohama, Kanagawa 223--8522 Japan}

\author{Makoto Nagai}
\affil{Advanced Technology Center, National Astronomical Observatory Japan, 2-21-1 Osawa, Mitaka, Tokyo 181-8588, Japan}
\author{Kazuhisa Kamegai}
\affil{Astronomy Data Center, National Astronomical Observatory Japan, 2-21-1 Osawa, Mitaka, Tokyo 181-8588, Japan}
\author{Takahiro Iino}
\affil{Information Technology Center, The University of Tokyo, 2-11-16, Yayoi, Bunkyo, Tokyo 113-8658, Japan}
\author{Takeshi Sakai}
\affil{Graduate School of Informatics and Engineering, The University of Electro-Communications, 1-5-1 Chofugaoka, Chofu, Tokyo 182-8585, Japan}


%somechange

\begin{abstract}
We report a statistical analysis exploring the origin of the overall low star formation efficiency (SFE) of the Galactic central molecular zone (CMZ) and the SFE diversity among the CMZ clouds using a wide-field HCN~$J$=4--3 map, whose \rv{optically thin} critical density ($\sim10^7\,\mathrm{cm}^{-3}$) is the highest among the tracers ever used in CMZ surveys.
Logistic regression is performed to empirically formulate star formation probability of 195 HCN clumps, 13 of which contain star formation signatures.
The explanatory parameters in the best-fit model are reduced into the virial parameter $\alpha_{\mathrm{vir}}$ without significant contribution from other parameters, whereas the performance of the model without $\alpha_{\mathrm{vir}}$ is no better than that using randomly generated data.
The threshold $\alpha_{\mathrm{vir}}$ is 6, which translates into a volume density ($n_{\mathrm{H_2}}$) of $10^{4.6}\,\mathrm{cm}^{-3}$ with the $n_{\mathrm{H_2}}$--$\alpha_{\mathrm{vir}}$ correlation.
The scarcity of the low-$\alpha_{\mathrm{vir}}$ clumps, whose fraction to all HCN clumps is 0.1, can be considered as one of the immediate causes of the suppressed SFE.
\rv{No correlation between the clump size or mass and star formation probability is found, implying that HCN~$J$=4--3 does not immediately trace the mass of star-forming gas above a threshold density.}
Meanwhile, star-forming and non-star-forming clouds are degenerate in the physical parameters of the CS~$\mathit{J}$=1--0 clouds, highlighting the efficacy of the HCN~$\mathit{J}$=4--3 line to probe star-forming regions in the CMZ.
The time scale of the high-$\alpha_{\mathrm{vir}}$ to low-$\alpha_{\mathrm{vir}}$ transition is $\lesssim2$ Myr, which is consistent with the tidal compression and X1/X2 orbit transition models but possibly does not fit the cloud--cloud collision picture.

\end{abstract}

%%%%%%%%%%%%%%%%%%%%%%%%%%%%%%%%%%
%
% INTRODUCTION
%
%%%%%%%%%%%%%%%%%%%%%%%%%%%%%%%%%%

\newcommand\nclumps{206}

\section{INTRODUCTION\label{section:introduction}}
The central molecular zone (CMZ) within 200 pc radius from the
Galactic nucleus provides a valuable opportunity to closely observe star formation (SF) under the harsh
environment of the galactic central region, which is characterized by elevated
gas temperature, density, and turbulent pressure.  Recent measurements
of the overall star formation rate (SFR) of the CMZ consistently
indicate inefficient SF in the CMZ, irrespective of the methods
used \citep[][and references therein]{Yusef-Zadeh2009,Longmore2013,Barnes2017}.
A majority of the giant molecular clouds (GMCs) in the CMZ lack signatures
of active SF despite their large mass and high density that are
comparable to massive SF regions in the Galactic disk regions \citep{Longmore2012b,Rathborne2014a,Kruijssen2014,Kauffmann2017a}.
Thus, the Galactic CMZ is considered as a region of suppressed SF that does not follow
the scaling relation between SFR and dense-gas mass \citep{Gao2004,Lada2012a}.

\begin{revision}
Meanwhile, individual regions in the CMZ exhibit a wide range of SF efficiency (SFE).
In contrast to the above-mentioned SF-less GMCs represented by the brick cloud \citep{Longmore2012b,Rathborne2014a},
the Sgr B2 complex harbors the most active cluster formation in the Galaxy \citep[e.g.,][]{Ginsburg2018}.
The presence of the three young massive clusters, the Arches, Quintuplet, and Central clusters \citep[e.g.,][]{Nagata1995,Morris1996,Figer1999,Lu2013}, suggests such burst-like
SF activity is not an exceptional event in the CMZ when viewed over the time scale of a few Myr, whereas such extreme SF is rare elsewhere in the Galaxy.
\end{revision}
This evokes a question about the origin of such SF diversity among 
the clouds under a relatively uniform environment of the CMZ.

Previous studies have revealed that the inefficient SF of the majority of the CMZ clouds
is likely to be related to their low degree of self-gravity that is atypical in solar
neighborhood SF regions.
Quiescent clouds in the CMZ are observed to have flat density profiles
\citep{Kauffmann2017a}, column density PDF without a pronounced power-law tail \citep{Rathborne2014a},
and generally high virial parameters ($\avir \equiv \Mvt/M$) of $\sim 5-10$ \citep{Miyazaki2000}.
In terms of the volumetric SF theory \citep{Krumholz2005,Padoan2011}, a low degree of self-gravity elevates the
threshold density for self-gravitational instability and lowers the
SFE per free-fall
time \citep{Rathborne2014a,Kruijssen2014,Schruba2019}.

As for the origin of the active SF sites, external triggering
mechanisms related to molecular gas dynamics in 100-pc
scale \citep{Sofue1995,Sawada2004,Henshaw2016} are frequently
discussed.
The tidally-triggered SF model \citep{Longmore2013,Kruijssen2015}\ assumes
that collapse of GMCs is induced through the tidal interaction with the stellar distribution near \sgras\ at the pericenter passage.
Accumulation of molecular clouds at the
X1--X2 orbit crowded region in the bar potential
\citep{Binney1991,Hasegawa1994} is also a popular theory for the formation
of the burst-like SF in Sgr B2 and Sgr C.
In this line of theories, the apparent lack of SF signatures may not
necessarily indicate low SFE, but could result from a short
duration time of the SF phase after the cloud experienced the
trigger \citep{Barnes2017}.

Frequent cloud--cloud collisions (CCCs) due to the crowdedness of
dense gas \citep{Hasegawa1994,Tan2000,Jeffreson2018} can also affect
the dynamical state and SFE.  CCC is argued to facilitate
SF in Sgr B2 and the 50-\kmps\ cloud
\citep{Hasegawa1994,Tsuboi2015,Uehara2019} and possibly future high-mass
star formation in the Brick cloud \citep{Higuchi2014}.  On the other
hand, recent observations have found that a significant fraction of
CCC candidate sites in the CMZ lacks on-going
SF \citep{Tanaka2018a,Enokiya2019}; in those clouds, CCC may even
enhance turbulent pressure and stabilize the clouds against
self-gravitational collapse.

Accurate knowledge of mass, distribution, and physical conditions of
dense molecular gas is critical to comprehensively understand the SF in the CMZ including the above pictures.
In the context of the SF study, the dense gas is usually defined as gas with $\nHH \gtrsim 10^4\ \pcc$, in the sense that
mass above this SF critical density (\rv{\ncritSF}) is known to be closely correlated with SFR in various scales \cite[e.g.,][]{Lada2012a} \rv{outside the Galactic central environment};
however, as readily known from the observational fact that $10^4\ \pcc$ is no higher than the mean density of CMZ clouds,
the \ncritSF\ is likely to be substantially elevated in the CMZ. 
Theories predict that the generally high turbulent pressure in the CMZ may increase \ncritSF\ to $\sim 10^7\ \pcc$ \citep{Rathborne2014a,Kruijssen2014}.
\begin{revision}
Although the HCN \JJ{1}{0}\ is often used as a tracer of high-density gas closely associated with SF \citep[e.g.,][]{Gao2004,Lada2012a}, the elevated \ncritSF\ in the CMZ could suggest
that molecular lines with higher critical density for excitation (\ncrit) than that of HCN \JJ{1}{0}\ are more advantageous to investigate SF gas in the CMZ.
\end{revision}

\begin{revision}
The dense gas distribution in the CMZ has been investigated using many millimeter to submillimeter molecular lines \citep{Tsuboi1999,Miyazaki2000,Nagai2007,Oka2012,Jones2012,Jones2013,Ott2014,Ginsburg2016,Mills2018}.
In this series of papers, we report a new CMZ survey observation using
the submillimeter HCN \JJ{4}{3}\ as the main tracer, whose 
optically-thin \ncrit
\footnote{\rv{
We define the optically-thin \ncrit\ by Equation 4
of \cite{Shirley2015}; $\ncrit \equiv A_{i,j} / \sum_{k \neq
i} \gamma_{i,j}$ for the $i \rightarrow j$ level transition, where $A$
and $\gamma$ are the Einstein $A$ coefficient and collisional
excitation rate, respectively.}}
of $2\times 10^7\ \pcc$.
This does not indicate that HCN \JJ{4}{3}\ exclusively traces the SF gas with $\nHH\gtrsim 10^7\ \pcc$, 
because the $J$=4 level of HCN is excited at $\nHH \sim 10^6\ \pcc$ by the photon-trapping effect at the high optical depth of $\sim 10$--$30$ typical in the CMZ environment \citep{Tanaka2018b}.
The efficiency as a high-density tracer is further limited by the beam-dilution effect.
In fact, the excitation analysis \citep{Tanaka2018b}\ shows that the HCN \JJ{4}{3}\ emission mostly originates from volumes with $\nHH\sim10^{3.4\mbox{--}4.8}\ \pcc$ when viewed with a resolution of $60''$ (=2.4 pc).
Nonetheless, the effective \ncrit\ of the HCN \JJ{4}{3}\ line is among the highest in those of the tracers that have been used in previous dense gas surveys toward the CMZ
\footnote{\rv{
The only comparable high-density tracer is \HCCCN\ \JJ{23}{22}\  used by \cite{Mills2018}, whose
optically thin \ncrit\ is $1\times10^6\ \pcc$;  however, its bright emission is detected only near the central regions of a few massive GMC complexes,
presumably due to the low column density of \HCCCN\ and the relatively high upper state energy of the $J=23$ level
($\Eu/\kB = 130\ \kelvin$).
}}.
As the HCN \JJ{4}{3}\ line is widespread over the major GMC complexes in the CMZ owing to the high HCN abundance, it serves as an efficient probe of high-density gas
that is not fully explored previously.
Although the HCN \JJ{4}{3}\ luminosity may not be used to trace the mass of dense gas with $\nHH \gtrsim \ncritSF$ for the abovementioned reasons,
we may able to identify dense gas structures that are associated with SF in the HCN \JJ{4}{3}\ image, and to probe their fundamental physical properties.
\end{revision}

In our first paper (\citealt{Tanaka2018b}; hereafter Paper I), we
presented the HCN \JJ{4}{3}\ map obtained with the Atacama
Submillimeter Telescope Experiment (ASTE), along with the maps of the
HNC, \HCNt, and \HCCCN\ lines in the 3-mm band obtained with the
Nobeyama Radio Observatory (NRO) 45-m telescope.  We reduced these new
data and other archival data into maps of column density, volume
density, temperature, and the fractional abundances of the molecular
species involved, by utilizing the excitation analysis code using the
hierarchical Bayesian framework.
Based on these results, we showed that the physical
condition and chemical composition in the CMZ clouds are predominantly
determined by the widespread shocks.

In this second paper, we present a statistical analysis investigating 
the effect of the physical conditions in the CMZ clouds on their SF activities. 
Our goal is to find the physical parameters that are
best correlated with the SF capability of individual molecular clumps
using a purely objective method. 
A similar analysis has been reported for the M33 galaxy by  
\cite{Komugi2018}, who successfully applied a principal component analysis (PCA)
 to formulate the SFR using other observed physical quantities of GMCs.
In this paper, we use the logistic regression (LR) analysis to relate SF activity
to the physial parameters of CMZ clouds.
The LR analysis parametrizes the SF activity using a binary flag representing the presence of SF signatures,
instead of SFR, which is difficult to measure accurately for th 0.1--1 pc scale clumps
identified in our HCN survey.
LR analysis gives the best parameter combination in discriminating two groups, whereby
we can explore the physical conditions responsible for the diverse SF activities among the CMZ clouds.

The rest of this paper is structured as below.  We first briefly summarize
the data and analysis in Paper I in Section \ref{section:data}.  In
Section \ref{section:clumpid}, we identify dense gas clumps in the
HCN \JJ{4}{3}\ cube, and their physical parameters are
calculated.  In Section \ref{section:pca}, PCA is applied to investigate
the intercorrelation among the clump parameters and 
to find the fundamental parameter combination.
We also show that many of the known observational laws interrelating the clump parameters are
immediately derived from the result of PCA.
Section \ref{section:logisticregression} describes the logistic
regression analysis to find the parameter that discriminates
star-forming clumps from quiescent clumps.
\begin{revision}
We apply the analysis to three data sets: the HCN \JJ{4}{3}\ clumps, the CS clouds with lower
density \citep{Miyazaki2000}, and control data generated using mock SF
signals.
\end{revision}
Upon these results, we discuss the SF in the CMZ in
Section \ref{section:discussion}.  A summary of the paper is given in
Section \ref{section:summary}.

\section{DATA\label{section:data}}
This paper uses the dense gas maps of the CMZ presented in Paper I.
The HCN \JJ{4}{3}\ data were obtained in 2010--2013 utilizing the CATS345 receiver installed on the ASTE 10-m telescope. 
The \HCNt\ \JJ{1}{0}, HNC \JJ{1}{0}, and \HCCCN\ \JJ{10}{9}\ data were simultaneously taken with the BEARS receiver array of the NRO 45-m telescope in 2010.
The ASTE and NRO observations were conducted in the On-The-Fly mode to produce Nyquist-sampled maps with $24''$ and $20''$ effective beamsizes, respectively.
They cover the four major GMC complexes in the CMZ, Sgr A--D, except for the gap region between the Sgr B and Sgr D complexes being not observed in HCN \JJ{4}{3}. 

We also use the results of the excitation analysis in Paper I.
Using the above and 9 other molecular line data taken from literature, we made \PPV\ three-dimensional maps of the hydrogen column density (\NHH), volume density (\nHH), gas kinetic temperature (\Tkin), and the fractional abundances of the molecular species involved.
The large velocity gradient (LVG) approximation \citep{Goldreich1974} was used in solving the non-LTE excitation equations.
The coefficients of the collisional and radiative transitions were taken from the Leiden Atomic and Molecular Database \citep[LAMDA;][]{Schoier2005}. 

As the framework of the parameter inference, we developed a method using the hierarchical Bayesian (HB) technique.
The HB method is able to consider errors from the calibration uncertainties and the deficiency of the one-zone model that cannot be treated with the standard maximum-likelihood (ML) method, by putting additional constraints from prior and hyper-prior probabilities.
The new HB method successfully suppressed the artificial correlation among the inferred parameters, which was found to affect the ML analysis critically.
We calculated the mean values and errors of the model parameters on the marginal posterior probabilities at individual voxels, and composed them into \PPV\ cubes where voxels with an uncertainty greater than 0.2 dex in the base-10 logarithms are blanked out.
For further details of the molecular line data and the analysis, we refer readers to Paper I.

\section{CLUMP IDENTIFICATION\label{section:clumpid}}
\renewcommand{\nclumps}{195}

\subsection{Method}
We identify molecular clumps in the HCN \JJ{4}{3}\ data cube, by employing the CLUMPFIND algorithm \citep{Williams1994}\ with a minor modification to adjust to the highly crowded environment of the CMZ.

The CLUMPFIND algorithm defines clumps as local peaks and contiguous voxels within contours enclosing them.  Contours enclosing more than one local peaks are partitioned and assigned to the peak at the closest \PPV\ distance.
The standard algorithm continues these steps starting from the highest contour level until all voxels within the lowest significant contour are assigned to either clump.
However, the partitioned contours often create irregular-shaped extensions in the identified clumps, due to the crowdedness of local peaks and high volume-filling factor in the CMZ;
this affects the accuracy of the estimates of the sizes and velocity dispersion assuming an ellipsoidal shape.

\begin{figure}
\ifdraft
\epsscale{0.5}
\else
\epsscale{1.}
\fi
\plotone{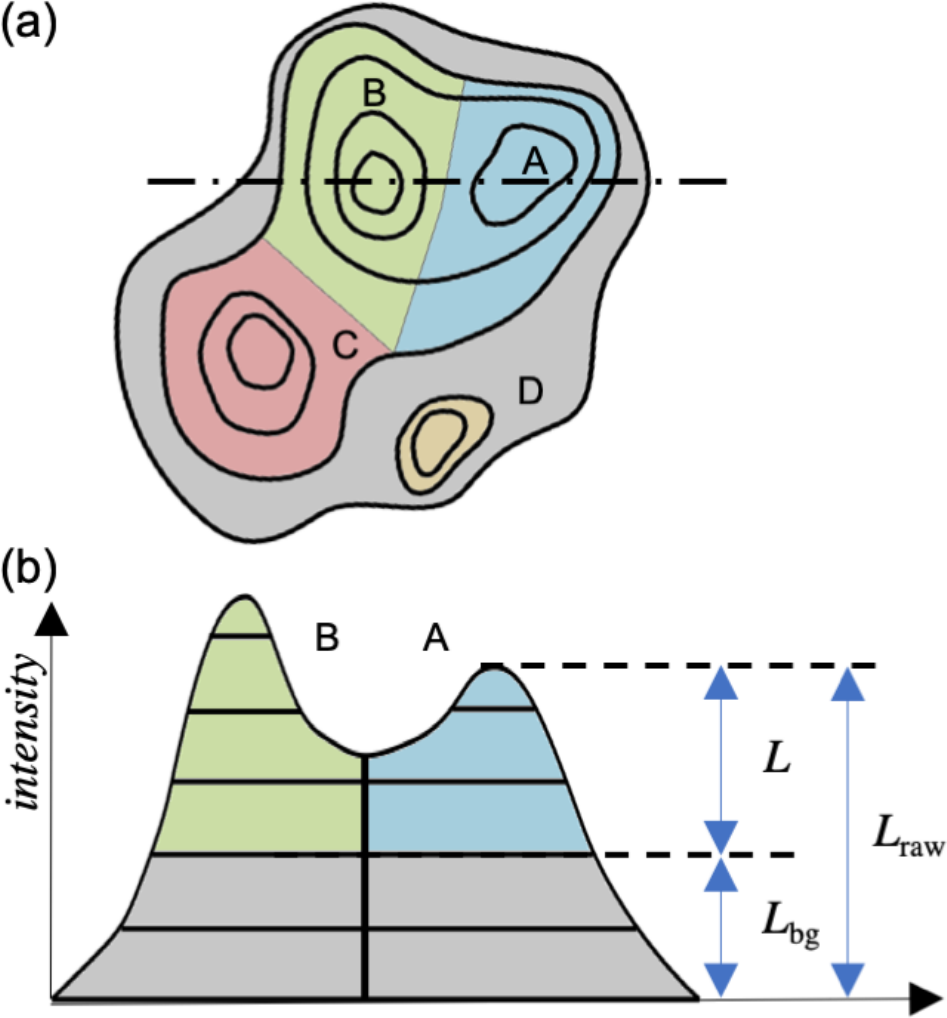}
\caption{ (a) Schematic drawing of the clump identification algorithm
  using an example of two-dimensional contours. (b) Definition of the
  clump luminosity ($L$) and the background luminosity ($L_{\rm bg}$)
  on the one-dimensional intensity profile along the dashed line on panel
  a. \label{fig:clid}}
\end{figure}

\begin{figure}
\begin{revision}
\ifdraft
\epsscale{0.5}
\else
\epsscale{1.}
\fi
\plotone{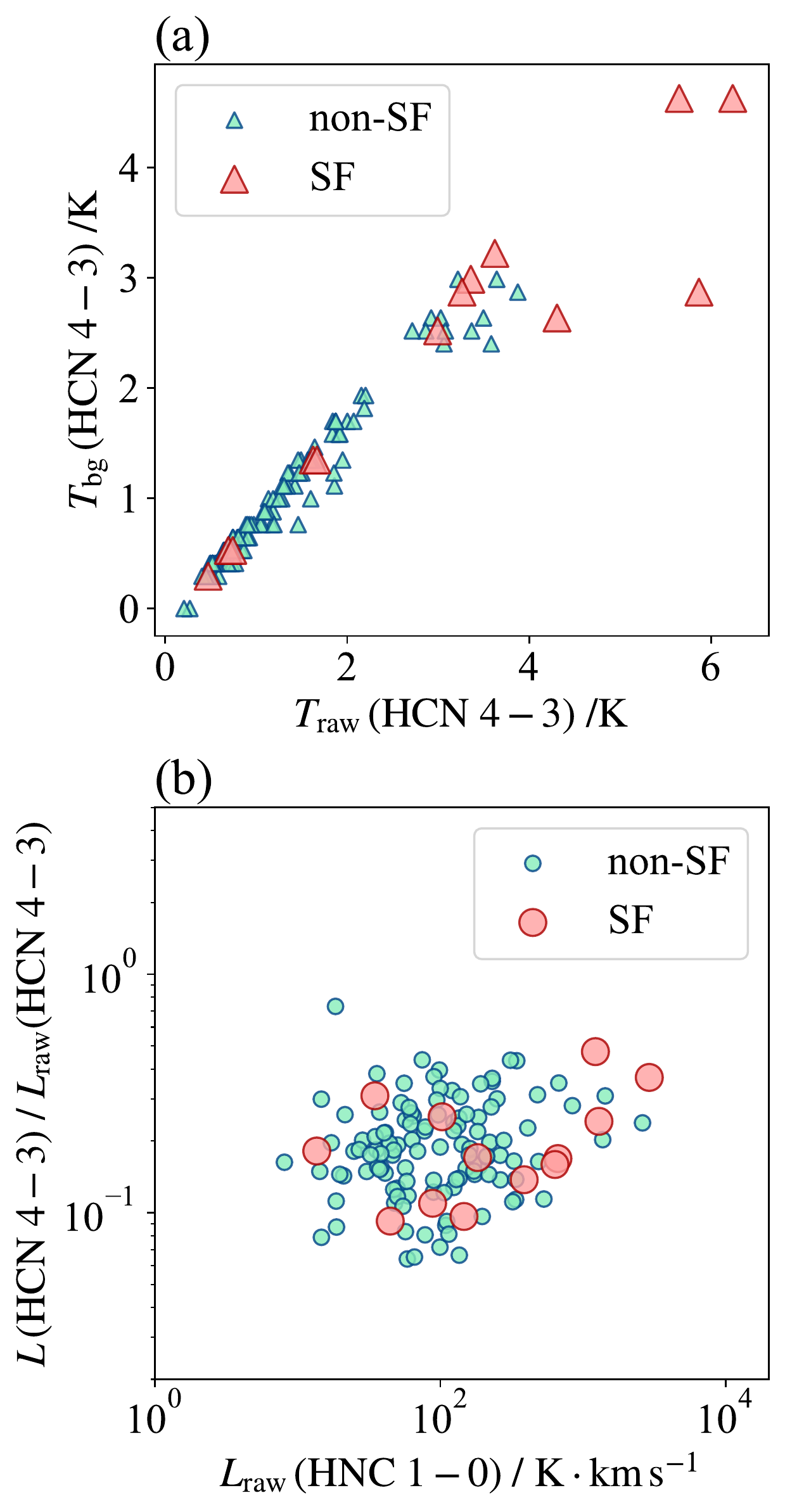}
\caption{\rv{ (a) Background emission levels ($T_{\rm bg}$) plotted against the average intensities ($T_{\rm raw}$) including both the background and clump emissions.   The smaller blue and larger red symbols are for the non-SF and SF clumps, respectively, whose definitions are given in \S\ref{section:SFsignatures}.  (b) Ratios of the clump HCN \JJ{4}{3}\ luminosity ($L\left(\rm HCN\,\JJ{4}{3}\right)$ defined by Equation \ref{eqn:clumplum}) to the raw luminosity $L_{\rm raw}\left(\rm HCN\,\JJ{4}{3}\right)$ plotted against clump luminosity.  The  luminosities are measured as summations of the integrated intensity within the lowest contour of the individual clumps.   The smaller blue and larger red symbols are for the non-SF and SF clumps, respectively, whose definitions are given in \S\ref{section:SFsignatures}.} \label{fig:bgratio}}
\end{revision}
\end{figure}

In order to avoid this problem, we made a minor modification so that the CLUMPFIND stops partitioning contours at a given background emission level.
We define the background level as the contour level at which the number of the enclosed peaks is more than three;
the emission below the threshold is regarded as a background that does not belong to either clump.
We give the schematic concept of the algorithm in Figure \ref{fig:clid}.
We confirmed that the modified algorithm effectively reduced the number of irregular-shaped clumps when it was applied to our HCN data.
We note that the above definition of the background level was found from trial-and-error without physical or statistical ground, and that the modified method has not been tested with other data. 
Nonetheless, as we will see \S\ref{section:pca}, the identified HCN clumps are found to follow the same observational scaling relations as those known from previous studies;
therefore, we can consider that our modified algorithm was able to extract physically meaningful structures with sufficient accuracy.
\rv{
The background intensity levels ($T_{\rm bg}$) of individual clumps are shown in Figure \ref{fig:bgratio}a, plotted against the average intensities ($T_{\rm raw}$) that include both the background and clump emissions. 
}

We apply the modified algorithm to the HCN \JJ{4}{3}\ cube  convolved with a $25''\times25''\times4\ \kmps$ three-dimensional Gaussian kernel to smoothe out spike noises.
The lowest contour level and contour inclement are chosen to be 3$\sigma$ and 2$\sigma$, respectively, where 1$\sigma = 0.08\ \kelvin$ for the kernel-smoothed data.
The automatic algorithm identified \rv{472} clumps, from which we finally adopt \nclumps\ clumps after manually filtering-out irregular-shaped clumps, spurious identifications of spectral spike-like features, and unnaturally broad-velocity features created by the spectrometer baseline that was not removed in the data reduction.
The identified clumps are listed in Table \ref{table:clumplist}.

The clump list includes a few Galactic disk sources, which are distinguished by the lack of velocity gradient in the Galactic longitude direction and the narrow line widths that are atypical in the CMZ clouds.
However, galactic disk clouds are mostly transparent in HCN \JJ{4}{3}, since density higher than $J=3$ critical density ($\sim 10^7\ \pcc$) is limited to SF cores in the Galactic disk region; indeed, our list includes only two Galactic disk sources (clump id \#37 and \#94).
We exclude them from the analysis in the following part of the paper.

\subsection{Physical Parameters\label{section:massEstimate}}
Table \ref{table:clumplist} tabulates the first moment \PPV\ positions, FWHM sizes in the major and minor axes (\rmin, \rmaj), position angles, velocity widths in FWHM (\Dv), averaged line intensities, masses ($M$), virial parameters (\avir), and the physical condition parameters (\Tkin\ and \nHH).
We define the position angle by the direction along which the intensity-weighted spatial dispersion is the largest.
The clumps sizes and velocity widths are measured with one-dimensional Gaussian fitting of the intensity distribution collapsed on the respective axes.
The \PPV\ sizes are measured in using the original cube (i.e., the data before the kernel smoothing) and corrected for the finite spatial and velocity resolutions by using the approximation
$x_{\rm corr} \sim \sqrt{x_{\rm raw}^2 - (\Delta x_{\rm FWHM}/2)^2 }$,
where $x_{\rm corr}$ and $x_{\rm raw}$ are corrected and observed radii, respectively, and $\Delta x_{\rm FWHM}$ is the resolution in FWHM.
In translating the angular size to the linear scale, \rv{we adopt a distance of 8.18 kpc \citep{GravityCollaboration2019}.}

The HCN \JJ{4}{3}\ luminosity is obtained by summing the voxel values enclosed in the clump, where the background intensity (i.e., the contour level at which the clump identification stopped; see Figure \ref{fig:clid}b) is subtracted.
Since the background emission levels are undefinable for the lines other than HCN \JJ{4}{3}, we adopt a crude approximation that the ratio of the background to the clump emission is the same for all
lines;
then the luminosity of line $X$, $L(X)$, is written as
\begin{eqnarray}
  L(X)&=& L_{\rm raw}(X) \cdot \frac { L(\HCN \JJ{4}{3}) } { L_{\rm
      raw}(\HCN \JJ{4}{3}) }, \label{eqn:clumplum}
\end{eqnarray}
where the subscript `raw' is for the raw luminosity that includes both of the clump and background components.
The averaged intensity is obtained by dividing the luminosity by the projected surface area $\sim \pi\,\rmaj\cdot\rmin$.

\begin{revision}
  Figure \ref{fig:bgratio}b\ shows the ratio of $L(\HCN \JJ{4}{3})$ to $L_{\rm raw}(\HCN \JJ{4}{3})$ plotted against $L_{\rm raw}(\HNC \JJ{1}{0})$.
  The average ratio is 0.28; namely, approximately 70\% of the HCN \JJ{4}{3}\ emission enclosed within the projected boundary of the clumps is attributed to spatially extended background emission.
  The sum of the emission within the boundary of any clumps constitutes 33\% of the entire observed luminosity.  
  In total, $\sim 10\%$ of the observed \HCN\ \JJ{4}{3}\ emission is judged as the clump emission by our algorithm,
  due to the exclusion of the extended emission.
  The clump emission area contains nearly all SF signatures known in the observed spatial and velocity ranges (\S\ref{section:SFsignatures}), indicating that our method was able to capture dense gas structure that is closely associated with SF activity in the CMZ.
  No systematic difference in the  $L(\HCN \JJ{4}{3})$/$L_{\rm raw}(\HCN \JJ{4}{3})$ ratio is found between SF and non-SF clumps in Figure \ref{fig:bgratio}b.
  By using the $r$--$M$ and $r$--$\Dv$ relationships (\S\ref{section:pc1}), the exclusion of the extended background emission results in the $r$ and $\Dv$ values that are approximately 0.6 and 0.8 times those that would be obtained with the standard method including the extended background emission, respectively.
\end{revision}

\newcommand{\rclfg}{\ifmmode R_{\rm HNC/HCN}\else$R_{\rm HNC/HCN}$\fi}
The cloud masses are calculated from\ the HNC \JJ{1}{0}\ luminosity, which we found is the best gas mass tracer in our dataset (Paper I).
We use the $X$-factor for the luminosity-to-mass conversion, $\NHH/I(\HNC) = 10^{21.1}\ \psc/(\kelvin\cdot\kmps)$, based on the dust-based mass estimate (Paper I).
The virial parameter \avir\ is calculated on the definition by \cite{Bertoldi1992} :
\begin{eqnarray}
  \avir &\equiv& \frac{5r\sigma_v^2}{GM} \nonumber \\ &=&
  210\cdot\frac{\sqrt{\rmaj\cdot\rmin}}{\mathrm{pc}}\cdot\biggl(\frac{\Dv}{\kmps}\biggr)^2
  \biggl(\frac{M_{\phm{\odot}}}{\Msol}\biggr)^{-1}.
\end{eqnarray}
\begin{revision}
  This mass and \avir\ estimates rely on the assumption of the equal clump-to-background emission ratio for HNC and HCN \JJ{4}{3}, which is equivalent to the assumption of the same HNC \JJ{1}{0}/HCN\ \JJ{4}{3}\ intensity ratio (\rclfg) between the clump and background emissions.
  We calculate \rclfg\ for the voxels inside and outside the clumps; the former contains both the clump emission and the overlapping extended background emission, while the latter purely consists of extended emission.   The averages and standard deviations of $\log_{10}\,\rclfg$ are $-0.0\pm0.3$ and $0.2\pm0.3$ for the voxels inside and outside the clumps, respectively.
  This indicates that the  HCN \JJ{4}{3}\ intensity is enhanced by 0.2 dex in the clumps, though the degree of enhancement is within the range of the overall \rclfg\ variation.
  We do not further explore the impacts of the \rclfg\ variation on the mass estimate, as the variation is not much greater than the uncertainty in the HNC luminosity-to-mass conversion factor \citep{Tanaka2018b}.  It is also unlikely that the omission of the \rclfg\ variation creates artificial correlations between the clump masses and other parameters in the analyses in \S\ref{section:pca} and \S\ref{section:logisticregression}, as the clump-to-clump variation in \rclfg\ is smaller than that in the HNC luminosity (i.e., the clump mass) and no systematic difference in \rclfg\ is present between SF and non-SF clumps.
  \end{revision}

\begin{revision}
  For the physical condition parameters (\Tkin\ and \nHH), we referred to Paper I, 
  because it calculates \PPV\ distribution of \nHH\ over the almost entire CMZ, which is unknown in other works regarding with the physical conditions of the CMZ clouds \cite[e.g.,][]{Ginsburg2016,Krieger2017}.
\end{revision}
  We adopted the values at the closest positions to the clump centers in the \Tkin\ and \nHH\ maps, as the excitation analysis is made with a different \PPV\ gridding from that of the clump identification.
The physical conditions are unknown for 65 clumps, which are outside the coverage of the excitation analysis;
their entries are left blank in Table \ref{table:clumplist}.

\begin{revision}
  The analysis in Paper I adopts the one-zone approximation, although observations have shown that the CMZ clouds generally consist of two components with different \Tkin\ and \nHH\ \citep{Huettemeister1993,Arai2016,Krieger2017,Mills2017,Tanaka2018b}:
  one with $\Tkin\sim 20$--$40\ \kelvin,\nHH\sim 10^{3-4}\ \pcc$ and the other with $\Tkin\sim 10^2\ \kelvin, \nHH\gtrsim 10^{4-5}\ \pcc$ \citep{Nagai2007,Arai2016,Mills2017,Tanaka2018b}.
  The \Tkin\ and \nHH\ values calculated in Paper I are averages over the multi-phase gas, with larger weights for the high-density (and high-temperature) component (\S6.2 of Paper I);  the ratios of the intensities of the high-density component to those of the low-density components are estimated to be $\gtrsim 3$.   This indicates that the values from Paper I predominantly traces the physical conditions of the high-density component.  However, they may somewhat underestimate \Tkin\ and \nHH\ of the HCN clumps, as the observed HCN \JJ{4}{3}\ emission originates almost purely from the high-density component.
\end{revision}

\subsection{Comparison with Other Catalogs\label{section:otherCats}}

We utilize two published cloud catalogs for comparison with our CMZ clumps.
The Galactic ring survey catalog (GRS;\citealt{Jackson2006,Rathborne2009}) is based on the FCRAO \COt\ \JJ{1}{0}\ survey covering an \gl\ range of $18^\circ$ -- $55^\circ.7$,
where the UMSB $^{12}{\rm CO}$ \JJ{1}{0}\ survey \citep{Sanders1986,Clemens1986} is incorporated for calculating the physical properties of the clouds.
We use the GRS catalog as a template of cloud in a typical Galactic disk environment whose \nHH\ is $\sim 10^{2-3}\ \pcc$.

\begin{revision}
  As a template of the CMZ clouds, we use the CS \JJ{1}{0}\ cloud catalog \citep[hereafter MT00 clouds;][]{Miyazaki2000} identified based on the NRO 45-m observations \citep{Tsuboi1999,Miyazaki2000}.
  \end{revision}
The CS \JJ{1}{0}\ critical density is $10^{4.8}\ \pcc$, which is closer to the typical density of the CMZ clouds than the HCN \JJ{4}{3}\ critical density is.
Hence, the MT00 clouds can be regarded as a standard of CMZ clouds with a `moderate' density, in comparison to the high-density clumps identified in the HCN \JJ{4}{3}\ map.
Out of the \nclumps\ HCN clumps, 62 have counterparts in the MT00 catalog;
\begin{revision}
  in the cross identification, HCN and MT00 clouds whose \PPV\ distance is less than the radius of the larger cloud are identified as counterparts to each other.   
  The other 133 HCN clumps are associated with spatially extended CS emission that is not identified as clump emission with their criteria (we refer readers to \S2.1 of \citealt{Miyazaki2000} for details of their clump identification).
  In all cases, the MT00 clouds are greater than the counterpart HCN clumps.  We find that 14 MT00 clouds encompass more than one HCN clumps.
\end{revision}

A comparison between the HCN clumps, GRS clouds, and MT00 clouds is made on the size--mass and size--velocity width diagrams in Figures and \ref{fig:sizemass} and \ref{fig:sizevel}, in which we find that the HCN clump list and the MT00 catalog cover different size scales of the molecular gas structure in the CMZ.
The radii of the HCN clumps span from a few 0.1 to 1 pc in radius, \rv{whereas those of the MT00 clouds span from 1 to 10 pc.  Their median radii differ by 0.8 dex.
  This difference is greater than the differences in the resolution beam size ($24''$ and $34'$ for the HCN and CS observations, respectively) even when the effect of the omission of the extended emission in the HCN data is considered.
  The cloud/clump sizes may also depend on the signal-to-noise levels of the maps, but
  the signal-to-noise ratio of the MT00 data (130 at the peak position of the 50-\kmps\ cloud) is close to or slightly higher than that ot the HCN data (90).
  Hence, the different sizes the MT00 clouds and the HCN clumps likely reflect the different size scales the HCN \JJ{4}{3}\ and CS \JJ{1}{0}\ observations are sensitive to.
}
We may regard the identified HCN \JJ{4}{3}\ structures as `clumps-scale' structures in comparison to the `cloud-scale' structures identified by MT00.
The GRS clouds cover a wider range of size scales from 0.1 to 10 pc, which contains those of the HCN clumps and the MT00 clouds.

\newif\ifcompilable
\compilablefalse
\ifcompilable
\fi

\newcommand\lunit{$\kelvin\,\kmps$}
\begin{longrotatetable}
\begin{deluxetable}{cccccccccccccccccc}
\tablecaption{HCN clumps \label{table:clumplist}}
\centering
\tablecolumns{18}
\tabletypesize{\scriptsize}

\tablehead{
 \colhead{}\vspace{-2.5ex} &  & &  & \colhead{} & \colhead{} & \colhead{} & \colhead{} & \multicolumn{4}{c}{$\int \Tmb\,\dv$} &  \colhead{\logt} & \colhead{\logt} & \colhead{\logt} & \colhead{\logt} & \colhead{} & \colhead{} \\
 \colhead{\#}\vspace{-4.ex}  & \colhead{$\gl$} & \colhead{$\gb$} & \colhead{$\vlsr$} & \colhead{$r_{\mathrm maj}$} & \colhead{$r_{\mathrm min}$} & \colhead{PA} & \colhead{$\Dv$}& &&& &  \colhead{} & \colhead{} & \colhead{} & \colhead{} & \colhead{\fsf\tablenotemark{\scriptsize a}} & \colhead{notes} \\
 &  \colhead{} & \colhead{} & \colhead{}  & & & & & \colhead{HCN} & \colhead{\HCNt} & \colhead{HNC}  & \colhead{\HCCCN} & \colhead{$M$} & \colhead{\avir}& \colhead{\nHH}& \colhead{\Tkin} && \\
\cline{9-12} 
\colhead{} & \colhead{(deg.)} & \colhead{(deg.)} & \colhead{($\kmps$)} & \colhead{($\mathrm{pc}$)} & \colhead{($\mathrm{pc}$)} & \colhead{(deg.)} & \colhead{($\kmps$)} & \multicolumn{4}{c}{(\lunit)} & \colhead{(\Msol)} & \colhead{} & \colhead{($\pcc$)} & \colhead{(K)}
}
\tablewidth{0pt}
\decimals
\colnumbers
\startdata
\phn\phn1\dotfill&${-0.587}$ & \phm{+}${0.017}$ & ${-106}$ & $0.15$ & $0.11$ & ${-27}$ & ${11.1}$ & ${\phn}{\phn}{3.5}$ & ${\phn}{1.3}{\phn}$ & ${\phn}{0.40}$ & ${\phn}{1.0}{\phn}$ & ${1.4}$ & ${2.1}$ & \nodata & \nodata & \nodata &  \\
\phn\phn2\dotfill&${-0.580}$ & \phm{+}${0.000}$ & ${\phn}{-86}$ & $0.72$ & $0.50$ & \phm{+}${\phn}{5}$ & ${\phn}{6.7}$ & ${\phn}{\phn}{4.2}$ & ${\phn}{2.6}{\phn}$ & ${\phn}{0.25}$ & ${\phn}{3.8}{\phn}$ & ${2.5}$ & ${1.3}$ & \nodata & \nodata & \nodata &  \\
\phn\phn3\dotfill&${-0.578}$ & ${-0.182}$ & ${\phn}{-32}$ & $0.43$ & $0.35$ & ${-46}$ & ${10.0}$ & ${\phn}{\phn}{3.7}$ & ${\phn}{1.7}{\phn}$ & ${\phn}{0.52}$ & ${\phn}{3.2}{\phn}$ & ${2.2}$ & ${1.7}$ & \nodata & \nodata & \nodata &  \\
\phn\phn4\dotfill&${-0.578}$ & \phm{+}${0.018}$ & ${\phn}{-95}$ & $0.74$ & $0.31$ & \phm{+}${76}$ & ${\phn}{9.1}$ & ${\phn}{\phn}{2.8}$ & ${\phn}{1.4}{\phn}$ & ${\phn}{0.19}$ & ${\phn}{0.76}$ & ${1.7}$ & ${2.3}$ & \nodata & \nodata & \nodata &  \\
\phn\phn5\dotfill&${-0.576}$ & ${-0.175}$ & ${\phn}{-75}$ & $0.49$ & $0.46$ & ${-40}$ & ${\phn}{9.5}$ & ${\phn}{\phn}{5.3}$ & ${\phn}{1.9}{\phn}$ & ${\phn}{1.3}{\phn}$ & ${\phn}{5.3}{\phn}$ & ${2.5}$ & ${1.5}$ & \nodata & \nodata & \nodata &  \\
\phn\phn6\dotfill&${-0.562}$ & ${-0.001}$ & ${\phn}{-80}$ & $1.33$ & $0.79$ & ${\phn}{-9}$ & ${10.6}$ & ${\phn}{10.0}$ & ${\phn}{7.1}{\phn}$ & ${\phn}{1.7}{\phn}$ & ${10.6}{\phn}$ & ${3.3}$ & ${1.1}$ & \nodata & \nodata & \nodata &  \\
\phn\phn7\dotfill&${-0.558}$ & ${-0.107}$ & ${\phn}{-52}$ & $0.68$ & $0.50$ & \phm{+}${21}$ & ${10.2}$ & ${\phn}{26.9}$ & ${12.0}{\phn}$ & ${\phn}{7.6}{\phn}$ & ${21.3}{\phn}$ & ${3.2}$ & ${0.9}$ & \nodata & \nodata & mgw & Sgr~C \\
\phn\phn8\dotfill&${-0.550}$ & ${-0.124}$ & \phm{+}${\phn}{71}$ & $0.91$ & $0.68$ & \phm{+}${41}$ & ${10.8}$ & ${\phn}{\phn}{9.5}$ & ${\phn}{2.6}{\phn}$ & ${\phn}{1.2}{\phn}$ & ${\phn}{1.7}{\phn}$ & ${2.3}$ & ${2.0}$ & \nodata & \nodata & \nodata &  \\
\phn\phn9\dotfill&${-0.548}$ & ${-0.193}$ & ${\phn}{-44}$ & $0.99$ & $0.54$ & \phm{+}${\phn}{0}$ & ${\phn}{8.6}$ & ${\phn}{\phn}{7.4}$ & ${\phn}{3.1}{\phn}$ & ${\phn}{1.2}{\phn}$ & ${\phn}{4.4}{\phn}$ & ${2.7}$ & ${1.4}$ & ${4.3}$ & ${1.9}$ & \nodata &  \\
\phn10\dotfill&${-0.543}$ & ${-0.051}$ & ${\phn}{-99}$ & $1.38$ & $1.28$ & ${-30}$ & ${11.7}$ & ${\phn}{\phn}{6.9}$ & ${\phn}{4.3}{\phn}$ & ${\phn}{2.0}{\phn}$ & ${10.4}{\phn}$ & ${3.5}$ & ${1.1}$ & ${4.1}$ & ${1.9}$ & \nodata &  \\
\phn11\dotfill&${-0.543}$ & \phm{+}${0.020}$ & ${-125}$ & $0.70$ & $0.61$ & \phm{+}${46}$ & ${\phn}{6.7}$ & ${\phn}{\phn}{4.4}$ & ${\phn}{1.8}{\phn}$ & ${\phn}{0.43}$ & ${\phn}{1.4}{\phn}$ & ${2.1}$ & ${1.7}$ & \nodata & \nodata & \nodata &  \\
\phn12\dotfill&${-0.539}$ & ${-0.161}$ & ${\phn}{-58}$ & $0.50$ & $0.39$ & ${-61}$ & ${\phn}{6.4}$ & ${\phn}{\phn}{9.4}$ & ${\phn}{3.7}{\phn}$ & ${\phn}{1.4}{\phn}$ & ${\phn}{6.5}{\phn}$ & ${2.6}$ & ${1.0}$ & ${4.3}$ & ${1.9}$ & \nodata &  \\
\phn13\dotfill&${-0.535}$ & \phm{+}${0.003}$ & ${\phn}{-89}$ & $1.26$ & $1.08$ & \phm{+}${88}$ & ${11.2}$ & ${\phn}{\phn}{8.0}$ & ${\phn}{4.7}{\phn}$ & ${\phn}{1.1}{\phn}$ & ${10.1}{\phn}$ & ${3.4}$ & ${1.1}$ & \nodata & \nodata & \nodata &  \\
\phn14\dotfill&${-0.518}$ & ${-0.211}$ & ${\phn}{-34}$ & $0.31$ & $0.20$ & \phm{+}${45}$ & ${\phn}{9.8}$ & ${\phn}{\phn}{6.2}$ & ${\phn}{1.7}{\phn}$ & ${\phn}{0.85}$ & ${\phn}{2.2}{\phn}$ & ${1.9}$ & ${1.8}$ & \nodata & \nodata & \nodata &  \\
\phn15\dotfill&${-0.518}$ & \phm{+}${0.007}$ & ${\phn}{-84}$ & $0.67$ & $0.49$ & ${-53}$ & ${14.8}$ & ${\phn}{\phn}{9.1}$ & ${\phn}{5.4}{\phn}$ & ${\phn}{1.2}{\phn}$ & ${\phn}{6.8}{\phn}$ & ${2.7}$ & ${1.7}$ & \nodata & \nodata & \nodata &  \\
\phn16\dotfill&${-0.516}$ & ${-0.252}$ & ${\phn}{-56}$ & $1.11$ & $1.02$ & \phm{+}${66}$ & ${\phn}{6.3}$ & ${\phn}{\phn}{1.7}$ & ${\phn}{0.70}$ & ${\phn}{0.24}$ & ${\phn}{0.87}$ & ${2.2}$ & ${1.7}$ & \nodata & \nodata & \nodata &  \\
\phn17\dotfill&${-0.515}$ & ${-0.171}$ & ${\phn}{-56}$ & $1.13$ & $0.71$ & \phm{+}${17}$ & ${\phn}{9.7}$ & ${\phn}{\phn}{5.9}$ & ${\phn}{2.8}{\phn}$ & ${\phn}{1.6}{\phn}$ & ${\phn}{6.5}{\phn}$ & ${3.0}$ & ${1.3}$ & ${4.5}$ & ${1.8}$ & \nodata &  \\
\phn18\dotfill&${-0.513}$ & ${-0.037}$ & ${-100}$ & $1.16$ & $0.77$ & \phm{+}${35}$ & ${\phn}{7.5}$ & ${\phn}{\phn}{1.7}$ & ${\phn}{1.1}{\phn}$ & ${\phn}{0.30}$ & ${\phn}{2.6}{\phn}$ & ${2.6}$ & ${1.4}$ & ${4.0}$ & ${1.8}$ & \nodata &  \\
\phn19\dotfill&${-0.512}$ & ${-0.038}$ & ${\phn}{-61}$ & $1.05$ & $0.84$ & \phm{+}${21}$ & ${12.3}$ & ${\phn}{\phn}{9.2}$ & ${\phn}{4.7}{\phn}$ & ${\phn}{0.67}$ & ${\phn}{7.7}{\phn}$ & ${3.1}$ & ${1.4}$ & \nodata & \nodata & \nodata &  \\
\phn20\dotfill&${-0.507}$ & ${-0.213}$ & ${-112}$ & $0.72$ & $0.48$ & \phm{+}${35}$ & ${10.3}$ & ${\phn}{\phn}{8.2}$ & ${\phn}{2.4}{\phn}$ & ${\phn}{1.8}{\phn}$ & ${\phn}{2.7}{\phn}$ & ${2.3}$ & ${1.8}$ & \nodata & \nodata & \nodata &  \\
\phn21\dotfill&${-0.494}$ & ${-0.058}$ & ${-103}$ & $0.72$ & $0.58$ & \phm{+}${\phn}{9}$ & ${\phn}{5.9}$ & ${\phn}{\phn}{2.7}$ & ${\phn}{1.8}{\phn}$ & ${\phn}{0.40}$ & ${\phn}{5.0}{\phn}$ & ${2.6}$ & ${1.0}$ & ${4.0}$ & ${1.8}$ & \nodata &  \\
\phn22\dotfill&${-0.483}$ & ${-0.066}$ & ${\phn}{-20}$ & $0.50$ & $0.44$ & \phm{+}${48}$ & ${\phn}{8.7}$ & ${\phn}{\phn}{4.8}$ & ${\phn}{1.6}{\phn}$ & \nodata & ${\phn}{3.0}{\phn}$ & ${2.2}$ & ${1.6}$ & \nodata & \nodata & \nodata &  \\
\phn23\dotfill&${-0.478}$ & ${-0.253}$ & ${-115}$ & $1.06$ & $0.94$ & \phm{+}${28}$ & ${15.7}$ & ${\phn}{20.4}$ & ${\phn}{7.2}{\phn}$ & ${\phn}{3.2}{\phn}$ & ${\phn}{8.0}{\phn}$ & ${3.1}$ & ${1.6}$ & \nodata & \nodata & \nodata &  \\
\phn24\dotfill&${-0.475}$ & ${-0.012}$ & \phm{+}${\phn}{20}$ & $0.47$ & $0.35$ & \phm{+}${73}$ & ${\phn}{8.5}$ & ${\phn}{\phn}{4.0}$ & ${\phn}{2.3}{\phn}$ & ${\phn}{0.84}$ & ${\phn}{3.9}{\phn}$ & ${2.3}$ & ${1.5}$ & \nodata & \nodata & \nodata &  \\
\phn25\dotfill&${-0.456}$ & ${-0.234}$ & \phm{+}${106}$ & $1.02$ & $0.67$ & ${-41}$ & ${15.5}$ & ${\phn}{\phn}{5.6}$ & ${\phn}{0.98}$ & ${\phn}{0.23}$ & ${\phn}{1.5}{\phn}$ & ${2.3}$ & ${2.3}$ & \nodata & \nodata & \nodata &  \\
\phn26\dotfill&${-0.436}$ & ${-0.082}$ & ${\phn}{\phn}{-9}$ & $0.81$ & $0.48$ & ${-70}$ & ${\phn}{9.9}$ & ${\phn}{\phn}{9.5}$ & ${\phn}{3.9}{\phn}$ & ${\phn}{2.4}{\phn}$ & ${\phn}{3.6}{\phn}$ & ${2.5}$ & ${1.6}$ & ${4.0}$ & ${1.9}$ & \nodata &  \\
\phn27\dotfill&${-0.427}$ & ${-0.257}$ & ${-113}$ & $0.31$ & $0.25$ & \phm{+}${86}$ & ${14.6}$ & ${\phn}{\phn}{6.6}$ & ${\phn}{1.7}{\phn}$ & ${\phn}{0.63}$ & ${\phn}{1.2}{\phn}$ & ${1.6}$ & ${2.4}$ & \nodata & \nodata & \nodata &  \\
\phn28\dotfill&${-0.427}$ & \phm{+}${0.002}$ & ${\phn}{-86}$ & $0.93$ & $0.41$ & ${\phn}{-8}$ & ${\phn}{9.9}$ & ${\phn}{\phn}{2.8}$ & ${\phn}{1.7}{\phn}$ & ${\phn}{0.62}$ & ${\phn}{3.3}{\phn}$ & ${2.4}$ & ${1.7}$ & ${3.9}$ & ${1.8}$ & \nodata &  \\
\phn29\dotfill&${-0.420}$ & ${-0.257}$ & \phm{+}${\phn}{31}$ & $0.58$ & $0.43$ & ${-48}$ & ${\phn}{8.4}$ & ${\phn}{\phn}{5.4}$ & ${\phn}{2.4}{\phn}$ & ${\phn}{2.1}{\phn}$ & ${\phn}{3.3}{\phn}$ & ${2.3}$ & ${1.6}$ & \nodata & \nodata & \nodata &  \\
\phn30\dotfill&${-0.415}$ & ${-0.073}$ & ${-174}$ & $0.82$ & $0.46$ & ${-67}$ & ${\phn}{7.6}$ & ${\phn}{\phn}{2.6}$ & ${\phn}{1.0}{\phn}$ & \nodata & ${\phn}{1.1}{\phn}$ & ${2.0}$ & ${1.9}$ & \nodata & \nodata & \nodata &  \\
\phn31\dotfill&${-0.415}$ & ${-0.024}$ & ${\phn}{-23}$ & $0.64$ & $0.52$ & ${-33}$ & ${\phn}{9.1}$ & ${\phn}{11.9}$ & ${\phn}{5.0}{\phn}$ & ${\phn}{1.4}{\phn}$ & ${\phn}{6.0}{\phn}$ & ${2.6}$ & ${1.4}$ & ${3.8}$ & ${2.0}$ & \nodata &  \\
\phn32\dotfill&${-0.402}$ & ${-0.223}$ & ${\phn}{-65}$ & $0.85$ & $0.75$ & ${-38}$ & ${48.4}$ & ${243.1}$ & ${65.9}{\phn}$ & ${30.2}{\phn}$ & ${51.5}{\phn}$ & ${3.8}$ & ${1.8}$ & \nodata & \nodata & \nodata & CO$-0.40$$-0.22$ \\
\phn33\dotfill&${-0.402}$ & ${-0.023}$ & ${\phn}{-13}$ & $0.96$ & $0.63$ & ${-10}$ & ${11.4}$ & ${\phn}{14.7}$ & ${\phn}{6.2}{\phn}$ & ${\phn}{2.6}{\phn}$ & ${\phn}{7.6}{\phn}$ & ${2.9}$ & ${1.4}$ & ${4.0}$ & ${2.0}$ & \nodata &  \\
\phn34\dotfill&${-0.392}$ & ${-0.132}$ & \phm{+}${\phn}{11}$ & $0.62$ & $0.62$ & \phm{+}${80}$ & ${15.4}$ & ${\phn}{21.6}$ & ${\phn}{6.2}{\phn}$ & ${\phn}{2.9}{\phn}$ & ${10.2}{\phn}$ & ${2.9}$ & ${1.6}$ & \nodata & \nodata & \nodata &  \\
\phn35\dotfill&${-0.392}$ & ${-0.036}$ & \phm{+}${\phn}{72}$ & $0.74$ & $0.49$ & \phm{+}${\phn}{1}$ & ${\phn}{6.0}$ & ${\phn}{\phn}{6.1}$ & ${\phn}{2.4}{\phn}$ & ${\phn}{1.1}{\phn}$ & ${\phn}{9.6}{\phn}$ & ${2.9}$ & ${0.8}$ & ${4.2}$ & ${1.9}$ & g &  \\
\phn36\dotfill&${-0.386}$ & \phm{+}${0.015}$ & ${\phn}{-78}$ & $1.45$ & $0.82$ & ${-22}$ & ${\phn}{7.6}$ & ${\phn}{\phn}{5.2}$ & ${\phn}{3.2}{\phn}$ & ${\phn}{1.1}{\phn}$ & ${\phn}{7.3}{\phn}$ & ${3.2}$ & ${0.9}$ & ${4.1}$ & ${1.8}$ & \nodata &  \\
\phn37\dotfill&${-0.385}$ & ${-0.245}$ & \phm{+}${\phn}{21}$ & $0.59$ & $0.55$ & \phm{+}${13}$ & ${\phn}{7.4}$ & ${\phn}{37.7}$ & ${11.8}{\phn}$ & ${\phn}{7.3}{\phn}$ & ${30.9}{\phn}$ & ${3.4}$ & ${0.5}$ & \nodata & \nodata & \nodata & G$359.62$$-0.24$ \\
\phn38\dotfill&${-0.374}$ & ${-0.184}$ & ${\phn}{-90}$ & $0.55$ & $0.39$ & ${-52}$ & ${\phn}{6.7}$ & ${\phn}{\phn}{2.6}$ & ${\phn}{0.37}$ & \nodata & ${\phn}{0.42}$ & ${1.4}$ & ${2.3}$ & \nodata & \nodata & \nodata &  \\
\phn39\dotfill&${-0.373}$ & ${-0.083}$ & \phm{+}${\phn}{22}$ & $0.62$ & $0.60$ & \phm{+}${11}$ & ${10.2}$ & ${\phn}{\phn}{9.7}$ & ${\phn}{2.3}{\phn}$ & ${\phn}{1.1}{\phn}$ & ${\phn}{4.4}{\phn}$ & ${2.6}$ & ${1.6}$ & \nodata & \nodata & \nodata &  \\
\phn40\dotfill&${-0.363}$ & ${-0.240}$ & ${\phn}{-57}$ & $0.17$ & $0.17$ & ${-82}$ & ${\phn}{7.8}$ & ${\phn}{\phn}{5.0}$ & ${\phn}{0.69}$ & ${\phn}{0.42}$ & ${\phn}{1.4}{\phn}$ & ${1.6}$ & ${1.7}$ & \nodata & \nodata & \nodata &  \\
\phn41\dotfill&${-0.363}$ & ${-0.133}$ & ${\phn}{-31}$ & $0.76$ & $0.66$ & ${-65}$ & ${\phn}{9.4}$ & ${\phn}{\phn}{6.7}$ & ${\phn}{3.4}{\phn}$ & ${\phn}{2.1}{\phn}$ & ${\phn}{7.6}{\phn}$ & ${2.9}$ & ${1.2}$ & ${4.2}$ & ${1.8}$ & \nodata &  \\
\phn42\dotfill&${-0.341}$ & ${-0.215}$ & \phm{+}${\phn}{38}$ & $0.21$ & $0.17$ & ${-22}$ & ${\phn}{6.5}$ & ${\phn}{\phn}{3.1}$ & ${\phn}{0.72}$ & ${\phn}{0.27}$ & ${\phn}{1.5}{\phn}$ & ${1.7}$ & ${1.6}$ & \nodata & \nodata & \nodata &  \\
\phn43\dotfill&${-0.340}$ & ${-0.181}$ & \phm{+}${\phn}{32}$ & $0.99$ & $0.84$ & \phm{+}${19}$ & ${\phn}{9.0}$ & ${\phn}{\phn}{3.6}$ & ${\phn}{1.9}{\phn}$ & ${\phn}{0.37}$ & ${\phn}{2.3}{\phn}$ & ${2.5}$ & ${1.6}$ & \nodata & \nodata & \nodata &  \\
\phn44\dotfill&${-0.327}$ & ${-0.086}$ & \phm{+}${\phn}{12}$ & $0.32$ & $0.29$ & \phm{+}${30}$ & ${\phn}{7.4}$ & ${\phn}{\phn}{3.1}$ & ${\phn}{1.8}{\phn}$ & ${\phn}{0.43}$ & ${\phn}{2.4}{\phn}$ & ${2.0}$ & ${1.6}$ & ${3.8}$ & ${2.0}$ & \nodata &  \\
\phn45\dotfill&${-0.326}$ & ${-0.181}$ & ${\phn}{-76}$ & $0.40$ & $0.40$ & \phm{+}${14}$ & ${11.3}$ & ${\phn}{\phn}{7.3}$ & ${\phn}{1.3}{\phn}$ & ${\phn}{0.45}$ & ${\phn}{2.6}{\phn}$ & ${2.1}$ & ${1.9}$ & \nodata & \nodata & \nodata &  \\
\phn46\dotfill&${-0.324}$ & ${-0.084}$ & \phm{+}${\phn}{21}$ & $0.28$ & $0.22$ & \phm{+}${51}$ & ${\phn}{6.0}$ & ${\phn}{\phn}{3.7}$ & ${\phn}{1.3}{\phn}$ & ${\phn}{0.78}$ & ${\phn}{3.4}{\phn}$ & ${2.1}$ & ${1.2}$ & ${3.9}$ & ${1.9}$ & \nodata &  \\
\phn47\dotfill&${-0.321}$ & ${-0.194}$ & ${\phn}{-74}$ & $0.29$ & $0.24$ & ${-18}$ & ${\phn}{9.5}$ & ${\phn}{\phn}{4.1}$ & ${\phn}{1.8}{\phn}$ & ${\phn}{0.43}$ & ${\phn}{3.7}{\phn}$ & ${2.1}$ & ${1.6}$ & \nodata & \nodata & \nodata &  \\
\phn48\dotfill&${-0.320}$ & ${-0.135}$ & \phm{+}${\phn}{82}$ & $0.54$ & $0.46$ & \phm{+}${31}$ & ${\phn}{5.7}$ & ${\phn}{\phn}{3.3}$ & ${\phn}{0.87}$ & ${\phn}{0.09}$ & ${\phn}{2.6}{\phn}$ & ${2.2}$ & ${1.3}$ & \nodata & \nodata & \nodata &  \\
\phn49\dotfill&${-0.318}$ & ${-0.089}$ & \phm{+}${\phn}{26}$ & $0.26$ & $0.23$ & ${-85}$ & ${\phn}{9.8}$ & ${\phn}{\phn}{7.2}$ & ${\phn}{2.3}{\phn}$ & ${\phn}{1.3}{\phn}$ & ${\phn}{3.4}{\phn}$ & ${2.1}$ & ${1.6}$ & ${3.9}$ & ${2.0}$ & \nodata &  \\
\phn50\dotfill&${-0.313}$ & ${-0.011}$ & ${\phn}{-43}$ & $1.59$ & $1.12$ & \phm{+}${86}$ & ${17.3}$ & ${\phn}{18.1}$ & ${\phn}{9.3}{\phn}$ & ${\phn}{4.0}{\phn}$ & ${10.8}{\phn}$ & ${3.5}$ & ${1.4}$ & ${3.8}$ & ${2.0}$ & \nodata &  \\
\phn51\dotfill&${-0.309}$ & ${-0.068}$ & \phm{+}${\phn}{54}$ & $0.45$ & $0.29$ & ${-90}$ & ${29.6}$ & ${\phn}{49.4}$ & ${10.3}{\phn}$ & ${\phn}{2.7}{\phn}$ & ${\phn}{5.8}{\phn}$ & ${2.4}$ & ${2.4}$ & ${4.0}$ & ${2.2}$ & \nodata & CO$-0.30$$-0.07$ \\
\phn52\dotfill&${-0.309}$ & \phm{+}${0.010}$ & ${\phn}{-34}$ & $0.76$ & $0.66$ & \phm{+}${38}$ & ${10.0}$ & ${\phn}{\phn}{7.1}$ & ${\phn}{3.9}{\phn}$ & ${\phn}{1.1}{\phn}$ & ${\phn}{4.7}{\phn}$ & ${2.7}$ & ${1.5}$ & ${3.9}$ & ${1.9}$ & \nodata &  \\
\phn53\dotfill&${-0.306}$ & ${-0.028}$ & ${\phn}{-30}$ & $0.83$ & $0.69$ & ${-43}$ & ${13.6}$ & ${\phn}{\phn}{7.0}$ & ${\phn}{3.5}{\phn}$ & ${\phn}{1.1}{\phn}$ & ${\phn}{4.3}{\phn}$ & ${2.7}$ & ${1.8}$ & ${3.8}$ & ${1.9}$ & \nodata &  \\
\phn54\dotfill&${-0.302}$ & ${-0.063}$ & \phm{+}${\phn}{\phn}{8}$ & $0.65$ & $0.64$ & \phm{+}${12}$ & ${27.6}$ & ${\phn}{93.5}$ & ${29.3}{\phn}$ & ${12.1}{\phn}$ & ${18.4}{\phn}$ & ${3.2}$ & ${1.8}$ & ${3.6}$ & ${2.1}$ & \nodata & CO$-0.30$$-0.07$ \\
\phn55\dotfill&${-0.291}$ & ${-0.204}$ & ${\phn}{-23}$ & $0.57$ & $0.37$ & ${-28}$ & ${\phn}{7.6}$ & ${\phn}{\phn}{3.9}$ & ${\phn}{1.9}{\phn}$ & ${\phn}{0.16}$ & ${\phn}{3.3}{\phn}$ & ${2.3}$ & ${1.5}$ & \nodata & \nodata & \nodata &  \\
\phn56\dotfill&${-0.289}$ & \phm{+}${0.002}$ & ${\phn}{-60}$ & $0.44$ & $0.41$ & ${-55}$ & ${\phn}{8.3}$ & ${\phn}{\phn}{6.3}$ & ${\phn}{3.4}{\phn}$ & ${\phn}{1.3}{\phn}$ & ${\phn}{5.1}{\phn}$ & ${2.4}$ & ${1.4}$ & ${4.0}$ & ${1.9}$ & \nodata &  \\
\phn57\dotfill&${-0.268}$ & ${-0.058}$ & ${\phn}{-54}$ & $0.69$ & $0.60$ & ${-42}$ & ${12.2}$ & ${\phn}{10.6}$ & ${\phn}{6.1}{\phn}$ & ${\phn}{1.7}{\phn}$ & ${\phn}{4.6}{\phn}$ & ${2.6}$ & ${1.7}$ & ${3.8}$ & ${2.0}$ & \nodata &  \\
\phn58\dotfill&${-0.268}$ & \phm{+}${0.006}$ & \phm{+}${\phn}{71}$ & $0.67$ & $0.50$ & ${-28}$ & ${11.7}$ & ${\phn}{\phn}{6.7}$ & ${\phn}{4.0}{\phn}$ & ${\phn}{1.7}{\phn}$ & ${10.8}{\phn}$ & ${2.9}$ & ${1.3}$ & ${4.1}$ & ${1.9}$ & \nodata &  \\
\phn59\dotfill&${-0.259}$ & \phm{+}${0.022}$ & ${\phn}{-90}$ & $0.79$ & $0.77$ & \phm{+}${60}$ & ${21.2}$ & ${\phn}{34.3}$ & ${17.9}{\phn}$ & ${\phn}{5.4}{\phn}$ & ${28.4}{\phn}$ & ${3.5}$ & ${1.3}$ & \nodata & \nodata & \nodata &  \\
\phn60\dotfill&${-0.246}$ & \phm{+}${0.012}$ & ${\phn}{-46}$ & $0.78$ & $0.73$ & \phm{+}${47}$ & ${10.8}$ & ${\phn}{\phn}{5.1}$ & ${\phn}{2.6}{\phn}$ & ${\phn}{0.43}$ & ${\phn}{3.6}{\phn}$ & ${2.6}$ & ${1.7}$ & \nodata & \nodata & \nodata &  \\
\phn61\dotfill&${-0.224}$ & ${-0.084}$ & ${\phn}{-36}$ & $0.30$ & $0.25$ & \phm{+}${64}$ & ${\phn}{6.7}$ & ${\phn}{\phn}{6.2}$ & ${\phn}{3.7}{\phn}$ & ${\phn}{1.2}{\phn}$ & ${\phn}{6.7}{\phn}$ & ${2.4}$ & ${1.0}$ & ${3.7}$ & ${1.9}$ & \nodata &  \\
\phn62\dotfill&${-0.221}$ & ${-0.150}$ & \phm{+}${\phn}{\phn}{6}$ & $0.15$ & $0.13$ & \phm{+}${63}$ & ${11.6}$ & ${\phn}{\phn}{5.2}$ & ${\phn}{2.5}{\phn}$ & ${\phn}{0.57}$ & ${\phn}{3.1}{\phn}$ & ${1.9}$ & ${1.6}$ & \nodata & \nodata & \nodata &  \\
\phn63\dotfill&${-0.218}$ & ${-0.010}$ & \phm{+}${\phn}{55}$ & $0.85$ & $0.72$ & ${-61}$ & ${14.8}$ & ${\phn}{25.6}$ & ${13.2}{\phn}$ & ${\phn}{2.2}{\phn}$ & ${20.2}{\phn}$ & ${3.4}$ & ${1.2}$ & ${4.1}$ & ${1.9}$ & \nodata &  \\
\phn64\dotfill&${-0.176}$ & \phm{+}${0.018}$ & \phm{+}${\phn}{59}$ & $0.96$ & $0.53$ & \phm{+}${43}$ & ${13.8}$ & ${\phn}{14.7}$ & ${\phn}{6.2}{\phn}$ & ${\phn}{2.6}{\phn}$ & ${10.7}{\phn}$ & ${3.0}$ & ${1.4}$ & ${4.1}$ & ${2.0}$ & \nodata &  \\
\phn65\dotfill&${-0.175}$ & ${-0.105}$ & \phm{+}${\phn}{89}$ & $0.74$ & $0.62$ & \phm{+}${75}$ & ${10.2}$ & ${\phn}{\phn}{2.4}$ & ${\phn}{1.0}{\phn}$ & ${\phn}{0.17}$ & ${\phn}{1.2}{\phn}$ & ${2.0}$ & ${2.1}$ & \nodata & \nodata & \nodata &  \\
\phn66\dotfill&${-0.149}$ & \phm{+}${0.019}$ & \phm{+}${\phn}{66}$ & $1.01$ & $0.30$ & \phm{+}${88}$ & ${\phn}{4.1}$ & ${\phn}{\phn}{1.6}$ & ${\phn}{0.57}$ & ${\phn}{0.23}$ & ${\phn}{0.89}$ & ${1.8}$ & ${1.5}$ & ${4.0}$ & ${2.0}$ & \nodata &  \\
\phn67\dotfill&${-0.148}$ & ${-0.084}$ & \phm{+}${\phn}{15}$ & $0.98$ & $0.49$ & \phm{+}${52}$ & ${\phn}{3.3}$ & ${\phn}{\phn}{4.1}$ & ${\phn}{1.4}{\phn}$ & ${\phn}{1.1}{\phn}$ & ${\phn}{3.0}{\phn}$ & ${2.5}$ & ${0.7}$ & ${4.3}$ & ${1.9}$ & w &  \\
\phn68\dotfill&${-0.142}$ & \phm{+}${0.043}$ & ${\phn}{-53}$ & $0.43$ & $0.42$ & ${-55}$ & ${13.3}$ & ${\phn}{10.3}$ & ${\phn}{5.5}{\phn}$ & ${\phn}{2.3}{\phn}$ & ${\phn}{7.6}{\phn}$ & ${2.6}$ & ${1.6}$ & \nodata & \nodata & \nodata &  \\
\phn69\dotfill&${-0.132}$ & ${-0.036}$ & \phm{+}${\phn}{44}$ & $0.58$ & $0.49$ & ${-18}$ & ${\phn}{9.5}$ & ${\phn}{\phn}{6.0}$ & ${\phn}{2.5}{\phn}$ & ${\phn}{1.1}{\phn}$ & ${\phn}{8.5}{\phn}$ & ${2.8}$ & ${1.2}$ & ${4.1}$ & ${1.8}$ & \nodata &  \\
\phn70\dotfill&${-0.129}$ & ${-0.074}$ & \phm{+}${\phn}{\phn}{2}$ & $0.76$ & $0.73$ & ${-30}$ & ${\phn}{7.6}$ & ${\phn}{30.1}$ & ${\phn}{9.4}{\phn}$ & ${\phn}{8.1}{\phn}$ & ${18.6}{\phn}$ & ${3.3}$ & ${0.7}$ & ${4.7}$ & ${2.0}$ & w & 20-\kmps\ \\
\phn71\dotfill&${-0.123}$ & \phm{+}${0.014}$ & \phm{+}${\phn}{63}$ & $0.38$ & $0.28$ & \phm{+}${52}$ & ${\phn}{6.6}$ & ${\phn}{\phn}{5.2}$ & ${\phn}{1.9}{\phn}$ & ${\phn}{0.87}$ & ${\phn}{3.8}{\phn}$ & ${2.2}$ & ${1.3}$ & ${4.1}$ & ${1.9}$ & \nodata &  \\
\phn72\dotfill&${-0.114}$ & \phm{+}${0.033}$ & \phm{+}${\phn}{15}$ & $0.63$ & $0.63$ & \phm{+}${43}$ & ${\phn}{9.2}$ & ${\phn}{\phn}{3.1}$ & ${\phn}{1.9}{\phn}$ & ${\phn}{0.62}$ & ${\phn}{3.2}{\phn}$ & ${2.4}$ & ${1.6}$ & \nodata & \nodata & \nodata &  \\
\phn73\dotfill&${-0.108}$ & ${-0.076}$ & \phm{+}${\phn}{12}$ & $1.23$ & $0.83$ & ${-37}$ & ${\phn}{8.4}$ & ${\phn}{47.8}$ & ${15.1}{\phn}$ & ${13.6}{\phn}$ & ${30.7}{\phn}$ & ${3.7}$ & ${0.4}$ & ${4.7}$ & ${1.9}$ & w & 20-\kmps\ \\
\phn74\dotfill&${-0.104}$ & \phm{+}${0.013}$ & \phm{+}${\phn}{64}$ & $0.52$ & $0.51$ & \phm{+}${60}$ & ${16.4}$ & ${\phn}{20.2}$ & ${\phn}{8.1}{\phn}$ & ${\phn}{3.7}{\phn}$ & ${10.0}{\phn}$ & ${2.8}$ & ${1.6}$ & ${4.3}$ & ${2.0}$ & \nodata &  \\
\phn75\dotfill&${-0.091}$ & ${-0.018}$ & ${-125}$ & $0.39$ & $0.28$ & \phm{+}${11}$ & ${\phn}{4.6}$ & ${\phn}{\phn}{3.4}$ & ${\phn}{0.74}$ & \nodata & ${\phn}{1.3}{\phn}$ & ${1.7}$ & ${1.4}$ & \nodata & \nodata & \nodata &  \\
\phn76\dotfill&${-0.078}$ & ${-0.123}$ & \phm{+}${\phn}{34}$ & $0.38$ & $0.37$ & \phm{+}${78}$ & ${\phn}{8.8}$ & ${\phn}{\phn}{9.6}$ & ${\phn}{4.6}{\phn}$ & ${\phn}{2.2}{\phn}$ & ${\phn}{8.2}{\phn}$ & ${2.6}$ & ${1.2}$ & ${4.1}$ & ${1.9}$ & \nodata &  \\
\phn77\dotfill&${-0.074}$ & ${-0.067}$ & \phm{+}${\phn}{25}$ & $0.11$ & $0.09$ & \phm{+}${87}$ & ${\phn}{2.9}$ & ${\phn}{\phn}{5.8}$ & ${\phn}{1.6}{\phn}$ & ${\phn}{1.2}{\phn}$ & ${\phn}{4.1}{\phn}$ & ${2.0}$ & ${0.2}$ & ${4.5}$ & ${1.9}$ & g &  \\
\phn78\dotfill&${-0.074}$ & ${-0.039}$ & ${\phn}{-17}$ & $0.33$ & $0.28$ & \phm{+}${25}$ & ${19.8}$ & ${\phn}{26.2}$ & ${\phn}{4.9}{\phn}$ & ${\phn}{1.8}{\phn}$ & ${\phn}{5.5}{\phn}$ & ${2.3}$ & ${2.1}$ & ${3.7}$ & ${2.1}$ & \nodata & CND \\
\phn79\dotfill&${-0.065}$ & ${-0.049}$ & ${\phn}{-46}$ & $0.56$ & $0.44$ & ${-40}$ & ${51.7}$ & ${122.2}$ & ${13.5}{\phn}$ & ${\phn}{1.5}{\phn}$ & ${13.0}{\phn}$ & ${2.9}$ & ${2.5}$ & \nodata & \nodata & \nodata & CND \\
\phn80\dotfill&${-0.051}$ & ${-0.044}$ & \phm{+}${\phn}{57}$ & $0.33$ & $0.27$ & ${-54}$ & ${13.1}$ & ${\phn}{11.5}$ & ${\phn}{1.9}{\phn}$ & ${\phn}{0.34}$ & ${\phn}{3.1}{\phn}$ & ${2.1}$ & ${2.0}$ & ${4.1}$ & ${2.0}$ & \nodata & CND \\
\phn81\dotfill&${-0.045}$ & ${-0.055}$ & \phm{+}${\phn}{62}$ & $0.34$ & $0.32$ & ${-80}$ & ${11.4}$ & ${\phn}{13.5}$ & ${\phn}{2.9}{\phn}$ & ${\phn}{1.0}{\phn}$ & ${\phn}{4.2}{\phn}$ & ${2.2}$ & ${1.7}$ & ${4.2}$ & ${2.0}$ & \nodata & CND \\
\phn82\dotfill&${-0.036}$ & ${-0.067}$ & ${\phn}{-70}$ & $0.47$ & $0.46$ & ${\phn}{-2}$ & ${34.5}$ & ${\phn}{18.3}$ & ${\phn}{4.0}{\phn}$ & ${\phn}{1.2}{\phn}$ & ${\phn}{4.7}{\phn}$ & ${2.4}$ & ${2.6}$ & \nodata & \nodata & \nodata &  \\
\phn83\dotfill&${-0.021}$ & \phm{+}${0.009}$ & ${\phn}{-30}$ & $0.69$ & $0.36$ & \phm{+}${62}$ & ${\phn}{7.3}$ & ${\phn}{\phn}{2.8}$ & ${\phn}{1.4}{\phn}$ & ${\phn}{0.21}$ & ${\phn}{2.6}{\phn}$ & ${2.2}$ & ${1.5}$ & \nodata & \nodata & \nodata &  \\
\phn84\dotfill&${-0.020}$ & ${-0.070}$ & \phm{+}${\phn}{60}$ & $2.09$ & $1.14$ & ${-42}$ & ${12.4}$ & ${\phn}{39.6}$ & ${16.9}{\phn}$ & ${12.6}{\phn}$ & ${21.9}{\phn}$ & ${3.9}$ & ${0.8}$ & ${4.4}$ & ${2.0}$ & \nodata & 50-\kmps\ \\
\phn85\dotfill&${-0.020}$ & ${-0.046}$ & ${\phn}{\phn}{-7}$ & $0.18$ & $0.16$ & ${\phn}{-0}$ & ${\phn}{6.9}$ & ${\phn}{\phn}{2.5}$ & ${\phn}{0.66}$ & ${\phn}{0.31}$ & ${\phn}{0.56}$ & ${1.2}$ & ${2.0}$ & ${4.0}$ & ${1.9}$ & \nodata &  \\
\phn86\dotfill&${-0.015}$ & ${-0.076}$ & \phm{+}${\phn}{36}$ & $1.48$ & $1.01$ & \phm{+}${48}$ & ${17.9}$ & ${131.7}$ & ${45.3}{\phn}$ & ${33.6}{\phn}$ & ${47.2}{\phn}$ & ${4.1}$ & ${0.8}$ & ${4.5}$ & ${2.1}$ & H & 50-\kmps\ \\
\phn87\dotfill&\phm{+}${0.000}$ & ${-0.030}$ & ${\phn}{\phn}{-7}$ & $0.89$ & $0.74$ & ${-89}$ & ${\phn}{9.6}$ & ${\phn}{\phn}{6.8}$ & ${\phn}{3.3}{\phn}$ & ${\phn}{1.1}{\phn}$ & ${\phn}{3.5}{\phn}$ & ${2.6}$ & ${1.6}$ & ${4.0}$ & ${1.9}$ & \nodata &  \\
\phn88\dotfill&\phm{+}${0.002}$ & \phm{+}${0.080}$ & ${\phn}{-21}$ & $0.72$ & $0.58$ & ${-81}$ & ${\phn}{7.3}$ & ${\phn}{\phn}{4.5}$ & ${\phn}{1.7}{\phn}$ & ${\phn}{0.48}$ & ${\phn}{2.7}{\phn}$ & ${2.4}$ & ${1.5}$ & \nodata & \nodata & \nodata &  \\
\phn89\dotfill&\phm{+}${0.014}$ & ${-0.010}$ & ${\phn}{-10}$ & $0.93$ & $0.75$ & ${-33}$ & ${15.5}$ & ${\phn}{11.9}$ & ${\phn}{6.5}{\phn}$ & ${\phn}{2.6}{\phn}$ & ${\phn}{6.9}{\phn}$ & ${3.0}$ & ${1.7}$ & ${4.1}$ & ${1.9}$ & \nodata &  \\
\phn90\dotfill&\phm{+}${0.017}$ & ${-0.018}$ & \phm{+}${\phn}{91}$ & $0.67$ & $0.60$ & ${-48}$ & ${24.6}$ & ${\phn}{84.6}$ & ${18.8}{\phn}$ & ${\phn}{7.5}{\phn}$ & ${17.5}{\phn}$ & ${3.2}$ & ${1.7}$ & ${4.1}$ & ${2.1}$ & \nodata & CO$+0.02$$-0.02$ \\
\phn91\dotfill&\phm{+}${0.018}$ & \phm{+}${0.036}$ & \phm{+}${\phn}{84}$ & $0.85$ & $0.80$ & \phm{+}${85}$ & ${20.3}$ & ${\phn}{50.9}$ & ${20.9}{\phn}$ & ${14.5}{\phn}$ & ${28.3}{\phn}$ & ${3.6}$ & ${1.3}$ & ${4.4}$ & ${2.0}$ & \nodata &  \\
\phn92\dotfill&\phm{+}${0.021}$ & ${-0.057}$ & ${\phn}{-13}$ & $1.04$ & $0.58$ & ${-85}$ & ${10.6}$ & ${\phn}{14.6}$ & ${\phn}{6.9}{\phn}$ & ${\phn}{3.8}{\phn}$ & ${10.3}{\phn}$ & ${3.1}$ & ${1.2}$ & ${4.0}$ & ${1.9}$ & \nodata &  \\
\phn93\dotfill&\phm{+}${0.050}$ & \phm{+}${0.026}$ & ${\phn}{-35}$ & $0.86$ & $0.54$ & \phm{+}${36}$ & ${17.1}$ & ${\phn}{18.8}$ & ${\phn}{9.7}{\phn}$ & ${\phn}{3.2}{\phn}$ & ${26.8}{\phn}$ & ${3.4}$ & ${1.2}$ & ${4.1}$ & ${1.9}$ & \nodata &  \\
\phn94\dotfill&\phm{+}${0.055}$ & ${-0.212}$ & \phm{+}${\phn}{14}$ & $0.70$ & $0.57$ & \phm{+}${26}$ & ${\phn}{6.3}$ & ${\phn}{\phn}{5.5}$ & ${\phn}{2.2}{\phn}$ & ${\phn}{1.2}{\phn}$ & ${\phn}{7.3}{\phn}$ & ${2.8}$ & ${0.9}$ & \nodata & \nodata & \nodata & non-GC~source \\
\phn95\dotfill&\phm{+}${0.061}$ & ${-0.070}$ & \phm{+}${\phn}{96}$ & $0.73$ & $0.58$ & \phm{+}${77}$ & ${\phn}{3.0}$ & ${\phn}{\phn}{1.4}$ & \nodata & \nodata & ${\phn}{0.04}$ & ${0.6}$ & ${2.5}$ & \nodata & \nodata & \nodata &  \\
\phn96\dotfill&\phm{+}${0.064}$ & ${-0.075}$ & \phm{+}${\phn}{50}$ & $1.25$ & $0.81$ & ${\phn}{-4}$ & ${15.5}$ & ${\phn}{35.0}$ & ${14.3}{\phn}$ & ${12.2}{\phn}$ & ${31.8}{\phn}$ & ${3.8}$ & ${0.9}$ & ${4.6}$ & ${1.9}$ & \nodata &  \\
\phn97\dotfill&\phm{+}${0.079}$ & ${-0.063}$ & ${\phn}{-45}$ & $0.66$ & $0.52$ & ${-66}$ & ${\phn}{7.0}$ & ${\phn}{\phn}{4.6}$ & ${\phn}{1.9}{\phn}$ & ${\phn}{0.45}$ & ${\phn}{3.2}{\phn}$ & ${2.4}$ & ${1.4}$ & ${3.6}$ & ${1.9}$ & \nodata &  \\
\phn98\dotfill&\phm{+}${0.097}$ & ${-0.017}$ & \phm{+}${\phn}{44}$ & $0.73$ & $0.60$ & ${-89}$ & ${\phn}{9.5}$ & ${\phn}{\phn}{5.5}$ & ${\phn}{2.6}{\phn}$ & ${\phn}{1.6}{\phn}$ & ${\phn}{4.4}{\phn}$ & ${2.6}$ & ${1.5}$ & ${4.5}$ & ${1.9}$ & \nodata &  \\
\phn99\dotfill&\phm{+}${0.101}$ & ${-0.134}$ & \phm{+}${\phn}{31}$ & $0.29$ & $0.28$ & \phm{+}${38}$ & ${\phn}{9.3}$ & ${\phn}{10.3}$ & ${\phn}{3.7}{\phn}$ & ${\phn}{2.3}{\phn}$ & ${\phn}{6.9}{\phn}$ & ${2.4}$ & ${1.3}$ & ${4.3}$ & ${1.9}$ & \nodata &  \\
100\dotfill&\phm{+}${0.103}$ & ${-0.082}$ & \phm{+}${\phn}{55}$ & $1.02$ & $0.67$ & \phm{+}${35}$ & ${12.9}$ & ${\phn}{85.5}$ & ${31.2}{\phn}$ & ${27.0}{\phn}$ & ${51.3}{\phn}$ & ${3.8}$ & ${0.6}$ & ${4.5}$ & ${2.0}$ & \nodata & 20-\kmps\ \\
101\dotfill&\phm{+}${0.106}$ & ${-0.005}$ & \phm{+}${\phn}{49}$ & $0.50$ & $0.48$ & ${-15}$ & ${10.2}$ & ${\phn}{12.2}$ & ${\phn}{5.7}{\phn}$ & ${\phn}{4.9}{\phn}$ & ${\phn}{9.5}{\phn}$ & ${2.8}$ & ${1.3}$ & ${4.4}$ & ${1.9}$ & \nodata &  \\
102\dotfill&\phm{+}${0.117}$ & ${-0.217}$ & \phm{+}${\phn}{\phn}{6}$ & $0.99$ & $0.68$ & \phm{+}${27}$ & ${12.9}$ & ${\phn}{\phn}{8.7}$ & ${\phn}{2.9}{\phn}$ & ${\phn}{1.1}{\phn}$ & ${\phn}{6.2}{\phn}$ & ${2.9}$ & ${1.6}$ & \nodata & \nodata & \nodata &  \\
103\dotfill&\phm{+}${0.119}$ & ${-0.115}$ & \phm{+}${\phn}{29}$ & $0.54$ & $0.48$ & \phm{+}${\phn}{2}$ & ${11.0}$ & ${\phn}{15.4}$ & ${\phn}{6.8}{\phn}$ & ${\phn}{4.4}{\phn}$ & ${12.0}{\phn}$ & ${2.9}$ & ${1.2}$ & ${4.4}$ & ${1.9}$ & \nodata &  \\
104\dotfill&\phm{+}${0.119}$ & \phm{+}${0.080}$ & ${\phn}{-27}$ & $0.04$ & $0.03$ & \phm{+}${84}$ & ${\phn}{5.0}$ & ${\phn}{\phn}{3.1}$ & ${\phn}{1.1}{\phn}$ & ${\phn}{0.95}$ & ${\phn}{0.96}$ & ${1.4}$ & ${0.9}$ & ${3.8}$ & ${1.9}$ & \nodata &  \\
105\dotfill&\phm{+}${0.128}$ & \phm{+}${0.021}$ & ${\phn}{-26}$ & $1.39$ & $1.33$ & \phm{+}${45}$ & ${13.8}$ & ${\phn}{\phn}{9.2}$ & ${\phn}{5.4}{\phn}$ & ${\phn}{2.2}{\phn}$ & ${\phn}{7.7}{\phn}$ & ${3.4}$ & ${1.4}$ & ${4.1}$ & ${1.9}$ & \nodata &  \\
106\dotfill&\phm{+}${0.133}$ & \phm{+}${0.003}$ & \phm{+}${\phn}{64}$ & $0.57$ & $0.55$ & \phm{+}${\phn}{0}$ & ${\phn}{8.0}$ & ${\phn}{11.6}$ & ${\phn}{4.1}{\phn}$ & ${\phn}{3.0}{\phn}$ & ${\phn}{7.4}{\phn}$ & ${2.7}$ & ${1.2}$ & ${4.4}$ & ${1.9}$ & \nodata &  \\
107\dotfill&\phm{+}${0.138}$ & \phm{+}${0.023}$ & ${\phn}{-13}$ & $0.91$ & $0.64$ & ${-37}$ & ${\phn}{8.8}$ & ${\phn}{\phn}{4.8}$ & ${\phn}{3.0}{\phn}$ & ${\phn}{1.0}{\phn}$ & ${\phn}{6.5}{\phn}$ & ${2.9}$ & ${1.2}$ & ${4.0}$ & ${1.8}$ & \nodata &  \\
108\dotfill&\phm{+}${0.158}$ & ${-0.042}$ & \phm{+}${\phn}{25}$ & $0.98$ & $0.85$ & ${-50}$ & ${10.2}$ & ${\phn}{\phn}{2.0}$ & ${\phn}{0.94}$ & ${\phn}{0.31}$ & ${\phn}{1.6}{\phn}$ & ${2.4}$ & ${1.9}$ & ${3.9}$ & ${1.9}$ & \nodata &  \\
109\dotfill&\phm{+}${0.164}$ & \phm{+}${0.076}$ & \phm{+}${\phn}{29}$ & $0.52$ & $0.46$ & \phm{+}${70}$ & ${\phn}{7.1}$ & ${\phn}{\phn}{4.4}$ & ${\phn}{1.3}{\phn}$ & ${\phn}{0.76}$ & ${\phn}{3.8}{\phn}$ & ${2.4}$ & ${1.3}$ & \nodata & \nodata & \nodata &  \\
110\dotfill&\phm{+}${0.165}$ & \phm{+}${0.063}$ & \phm{+}${\phn}{31}$ & $0.51$ & $0.48$ & \phm{+}${69}$ & ${11.8}$ & ${\phn}{\phn}{8.5}$ & ${\phn}{2.0}{\phn}$ & ${\phn}{0.53}$ & ${\phn}{2.8}{\phn}$ & ${2.2}$ & ${1.9}$ & \nodata & \nodata & \nodata &  \\
111\dotfill&\phm{+}${0.171}$ & ${-0.082}$ & \phm{+}${\phn}{\phn}{3}$ & $0.47$ & $0.39$ & \phm{+}${32}$ & ${17.2}$ & ${\phn}{33.9}$ & ${12.2}{\phn}$ & ${\phn}{6.3}{\phn}$ & ${13.8}{\phn}$ & ${2.9}$ & ${1.6}$ & ${4.0}$ & ${2.0}$ & \nodata & N3 \\
112\dotfill&\phm{+}${0.180}$ & ${-0.042}$ & \phm{+}${\phn}{22}$ & $0.41$ & $0.40$ & ${-48}$ & ${10.3}$ & ${\phn}{\phn}{7.8}$ & ${\phn}{3.7}{\phn}$ & ${\phn}{1.3}{\phn}$ & ${\phn}{5.9}{\phn}$ & ${2.5}$ & ${1.5}$ & ${3.9}$ & ${1.9}$ & \nodata &  \\
113\dotfill&\phm{+}${0.202}$ & \phm{+}${0.052}$ & \phm{+}${\phn}{21}$ & $0.85$ & $0.76$ & \phm{+}${80}$ & ${13.9}$ & ${\phn}{14.1}$ & ${\phn}{5.2}{\phn}$ & ${\phn}{2.1}{\phn}$ & ${13.3}{\phn}$ & ${3.2}$ & ${1.3}$ & ${4.0}$ & ${1.9}$ & \nodata &  \\
114\dotfill&\phm{+}${0.203}$ & ${-0.037}$ & \phm{+}${\phn}{25}$ & $1.79$ & $0.84$ & ${-46}$ & ${\phn}{9.0}$ & ${\phn}{\phn}{9.2}$ & ${\phn}{6.2}{\phn}$ & ${\phn}{1.9}{\phn}$ & ${12.8}{\phn}$ & ${3.5}$ & ${0.8}$ & ${4.1}$ & ${1.8}$ & \nodata &  \\
115\dotfill&\phm{+}${0.207}$ & ${-0.020}$ & \phm{+}${\phn}{83}$ & $0.60$ & $0.39$ & ${-43}$ & ${\phn}{4.5}$ & ${\phn}{\phn}{4.2}$ & ${\phn}{2.1}{\phn}$ & ${\phn}{0.81}$ & ${\phn}{4.2}{\phn}$ & ${2.4}$ & ${0.9}$ & ${4.4}$ & ${1.8}$ & \nodata &  \\
116\dotfill&\phm{+}${0.208}$ & ${-0.002}$ & \phm{+}${\phn}{44}$ & $0.30$ & $0.23$ & ${-51}$ & ${\phn}{5.6}$ & ${\phn}{\phn}{3.8}$ & ${\phn}{1.8}{\phn}$ & ${\phn}{0.55}$ & ${\phn}{4.0}{\phn}$ & ${2.2}$ & ${1.1}$ & ${4.1}$ & ${1.8}$ & mw &  \\
117\dotfill&\phm{+}${0.210}$ & ${-0.061}$ & \phm{+}${\phn}{31}$ & $1.35$ & $0.59$ & ${-49}$ & ${\phn}{6.0}$ & ${\phn}{\phn}{1.4}$ & ${\phn}{0.99}$ & ${\phn}{0.32}$ & ${\phn}{2.3}{\phn}$ & ${2.5}$ & ${1.3}$ & ${4.2}$ & ${1.8}$ & \nodata &  \\
118\dotfill&\phm{+}${0.227}$ & \phm{+}${0.070}$ & ${\phn}{-42}$ & $0.56$ & $0.45$ & ${-12}$ & ${13.9}$ & ${\phn}{\phn}{8.0}$ & ${\phn}{3.1}{\phn}$ & ${\phn}{1.1}{\phn}$ & ${\phn}{6.0}{\phn}$ & ${2.6}$ & ${1.7}$ & \nodata & \nodata & \nodata &  \\
119\dotfill&\phm{+}${0.231}$ & \phm{+}${0.012}$ & \phm{+}${\phn}{32}$ & $0.46$ & $0.33$ & \phm{+}${33}$ & ${\phn}{8.4}$ & ${\phn}{11.8}$ & ${\phn}{4.5}{\phn}$ & ${\phn}{5.9}{\phn}$ & ${\phn}{8.0}{\phn}$ & ${2.6}$ & ${1.2}$ & ${4.2}$ & ${2.0}$ & \nodata & brick \\
120\dotfill&\phm{+}${0.241}$ & \phm{+}${0.006}$ & \phm{+}${\phn}{39}$ & $0.80$ & $0.68$ & \phm{+}${64}$ & ${15.0}$ & ${\phn}{33.4}$ & ${12.2}{\phn}$ & ${17.3}{\phn}$ & ${26.0}{\phn}$ & ${3.4}$ & ${1.1}$ & ${4.5}$ & ${2.0}$ & \nodata & brick \\
121\dotfill&\phm{+}${0.255}$ & ${-0.059}$ & \phm{+}${\phn}{29}$ & $0.69$ & $0.48$ & \phm{+}${12}$ & ${\phn}{6.9}$ & ${\phn}{\phn}{2.8}$ & ${\phn}{1.8}{\phn}$ & ${\phn}{0.59}$ & ${\phn}{3.8}{\phn}$ & ${2.5}$ & ${1.3}$ & ${4.1}$ & ${1.8}$ & \nodata &  \\
122\dotfill&\phm{+}${0.256}$ & \phm{+}${0.020}$ & \phm{+}${\phn}{36}$ & $1.01$ & $0.73$ & \phm{+}${18}$ & ${12.1}$ & ${\phn}{10.3}$ & ${\phn}{4.8}{\phn}$ & ${\phn}{5.8}{\phn}$ & ${11.3}{\phn}$ & ${3.2}$ & ${1.2}$ & ${4.5}$ & ${1.9}$ & w & brick \\
123\dotfill&\phm{+}${0.272}$ & ${-0.098}$ & \phm{+}${\phn}{62}$ & $0.52$ & $0.50$ & ${-54}$ & ${\phn}{9.3}$ & ${\phn}{\phn}{3.1}$ & ${\phn}{1.3}{\phn}$ & ${\phn}{0.47}$ & ${\phn}{3.1}{\phn}$ & ${2.3}$ & ${1.7}$ & \nodata & \nodata & \nodata &  \\
124\dotfill&\phm{+}${0.275}$ & ${-0.061}$ & \phm{+}${\phn}{38}$ & $0.65$ & $0.43$ & ${-13}$ & ${\phn}{7.1}$ & ${\phn}{\phn}{4.6}$ & ${\phn}{2.8}{\phn}$ & ${\phn}{0.85}$ & ${\phn}{6.9}{\phn}$ & ${2.7}$ & ${1.1}$ & ${4.1}$ & ${1.8}$ & \nodata &  \\
125\dotfill&\phm{+}${0.286}$ & \phm{+}${0.008}$ & \phm{+}${114}$ & $0.40$ & $0.32$ & ${\phn}{-8}$ & ${21.4}$ & ${\phn}{\phn}{6.2}$ & ${\phn}{1.4}{\phn}$ & ${\phn}{0.93}$ & ${\phn}{1.7}{\phn}$ & ${1.9}$ & ${2.7}$ & \nodata & \nodata & \nodata &  \\
126\dotfill&\phm{+}${0.287}$ & \phm{+}${0.038}$ & \phm{+}${\phn}{\phn}{4}$ & $0.53$ & $0.43$ & \phm{+}${63}$ & ${\phn}{8.6}$ & ${\phn}{\phn}{9.8}$ & ${\phn}{2.8}{\phn}$ & ${\phn}{4.0}{\phn}$ & ${\phn}{5.9}{\phn}$ & ${2.5}$ & ${1.3}$ & ${4.1}$ & ${2.0}$ & \nodata &  \\
127\dotfill&\phm{+}${0.301}$ & \phm{+}${0.045}$ & ${\phn}{-28}$ & $0.32$ & $0.29$ & ${-12}$ & ${\phn}{8.7}$ & ${\phn}{\phn}{6.6}$ & ${\phn}{2.3}{\phn}$ & ${\phn}{1.7}{\phn}$ & ${\phn}{2.5}{\phn}$ & ${2.0}$ & ${1.7}$ & ${4.0}$ & ${2.0}$ & \nodata &  \\
128\dotfill&\phm{+}${0.311}$ & \phm{+}${0.013}$ & \phm{+}${\phn}{25}$ & $0.64$ & $0.39$ & \phm{+}${43}$ & ${\phn}{9.6}$ & ${\phn}{\phn}{5.4}$ & ${\phn}{3.4}{\phn}$ & ${\phn}{1.1}{\phn}$ & ${\phn}{6.9}{\phn}$ & ${2.6}$ & ${1.3}$ & ${4.0}$ & ${1.8}$ & \nodata &  \\
129\dotfill&\phm{+}${0.319}$ & \phm{+}${0.035}$ & \phm{+}${\phn}{\phn}{5}$ & $0.52$ & $0.42$ & ${\phn}{-3}$ & ${13.9}$ & ${\phn}{\phn}{5.0}$ & ${\phn}{1.9}{\phn}$ & ${\phn}{0.78}$ & ${\phn}{3.4}{\phn}$ & ${2.3}$ & ${2.0}$ & ${4.0}$ & ${1.8}$ & \nodata &  \\
130\dotfill&\phm{+}${0.334}$ & ${-0.073}$ & \phm{+}${\phn}{15}$ & $0.79$ & $0.71$ & ${-61}$ & ${\phn}{8.0}$ & ${\phn}{\phn}{4.5}$ & ${\phn}{2.8}{\phn}$ & ${\phn}{0.60}$ & ${\phn}{8.0}{\phn}$ & ${2.9}$ & ${1.1}$ & \nodata & \nodata & \nodata &  \\
131\dotfill&\phm{+}${0.334}$ & \phm{+}${0.056}$ & ${\phn}{\phn}{-1}$ & $1.13$ & $0.75$ & \phm{+}${\phn}{5}$ & ${10.8}$ & ${\phn}{\phn}{8.2}$ & ${\phn}{3.9}{\phn}$ & ${\phn}{2.4}{\phn}$ & ${\phn}{7.6}{\phn}$ & ${3.1}$ & ${1.3}$ & ${4.1}$ & ${1.8}$ & \nodata &  \\
132\dotfill&\phm{+}${0.335}$ & ${-0.079}$ & \phm{+}${\phn}{82}$ & $0.52$ & $0.51$ & ${-42}$ & ${10.0}$ & ${\phn}{\phn}{3.0}$ & ${\phn}{2.0}{\phn}$ & ${\phn}{0.84}$ & ${\phn}{2.4}{\phn}$ & ${2.2}$ & ${1.9}$ & ${4.1}$ & ${1.8}$ & \nodata &  \\
133\dotfill&\phm{+}${0.364}$ & ${-0.083}$ & \phm{+}${\phn}{26}$ & $1.18$ & $0.76$ & \phm{+}${13}$ & ${16.1}$ & ${\phn}{\phn}{7.5}$ & ${\phn}{4.7}{\phn}$ & ${\phn}{1.6}{\phn}$ & ${10.2}{\phn}$ & ${3.2}$ & ${1.5}$ & ${3.9}$ & ${1.8}$ & \nodata &  \\
134\dotfill&\phm{+}${0.364}$ & ${-0.012}$ & \phm{+}${\phn}{69}$ & $0.31$ & $0.17$ & \phm{+}${66}$ & ${\phn}{6.1}$ & ${\phn}{\phn}{3.3}$ & ${\phn}{1.0}{\phn}$ & ${\phn}{0.31}$ & ${\phn}{1.4}{\phn}$ & ${1.7}$ & ${1.6}$ & ${3.9}$ & ${1.9}$ & \nodata &  \\
135\dotfill&\phm{+}${0.370}$ & \phm{+}${0.038}$ & \phm{+}${\phn}{36}$ & $0.50$ & $0.45$ & ${\phn}{-9}$ & ${\phn}{8.3}$ & ${\phn}{13.5}$ & ${\phn}{6.5}{\phn}$ & ${\phn}{3.4}{\phn}$ & ${17.9}{\phn}$ & ${3.0}$ & ${0.8}$ & ${4.1}$ & ${1.8}$ & mw & cloud~c \\
136\dotfill&\phm{+}${0.393}$ & ${-0.080}$ & \phm{+}${100}$ & $0.77$ & $0.61$ & \phm{+}${62}$ & ${15.2}$ & ${\phn}{20.2}$ & ${12.2}{\phn}$ & ${\phn}{5.4}{\phn}$ & ${13.8}{\phn}$ & ${3.1}$ & ${1.4}$ & ${4.1}$ & ${2.0}$ & \nodata &  \\
137\dotfill&\phm{+}${0.399}$ & ${-0.100}$ & \phm{+}${\phn}{29}$ & $0.44$ & $0.35$ & \phm{+}${\phn}{7}$ & ${16.1}$ & ${\phn}{\phn}{6.4}$ & ${\phn}{2.4}{\phn}$ & ${\phn}{1.4}{\phn}$ & ${\phn}{6.1}{\phn}$ & ${2.5}$ & ${1.9}$ & ${3.8}$ & ${1.9}$ & \nodata &  \\
138\dotfill&\phm{+}${0.401}$ & ${-0.034}$ & \phm{+}${\phn}{85}$ & $0.85$ & $0.69$ & ${-16}$ & ${12.8}$ & ${\phn}{\phn}{9.8}$ & ${\phn}{5.4}{\phn}$ & ${\phn}{1.9}{\phn}$ & ${\phn}{6.3}{\phn}$ & ${2.9}$ & ${1.6}$ & ${3.8}$ & ${2.0}$ & \nodata &  \\
139\dotfill&\phm{+}${0.408}$ & ${-0.079}$ & \phm{+}${100}$ & $0.71$ & $0.54$ & ${-22}$ & ${\phn}{7.8}$ & ${\phn}{\phn}{2.7}$ & ${\phn}{1.4}{\phn}$ & ${\phn}{0.47}$ & ${\phn}{1.4}{\phn}$ & ${2.1}$ & ${1.9}$ & ${4.0}$ & ${2.0}$ & \nodata &  \\
140\dotfill&\phm{+}${0.411}$ & \phm{+}${0.042}$ & \phm{+}${\phn}{23}$ & $1.63$ & $0.58$ & \phm{+}${75}$ & ${\phn}{8.1}$ & ${\phn}{\phn}{2.8}$ & ${\phn}{1.4}{\phn}$ & ${\phn}{1.7}{\phn}$ & ${\phn}{4.8}{\phn}$ & ${2.9}$ & ${1.2}$ & ${4.7}$ & ${1.9}$ & \nodata & cloud~d \\
141\dotfill&\phm{+}${0.472}$ & ${-0.099}$ & \phm{+}${\phn}{28}$ & $0.62$ & $0.56$ & \phm{+}${23}$ & ${\phn}{9.5}$ & ${\phn}{\phn}{3.6}$ & ${\phn}{2.2}{\phn}$ & ${\phn}{0.69}$ & ${\phn}{3.6}{\phn}$ & ${2.5}$ & ${1.6}$ & ${4.0}$ & ${1.8}$ & \nodata &  \\
142\dotfill&\phm{+}${0.478}$ & ${-0.011}$ & \phm{+}${\phn}{26}$ & $0.45$ & $0.42$ & \phm{+}${67}$ & ${\phn}{5.2}$ & ${\phn}{\phn}{7.6}$ & ${\phn}{3.3}{\phn}$ & ${\phn}{4.5}{\phn}$ & ${10.9}{\phn}$ & ${2.8}$ & ${0.6}$ & ${4.4}$ & ${1.9}$ & m & cloud~e \\
143\dotfill&\phm{+}${0.482}$ & ${-0.013}$ & ${\phn}{-22}$ & $0.43$ & $0.42$ & ${-12}$ & ${\phn}{5.7}$ & ${\phn}{\phn}{2.6}$ & \nodata & \nodata & ${\phn}{0.12}$ & ${0.8}$ & ${2.7}$ & \nodata & \nodata & \nodata &  \\
144\dotfill&\phm{+}${0.487}$ & \phm{+}${0.066}$ & \phm{+}${\phn}{37}$ & $0.51$ & $0.47$ & ${-12}$ & ${\phn}{8.0}$ & ${\phn}{\phn}{2.3}$ & ${\phn}{1.2}{\phn}$ & ${\phn}{0.36}$ & ${\phn}{2.7}{\phn}$ & ${2.2}$ & ${1.6}$ & ${4.1}$ & ${1.9}$ & \nodata &  \\
145\dotfill&\phm{+}${0.491}$ & ${-0.108}$ & \phm{+}${\phn}{35}$ & $0.58$ & $0.50$ & ${-76}$ & ${11.5}$ & ${\phn}{\phn}{5.4}$ & ${\phn}{3.0}{\phn}$ & ${\phn}{1.4}{\phn}$ & ${\phn}{7.2}{\phn}$ & ${2.7}$ & ${1.5}$ & ${4.1}$ & ${1.8}$ & \nodata &  \\
146\dotfill&\phm{+}${0.509}$ & ${-0.088}$ & \phm{+}${\phn}{34}$ & $0.57$ & $0.56$ & \phm{+}${25}$ & ${\phn}{6.4}$ & ${\phn}{\phn}{2.3}$ & ${\phn}{0.79}$ & ${\phn}{0.32}$ & ${\phn}{2.5}{\phn}$ & ${2.3}$ & ${1.4}$ & \nodata & \nodata & \nodata &  \\
147\dotfill&\phm{+}${0.542}$ & \phm{+}${0.052}$ & \phm{+}${\phn}{30}$ & $0.92$ & $0.58$ & ${-62}$ & ${\phn}{8.2}$ & ${\phn}{\phn}{5.0}$ & ${\phn}{2.4}{\phn}$ & ${\phn}{2.4}{\phn}$ & ${\phn}{5.0}{\phn}$ & ${2.7}$ & ${1.3}$ & ${4.3}$ & ${1.9}$ & \nodata &  \\
148\dotfill&\phm{+}${0.551}$ & ${-0.065}$ & \phm{+}${\phn}{83}$ & $0.85$ & $0.41$ & ${-13}$ & ${\phn}{6.4}$ & ${\phn}{\phn}{2.6}$ & ${\phn}{1.4}{\phn}$ & ${\phn}{0.59}$ & ${\phn}{2.1}{\phn}$ & ${2.2}$ & ${1.5}$ & ${4.0}$ & ${1.9}$ & \nodata &  \\
149\dotfill&\phm{+}${0.551}$ & ${-0.022}$ & \phm{+}${\phn}{56}$ & $0.40$ & $0.32$ & ${-40}$ & ${\phn}{4.8}$ & ${\phn}{\phn}{1.7}$ & ${\phn}{0.57}$ & ${\phn}{0.67}$ & ${\phn}{1.4}{\phn}$ & ${1.8}$ & ${1.4}$ & ${4.3}$ & ${1.9}$ & \nodata &  \\
150\dotfill&\phm{+}${0.557}$ & \phm{+}${0.038}$ & ${\phn}{-24}$ & $0.71$ & $0.55$ & ${-63}$ & ${10.3}$ & ${\phn}{\phn}{6.7}$ & ${\phn}{3.2}{\phn}$ & ${\phn}{2.6}{\phn}$ & ${\phn}{2.4}{\phn}$ & ${2.3}$ & ${1.8}$ & ${3.8}$ & ${2.0}$ & \nodata &  \\
151\dotfill&\phm{+}${0.570}$ & ${-0.028}$ & ${\phn}{\phn}{-3}$ & $0.93$ & $0.73$ & \phm{+}${45}$ & ${\phn}{8.9}$ & ${\phn}{\phn}{6.3}$ & ${\phn}{2.4}{\phn}$ & ${\phn}{1.1}{\phn}$ & ${\phn}{4.4}{\phn}$ & ${2.8}$ & ${1.4}$ & \nodata & \nodata & \nodata &  \\
152\dotfill&\phm{+}${0.573}$ & ${-0.033}$ & \phm{+}${\phn}{52}$ & $0.48$ & $0.48$ & ${-48}$ & ${\phn}{6.7}$ & ${\phn}{\phn}{3.0}$ & ${\phn}{1.3}{\phn}$ & ${\phn}{0.83}$ & ${\phn}{3.5}{\phn}$ & ${2.3}$ & ${1.3}$ & ${4.4}$ & ${1.8}$ & \nodata &  \\
153\dotfill&\phm{+}${0.583}$ & ${-0.071}$ & \phm{+}${\phn}{88}$ & $1.19$ & $0.77$ & \phm{+}${27}$ & ${11.1}$ & ${\phn}{\phn}{9.7}$ & ${\phn}{4.6}{\phn}$ & ${\phn}{1.9}{\phn}$ & ${\phn}{6.4}{\phn}$ & ${3.0}$ & ${1.4}$ & ${4.0}$ & ${2.0}$ & \nodata &  \\
154\dotfill&\phm{+}${0.602}$ & ${-0.150}$ & \phm{+}${\phn}{94}$ & $0.82$ & $0.54$ & ${-29}$ & ${14.7}$ & ${\phn}{\phn}{9.5}$ & ${\phn}{4.1}{\phn}$ & ${\phn}{1.2}{\phn}$ & ${\phn}{8.5}{\phn}$ & ${2.9}$ & ${1.6}$ & ${3.9}$ & ${2.0}$ & \nodata &  \\
155\dotfill&\phm{+}${0.603}$ & \phm{+}${0.003}$ & \phm{+}${136}$ & $0.39$ & $0.35$ & \phm{+}${39}$ & ${20.1}$ & ${\phn}{\phn}{4.8}$ & ${\phn}{0.88}$ & ${\phn}{0.55}$ & ${\phn}{0.30}$ & ${1.1}$ & ${3.4}$ & \nodata & \nodata & \nodata &  \\
156\dotfill&\phm{+}${0.615}$ & ${-0.060}$ & \phm{+}${\phn}{52}$ & $0.86$ & $0.73$ & ${-68}$ & ${\phn}{9.5}$ & ${\phn}{11.8}$ & ${\phn}{3.5}{\phn}$ & ${\phn}{4.0}{\phn}$ & ${\phn}{4.1}{\phn}$ & ${2.7}$ & ${1.5}$ & ${4.3}$ & ${1.9}$ & \nodata &  \\
157\dotfill&\phm{+}${0.623}$ & ${-0.100}$ & \phm{+}${\phn}{65}$ & $0.51$ & $0.46$ & \phm{+}${24}$ & ${\phn}{7.0}$ & ${\phn}{10.6}$ & ${\phn}{4.0}{\phn}$ & ${\phn}{3.1}{\phn}$ & ${\phn}{6.7}{\phn}$ & ${2.6}$ & ${1.1}$ & ${4.5}$ & ${1.9}$ & \nodata &  \\
158\dotfill&\phm{+}${0.626}$ & ${-0.093}$ & \phm{+}${\phn}{26}$ & $0.14$ & $0.14$ & \phm{+}${72}$ & ${\phn}{7.0}$ & ${\phn}{10.9}$ & ${\phn}{2.5}{\phn}$ & ${\phn}{2.2}{\phn}$ & ${\phn}{3.1}{\phn}$ & ${1.9}$ & ${1.2}$ & ${4.1}$ & ${2.0}$ & \nodata &  \\
159\dotfill&\phm{+}${0.628}$ & ${-0.031}$ & \phm{+}${114}$ & $0.51$ & $0.42$ & ${-34}$ & ${10.0}$ & ${\phn}{\phn}{6.9}$ & ${\phn}{2.7}{\phn}$ & ${\phn}{1.7}{\phn}$ & ${\phn}{3.9}{\phn}$ & ${2.3}$ & ${1.6}$ & ${4.3}$ & ${2.0}$ & \nodata &  \\
160\dotfill&\phm{+}${0.631}$ & ${-0.063}$ & \phm{+}${\phn}{43}$ & $0.95$ & $0.66$ & \phm{+}${30}$ & ${\phn}{6.1}$ & ${\phn}{19.6}$ & ${\phn}{5.3}{\phn}$ & ${\phn}{6.0}{\phn}$ & ${\phn}{6.5}{\phn}$ & ${2.9}$ & ${0.9}$ & ${4.4}$ & ${2.0}$ & \nodata &  \\
161\dotfill&\phm{+}${0.636}$ & \phm{+}${0.056}$ & \phm{+}${\phn}{47}$ & $0.69$ & $0.43$ & ${-79}$ & ${17.1}$ & ${\phn}{11.6}$ & ${\phn}{4.0}{\phn}$ & ${\phn}{5.2}{\phn}$ & ${\phn}{7.7}{\phn}$ & ${2.7}$ & ${1.8}$ & ${4.2}$ & ${2.0}$ & \nodata &  \\
162\dotfill&\phm{+}${0.651}$ & ${-0.142}$ & \phm{+}${\phn}{19}$ & $0.33$ & $0.26$ & \phm{+}${10}$ & ${\phn}{4.0}$ & ${\phn}{\phn}{3.0}$ & ${\phn}{1.0}{\phn}$ & ${\phn}{0.89}$ & ${\phn}{1.2}{\phn}$ & ${1.6}$ & ${1.4}$ & ${4.3}$ & ${1.9}$ & \nodata &  \\
163\dotfill&\phm{+}${0.662}$ & ${-0.036}$ & \phm{+}${\phn}{47}$ & $1.08$ & $0.83$ & ${-47}$ & ${\phn}{8.6}$ & ${141.8}$ & ${20.1}{\phn}$ & ${28.2}{\phn}$ & ${34.7}{\phn}$ & ${3.7}$ & ${0.4}$ & ${4.1}$ & ${2.5}$ & m & Sgr~B2 \\
164\dotfill&\phm{+}${0.665}$ & ${-0.052}$ & \phm{+}${\phn}{76}$ & $0.59$ & $0.48$ & ${-41}$ & ${\phn}{4.1}$ & ${\phn}{\phn}{6.7}$ & ${\phn}{1.6}{\phn}$ & ${\phn}{1.1}{\phn}$ & ${\phn}{2.6}{\phn}$ & ${2.2}$ & ${1.0}$ & ${4.2}$ & ${2.1}$ & g &  \\
165\dotfill&\phm{+}${0.666}$ & ${-0.037}$ & \phm{+}${\phn}{85}$ & $0.44$ & $0.35$ & ${-22}$ & ${14.7}$ & ${\phn}{62.2}$ & ${12.8}{\phn}$ & ${\phn}{6.5}{\phn}$ & ${25.2}{\phn}$ & ${3.1}$ & ${1.2}$ & ${4.2}$ & ${2.1}$ & \nodata & Sgr~B2 \\
166\dotfill&\phm{+}${0.667}$ & ${-0.036}$ & ${-100}$ & $0.55$ & $0.43$ & ${-35}$ & ${12.4}$ & ${\phn}{10.8}$ & \nodata & ${\phn}{0.29}$ & \nodata & \nodata & \nodata & \nodata & \nodata & \nodata &  \\
167\dotfill&\phm{+}${0.676}$ & ${-0.029}$ & ${\phn}{-22}$ & $0.61$ & $0.46$ & \phm{+}${\phn}{9}$ & ${\phn}{8.6}$ & ${\phn}{\phn}{8.1}$ & ${\phn}{0.82}$ & ${\phn}{0.48}$ & \nodata & \nodata & \nodata & \nodata & \nodata & \nodata &  \\
168\dotfill&\phm{+}${0.677}$ & ${-0.064}$ & \phm{+}${\phn}{38}$ & $0.96$ & $0.68$ & ${-77}$ & ${\phn}{8.5}$ & ${\phn}{17.9}$ & ${\phn}{5.2}{\phn}$ & ${\phn}{3.4}{\phn}$ & ${\phn}{5.5}{\phn}$ & ${2.8}$ & ${1.3}$ & ${4.4}$ & ${2.0}$ & \nodata &  \\
169\dotfill&\phm{+}${0.677}$ & ${-0.028}$ & ${\phn}{-92}$ & $0.55$ & $0.40$ & ${-34}$ & ${13.3}$ & ${\phn}{22.1}$ & ${\phn}{0.82}$ & ${\phn}{0.84}$ & \nodata & \nodata & \nodata & \nodata & \nodata & \nodata &  \\
170\dotfill&\phm{+}${0.678}$ & ${-0.109}$ & \phm{+}${\phn}{11}$ & $0.40$ & $0.28$ & ${-53}$ & ${\phn}{5.3}$ & ${\phn}{\phn}{5.7}$ & ${\phn}{1.8}{\phn}$ & ${\phn}{1.3}{\phn}$ & ${\phn}{2.0}{\phn}$ & ${1.9}$ & ${1.4}$ & \nodata & \nodata & \nodata &  \\
171\dotfill&\phm{+}${0.681}$ & ${-0.081}$ & \phm{+}${\phn}{28}$ & $0.81$ & $0.41$ & ${-15}$ & ${\phn}{7.0}$ & ${\phn}{\phn}{3.5}$ & ${\phn}{0.92}$ & ${\phn}{0.94}$ & ${\phn}{0.63}$ & ${1.7}$ & ${2.1}$ & ${4.2}$ & ${2.0}$ & \nodata &  \\
172\dotfill&\phm{+}${0.684}$ & ${-0.172}$ & \phm{+}${\phn}{20}$ & $0.52$ & $0.36$ & ${-24}$ & ${\phn}{5.6}$ & ${\phn}{\phn}{3.3}$ & ${\phn}{1.6}{\phn}$ & ${\phn}{1.4}{\phn}$ & ${\phn}{2.0}{\phn}$ & ${2.0}$ & ${1.4}$ & ${4.3}$ & ${1.9}$ & \nodata &  \\
173\dotfill&\phm{+}${0.692}$ & ${-0.028}$ & \phm{+}${\phn}{80}$ & $0.58$ & $0.48$ & ${-86}$ & ${\phn}{6.4}$ & ${\phn}{\phn}{9.8}$ & ${\phn}{2.8}{\phn}$ & ${\phn}{6.3}{\phn}$ & ${\phn}{4.6}{\phn}$ & ${2.5}$ & ${1.2}$ & ${4.2}$ & ${2.3}$ & \nodata &  \\
174\dotfill&\phm{+}${0.695}$ & ${-0.171}$ & \phm{+}${\phn}{21}$ & $0.38$ & $0.31$ & ${-75}$ & ${\phn}{6.4}$ & ${\phn}{\phn}{7.4}$ & ${\phn}{3.2}{\phn}$ & ${\phn}{2.9}{\phn}$ & ${\phn}{3.7}{\phn}$ & ${2.2}$ & ${1.3}$ & ${4.3}$ & ${1.9}$ & \nodata &  \\
175\dotfill&\phm{+}${0.700}$ & \phm{+}${0.033}$ & \phm{+}${\phn}{39}$ & $0.26$ & $0.18$ & \phm{+}${53}$ & ${\phn}{6.6}$ & ${\phn}{\phn}{4.4}$ & ${\phn}{1.6}{\phn}$ & ${\phn}{1.2}{\phn}$ & ${\phn}{1.3}{\phn}$ & ${1.6}$ & ${1.7}$ & ${4.3}$ & ${1.9}$ & \nodata &  \\
176\dotfill&\phm{+}${0.704}$ & ${-0.023}$ & \phm{+}${\phn}{49}$ & $0.58$ & $0.43$ & ${-22}$ & ${12.9}$ & ${\phn}{17.5}$ & ${\phn}{5.5}{\phn}$ & ${10.0}{\phn}$ & ${11.2}{\phn}$ & ${2.8}$ & ${1.4}$ & ${4.4}$ & ${2.1}$ & \nodata &  \\
177\dotfill&\phm{+}${0.716}$ & ${-0.106}$ & \phm{+}${\phn}{87}$ & $0.60$ & $0.55$ & \phm{+}${47}$ & ${\phn}{4.9}$ & ${\phn}{\phn}{4.3}$ & ${\phn}{1.9}{\phn}$ & ${\phn}{0.60}$ & ${\phn}{3.3}{\phn}$ & ${2.4}$ & ${1.1}$ & ${4.1}$ & ${1.9}$ & \nodata &  \\
178\dotfill&\phm{+}${0.719}$ & \phm{+}${0.039}$ & \phm{+}${\phn}{\phn}{0}$ & $0.45$ & $0.38$ & ${-74}$ & ${11.7}$ & ${\phn}{11.5}$ & ${\phn}{4.5}{\phn}$ & ${\phn}{2.1}{\phn}$ & ${\phn}{2.8}{\phn}$ & ${2.1}$ & ${1.9}$ & \nodata & \nodata & \nodata &  \\
179\dotfill&\phm{+}${0.723}$ & ${-0.136}$ & \phm{+}${\phn}{38}$ & $0.35$ & $0.30$ & ${-44}$ & ${\phn}{7.9}$ & ${\phn}{\phn}{4.3}$ & ${\phn}{1.9}{\phn}$ & ${\phn}{1.4}{\phn}$ & ${\phn}{1.5}{\phn}$ & ${1.8}$ & ${1.9}$ & ${4.2}$ & ${2.0}$ & \nodata &  \\
180\dotfill&\phm{+}${0.733}$ & ${-0.017}$ & \phm{+}${\phn}{\phn}{9}$ & $0.16$ & $0.16$ & ${-60}$ & ${\phn}{6.0}$ & ${\phn}{\phn}{6.7}$ & ${\phn}{2.1}{\phn}$ & ${\phn}{1.8}{\phn}$ & ${\phn}{1.9}{\phn}$ & ${1.7}$ & ${1.4}$ & \nodata & \nodata & \nodata &  \\
181\dotfill&\phm{+}${0.733}$ & ${-0.000}$ & \phm{+}${\phn}{75}$ & $0.96$ & $0.52$ & ${-66}$ & ${13.8}$ & ${\phn}{\phn}{5.7}$ & ${\phn}{2.8}{\phn}$ & ${\phn}{1.5}{\phn}$ & ${\phn}{4.1}{\phn}$ & ${2.6}$ & ${1.8}$ & ${4.2}$ & ${1.9}$ & \nodata &  \\
182\dotfill&\phm{+}${0.734}$ & ${-0.158}$ & \phm{+}${\phn}{42}$ & $0.84$ & $0.43$ & \phm{+}${\phn}{7}$ & ${\phn}{8.6}$ & ${\phn}{\phn}{9.8}$ & ${\phn}{3.6}{\phn}$ & ${\phn}{4.6}{\phn}$ & ${\phn}{4.1}{\phn}$ & ${2.5}$ & ${1.5}$ & ${4.2}$ & ${2.0}$ & \nodata &  \\
183\dotfill&\phm{+}${0.737}$ & ${-0.151}$ & \phm{+}${\phn}{20}$ & $0.62$ & $0.45$ & \phm{+}${33}$ & ${\phn}{5.6}$ & ${\phn}{\phn}{3.5}$ & ${\phn}{1.2}{\phn}$ & ${\phn}{1.2}{\phn}$ & ${\phn}{1.2}{\phn}$ & ${1.9}$ & ${1.6}$ & ${4.2}$ & ${2.0}$ & \nodata &  \\
184\dotfill&\phm{+}${0.744}$ & \phm{+}${0.078}$ & \phm{+}${\phn}{25}$ & $0.73$ & $0.25$ & ${-77}$ & ${11.2}$ & ${\phn}{\phn}{4.0}$ & ${\phn}{1.8}{\phn}$ & ${\phn}{0.89}$ & ${\phn}{1.5}{\phn}$ & ${1.9}$ & ${2.2}$ & \nodata & \nodata & \nodata &  \\
185\dotfill&\phm{+}${0.752}$ & ${-0.074}$ & \phm{+}${\phn}{\phn}{9}$ & $0.96$ & $0.53$ & ${\phn}{-9}$ & ${10.6}$ & ${\phn}{37.3}$ & ${\phn}{8.9}{\phn}$ & ${11.6}{\phn}$ & ${\phn}{9.3}{\phn}$ & ${3.0}$ & ${1.2}$ & ${4.1}$ & ${2.2}$ & \nodata &  \\
186\dotfill&\phm{+}${0.757}$ & ${-0.159}$ & \phm{+}${\phn}{23}$ & $0.76$ & $0.30$ & ${-39}$ & ${\phn}{4.6}$ & ${\phn}{\phn}{2.7}$ & ${\phn}{0.92}$ & ${\phn}{1.2}{\phn}$ & ${\phn}{1.1}{\phn}$ & ${1.8}$ & ${1.5}$ & ${4.2}$ & ${2.0}$ & \nodata &  \\
187\dotfill&\phm{+}${0.764}$ & ${-0.064}$ & \phm{+}${\phn}{\phn}{1}$ & $0.56$ & $0.54$ & ${-36}$ & ${\phn}{7.9}$ & ${\phn}{\phn}{7.8}$ & ${\phn}{2.5}{\phn}$ & ${\phn}{2.9}{\phn}$ & ${\phn}{2.8}{\phn}$ & ${2.3}$ & ${1.6}$ & ${4.1}$ & ${2.0}$ & \nodata &  \\
188\dotfill&\phm{+}${0.770}$ & ${-0.041}$ & \phm{+}${\phn}{87}$ & $0.64$ & $0.52$ & ${\phn}{-1}$ & ${\phn}{6.8}$ & ${\phn}{\phn}{6.3}$ & ${\phn}{3.9}{\phn}$ & ${\phn}{1.9}{\phn}$ & ${\phn}{7.5}{\phn}$ & ${2.8}$ & ${1.0}$ & ${4.2}$ & ${1.8}$ & \nodata &  \\
189\dotfill&\phm{+}${0.774}$ & ${-0.060}$ & ${\phn}{-49}$ & $0.55$ & $0.47$ & \phm{+}${35}$ & ${12.2}$ & ${\phn}{11.9}$ & ${\phn}{4.1}{\phn}$ & ${\phn}{3.1}{\phn}$ & ${\phn}{2.0}{\phn}$ & ${2.1}$ & ${2.1}$ & ${3.9}$ & ${2.1}$ & \nodata &  \\
190\dotfill&\phm{+}${0.780}$ & ${-0.115}$ & \phm{+}${\phn}{46}$ & $0.64$ & $0.59$ & \phm{+}${\phn}{5}$ & ${\phn}{7.2}$ & ${\phn}{\phn}{9.5}$ & ${\phn}{3.7}{\phn}$ & ${\phn}{1.8}{\phn}$ & ${\phn}{6.9}{\phn}$ & ${2.8}$ & ${1.1}$ & ${4.3}$ & ${1.9}$ & \nodata &  \\
191\dotfill&\phm{+}${0.781}$ & ${-0.021}$ & \phm{+}${\phn}{24}$ & $0.55$ & $0.52$ & \phm{+}${53}$ & ${\phn}{8.9}$ & ${\phn}{\phn}{8.8}$ & ${\phn}{4.8}{\phn}$ & ${\phn}{4.2}{\phn}$ & ${\phn}{9.4}{\phn}$ & ${2.8}$ & ${1.1}$ & ${4.4}$ & ${1.9}$ & \nodata &  \\
192\dotfill&\phm{+}${0.784}$ & ${-0.169}$ & \phm{+}${\phn}{33}$ & $0.72$ & $0.39$ & ${-16}$ & ${\phn}{5.3}$ & ${\phn}{\phn}{3.5}$ & ${\phn}{1.6}{\phn}$ & ${\phn}{1.5}{\phn}$ & ${\phn}{2.7}{\phn}$ & ${2.3}$ & ${1.2}$ & ${4.3}$ & ${1.9}$ & \nodata &  \\
193\dotfill&\phm{+}${0.786}$ & ${-0.154}$ & \phm{+}${\phn}{39}$ & $0.60$ & $0.53$ & \phm{+}${20}$ & ${\phn}{6.7}$ & ${\phn}{\phn}{5.6}$ & ${\phn}{2.2}{\phn}$ & ${\phn}{1.9}{\phn}$ & ${\phn}{3.7}{\phn}$ & ${2.4}$ & ${1.3}$ & ${4.2}$ & ${2.0}$ & \nodata &  \\
194\dotfill&\phm{+}${0.794}$ & ${-0.025}$ & \phm{+}${\phn}{19}$ & $0.78$ & $0.60$ & ${-67}$ & ${10.1}$ & ${\phn}{\phn}{5.9}$ & ${\phn}{3.0}{\phn}$ & ${\phn}{2.3}{\phn}$ & ${\phn}{4.1}{\phn}$ & ${2.6}$ & ${1.6}$ & ${4.3}$ & ${1.9}$ & \nodata &  \\
195\dotfill&\phm{+}${0.803}$ & ${-0.218}$ & \phm{+}${\phn}{88}$ & $0.53$ & $0.48$ & ${-52}$ & ${\phn}{7.2}$ & ${\phn}{\phn}{2.2}$ & ${\phn}{1.2}{\phn}$ & ${\phn}{0.73}$ & ${\phn}{2.8}{\phn}$ & ${2.2}$ & ${1.5}$ & ${4.5}$ & ${1.8}$ & \nodata &  \\
\enddata
\tablenotetext{a}{abbreviations for the star formation flag:  g = {\it Spitzer} green object,  w = water maser, m = class-II methanol maser, c = cluster-forming region}
\end{deluxetable}
\end{longrotatetable}

\ifcompilable
\fi

\section{PRINCIPAL COMPONENT ANALYSIS\label{section:pca}}

This section describes the PCA that we applied to the HCN clumps.
By using the PCA, we attempt to investigate the intercorrelation among the clump parameters and thereby find the fundamental parameter combinations that characterize the physical properties of the clumps.

PCA converts given $N$ variables, each represented by a vector of the dimension of the sample size, into the same number of linear combinations of the parameters that are orthogonal to each other.
We use a normalized parameter space $\pvt_i = \left(\pvi - \bar{\pvi}\right)/\sigma_i$, where $\pvi$ is a vector whose elements are the sample values of the $i^{\mathrm{th}}$ variable, and $\bar{\pvi}$ and $\sigma_i$ are its average and standard deviation, respectively.
The diagonalized form of the covariance matrix of the normalized parameters is expressed using matrix $\tilde{P} \equiv \left(\pvt_1, \pvt_2, ..., \pvt_N\right)$ as
\begin{eqnarray}
\left(
\begin{array}{cccc}
\lambda_1 & & & 0 \\ & \lambda_2 & & \\ & & \ddots & \\ 0 & & &
\lambda_N \end{array} \right) &=&
\transverse{V}\transverse{\tilde{P}}\tilde{P}V, \label{eqn:eigeneq}
\end{eqnarray}
where $\lambda$ and $V$ are eigenvalues and $N\times N$ square matrix whose columns are eigenvectors of the covariance matrix $\transverse{\tilde{P}}\tilde{P}$, respectively.
Equation \ref{eqn:eigeneq} means that the new orthogonal parameters are given as the columns of the matrix $\tilde{P}V \equiv \left(\qv_1, \qv_2, ..., \qv_N\right)$.
The vectors \qvi\ are the principal components (PCs), which are labeled as PC$i$ ($i$=1, 2, ..., $N$) in the decreasing order of the corresponding eigenvalues $\lambda_i$.
The proportions of individual $\lambda_i$ to the total variance of the normalized parameter space ($\sum \lambda_i = N^2$) are referred to as the contribution ratios.
The eigenvectors which convert the \pvt\ space into the PC space are called the loading vectors.

We apply PCA to the parameter space of base-10 logarithms of radius $r\equiv \sqrt{\rmaj\cdot\rmin}$, \Dv, $M$, \Tkin, and \nHH\ of the 130 clumps for which all these parameters are obtained.
Values of $\bar{\pvi}$ and $\sigma_i$ are listed in Table \ref{table:pcadata}.
Table \ref{table:pca}\ tabulates the loading vectors and the contribution ratios, with errors estimated using the Jack-Knife (JK) method.
We explore the physical interpretation of the PCs in the following subsections.

\begin{deluxetable}{lccccc}
\tablecaption{Averages and Standard Deviation of Parameters Used For
  PCA\label{table:pcadata}} \input{table_pcadata.etex}
\end{deluxetable}

\begin{deluxetable*}{lcccccc}
\tablecaption{Results of PCA\label{table:pca}} \input{table_pca.etex}
\end{deluxetable*}

\subsection{PC1: Size--Mass and Size-Linewidth Relations\label{section:pc1}}
The PC1 loading vector indicates the direction of the largest dispersion in the normalized parameter space.
Hence, PC1 represents scaling relations among $r$, $M$, and \dv, whose PC1 loadings are large.

\begin{figure}
\ifdraft
\epsscale{0.5}
\else
\epsscale{1.}
\fi
\plotone{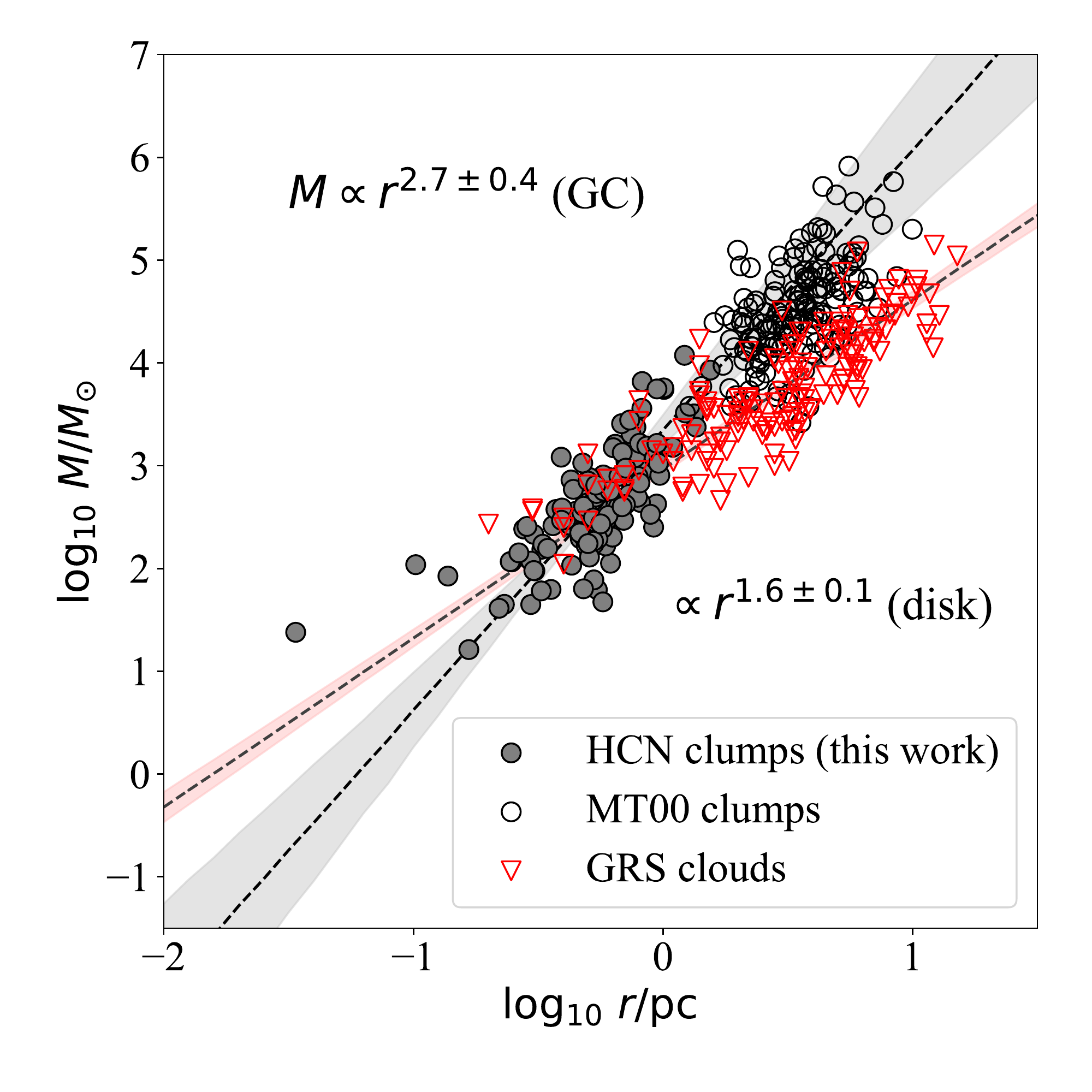}
\caption{ Size-mass relationship for the HCN clumps (this work), MT00
  clouds \citep{Miyazaki2000}, and GRS clouds
  \citep{Jackson2006,Roman-Duval2010}, denoted by filled circles, open
  circles, and triangles, respectively.  The dashed lines are the
  best-fit curves for the GC clouds (HCN clumps and MT00 clouds) and
  the GRS clouds, with uncertainties represented by hatched
  regions.}\label{fig:sizemass}
\end{figure}

Figure \ref{fig:sizemass} compares $r$--$M$ plots for the GC clouds (the HCN clumps and MT00 clouds) and the GRS clouds.
The HCN clumps and MT00 clouds are plotted on a consistent power-law relation with the index of $2.7\pm0.4$, which is steeper than the relation for the GRS clouds whose index is $1.6\pm0.1$.

The steep $r$--$M$ relation {\it among} the CMZ clouds may translate into generally steep $r$--$M$ relations {\it within} individual clouds indicative of their flat density profiles \citep{Kauffmann2017b}, by assuming that they have density profiles independent of size, mass, and mean density.
This assumption is corroborated by the presence of a consistent $r$--$M$ relation across the HCN clumps and the MT00 clouds, as a significant fraction of the CMZ clouds are identified by both of them but in different size scales.
Indeed, the $r$--$M$ index of 2.7 is close to the $r$--$M$ slopes of the mass distribution inside many CMZ clouds \citep{Kauffmann2017b}.

Figure \ref{fig:sizevel} presents the $r$--\Dv\ plots for the HCN clumps and MT00 clouds.
No significant correlation is found in the HCN clumps alone, reflecting the larger PC2 loading for \Dv\ than those for $r$ and $M$;
however, the HCN clumps and MT00 cloud can be placed on a consistent relationship with the power-law index of $\sim 0.7$ that is obtained from the PC1 loadings.
The $r$--\Dv\ relation steeper than the Galactic disk clouds with an index of $\sim 0.5$ is consistent with previous studies \citep{Shetty2012,Kauffmann2017a}.

\begin{revision}
The universality of the steep $r$--\Dv\ index in the CMZ is recently questioned by \cite{Henshaw2019}, who measured velocity dispersion at $0.07$-pc spatial scale to be $\sim 4\ \kmps$ (corresponding to $\sim 10\ \kmps$ in FWHM), which lies a factor of 4--9 above the steep size--line width relationships in \cite{Shetty2012} and \cite{Kauffmann2017a}.  \cite{Henshaw2019} argue that this inconsistency may imply a shallower size--line width index, or originates from uncertainty in the cloud extent in the line-of-sight dimension.
We confirm this inconsistency with our results; 
the mean FWHM velocity width of the HCN clumps, which is $\sim 10\ \kmps$, is similar to the value in \cite{Henshaw2019} despite the nearly an order of magnitude difference in the spatial scale between the two measurements.
This could imply an almost flat size--velocity width index below a few 0.1 pc spatial scale; however, a more elaborate comparison based on the same measurement method using higher-resolution images of the HCN clumps will be necessary to conclude.
\end{revision}

\subsection{PC2: Turbulent Heating\label{subsection:turbheat}}
PC2 creates scatter around the scaling relations represented by PC1.
The PC2 loading vector has negative elements for \Tkin\ and \Dv, and a large positive element for \nHH;
this means that larger \Dv\ from the $r$--\Dv\ relation tends to be accompanied by high \Tkin\ and low \nHH,
which is qualitatively consistent with the mechanical heating by shock waves \citep{Ao2013,Ginsburg2016}.

\begin{figure*}
\epsscale{1.}  \plotone{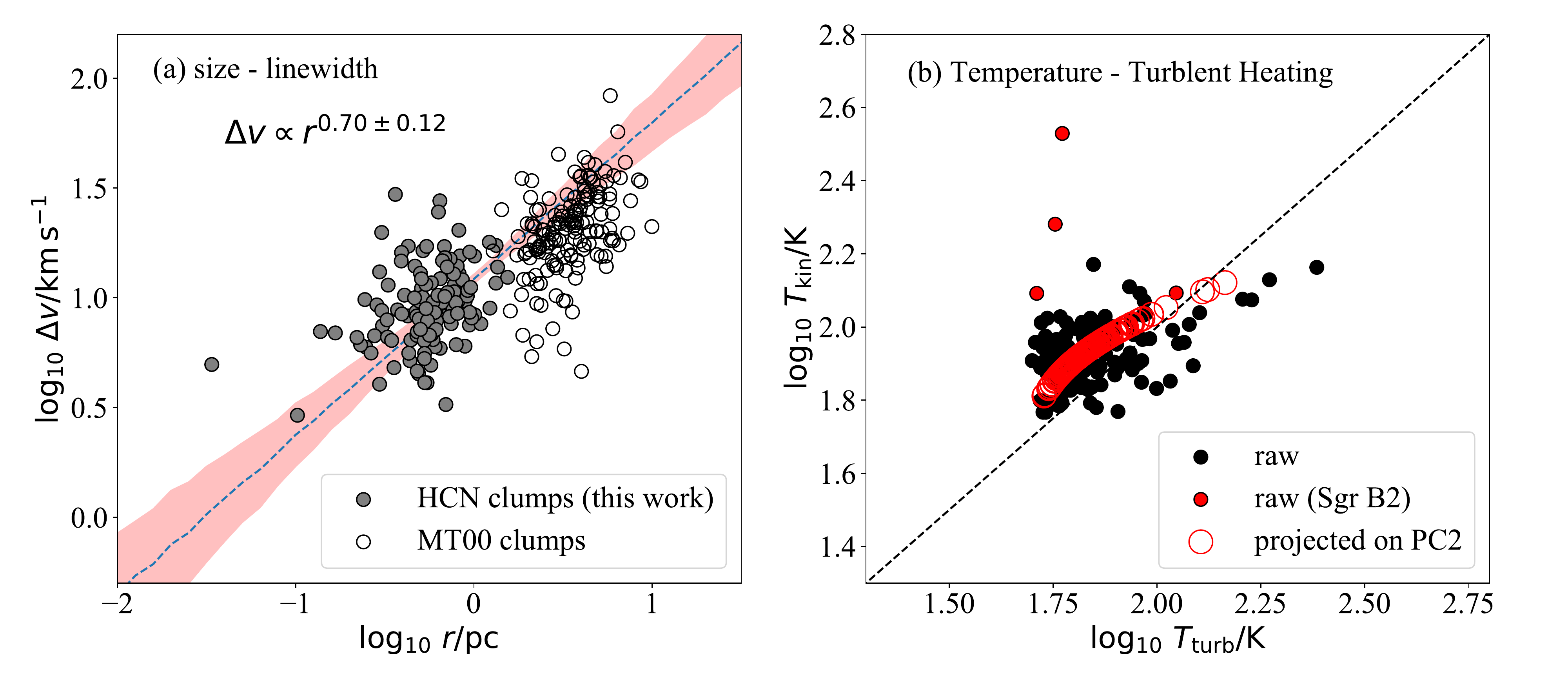}
\caption{ (a) Radius (r) -- linewidth (\Dv) plots for the HCN clumps
  (this work) and MT00 clouds \citep{Miyazaki2000}, denoted by filled
  and open circles, respectively.  (b) Temperatures calculated from
  turbulence-heating model (\Tturb) vs the kinetic temperatures
  (\Tkin) obtained from excitation analysis of multi-line molecular
  line data \citep{Tanaka2018b}.  The dashed line is the $\Tv = \Tkin$
  relation. \label{fig:sizevel}}
\end{figure*}

Figure \ref{fig:sizevel}b compares \Tkin\ and turbulence-heated temperature \Tturb\ obtained by using the formula given by \cite{Ao2013}, which is the solution of the thermal balance equation between the molecular line cooling and the turbulent heating, with a minor contribution from the dust cooling and the cosmic-ray heating terms.
For the dust temperature and cosmic-ray flux, we adopt 20 K \citep{Pierce-Price2000,Molinari2011} and $4\times 10^{-15}\ \mathrm{s}^{-1}$ \citep{OkaTakeshi2005,Goto2008}, respectively, for all clumps.
Figure \ref{fig:sizevel}b shows that the calculated \Tturb\ values are approximately similar to \Tkin, distributed with a $\sim0.2$ dex scatter around the $\Tturb = \Tkin$ correlation.
Although a few clumps have substantially lower \Tturb\ than \Tkin, they are all located near the Sgr~B2 cluster-forming region, where extremely intense UV heating is present or \Tkin\ measurement is possibly inaccurate due to the intense radio/infrared continuum \citep{Tanaka2018b}.

We repeat the same calculation but using the parameters that are projected onto the PC2 axis, i.e., the parameters from which contributions from all PCs other than PC2 are removed; with this calculation, we approximately extract the contribution of the PC2 in the thermal balance equation.
The result is shown in Figure \ref{fig:sizevel}b, in which the $\Tturb$--$\Tkin$ correlation is improved from that with the original parameters.
This confirms that the \Tturb--\Tkin\ correlation is created by PC2.

Thus, PC2 may be roughly understood as the effects of local shock events that heats molecular clouds and disturb the $r$--\Dv\ relation that is created by large-scale shocks.
Indeed, one of the largest negative PC2 scores in our sample is obtained for the CCC candidate cloud CO$-0.30$$-0.07$ (clumps \#51 and \#54; \citealt{Tanaka2014,Tanaka2015}), which has an extremely broad velocity width of $\sim 50\ \kmps$ indicative of violent shock.
Interferometric observations of CO$-0.30$$-0.07$ and another CCC candidate cloud CO$-0.40$$-0.22$ have detected intense temperature elevation in broad-velocity filaments, indicative of the correlation between molecular gas heating and local shock events in $\lesssim 0.1$ pc scale \citep{Tanaka2015,Tanaka2018a}.

\subsection{PC5: Virial Theorem\label{subsection:VT}}
PC5 is the direction of the least dispersion in the normalized parameter space, indicating that the clumps are aligned on the plane normal to the PC5 loading vector.
This translates into a multiplicative relationship among the parameters by converting the PC5 loadings into indices in the original parameter space:
\begin{eqnarray}
\biggl(\frac{r}{\pc}\biggr)^{2.0\pm0.5}\cdot\biggl(\frac{\Dv}{\kmps}\biggr)^{1.7\pm0.4}\cdot\biggl(\frac{M}{\Msol}\biggr)^{-1.3\pm0.1}\cdot
\nonumber \\ \biggl(\frac{\nHH}{\rm cm^{-3}}\biggr)^{0.9\pm0.3} =
10^{1.6\pm1.5}, \label{eqn:pc5}
\end{eqnarray}
where the \Tkin\ term is omitted as its PC5 loading is small and non-significant.

The signs of the indices of $r$, \Dv, and $M$ in Eq.\ref{eqn:pc5} (i.e., +, +, and $-$) are consistent with those of respective parameters in the expression of $\alpha \propto r\,\dv^2\,M^{-1}$, which infers that PC5 is related with the virial relation among the GC clumps.
By replacing \Dv\ with $\left(\frac{{8\,\mathrm{ln}2}\cdot\avir GM}{5R}\right)^{0.5}$ and using $M\propto r^{2.7}$, we obtain a virial theorem law dependent on \nHH;
\begin{eqnarray}
\avir \propto \nHH^{-0.7\pm0.2} \label{eq:vtlaw1}.
\end{eqnarray}
A more accurate formulation of \avir\ is directly obtained from multiple linear regression in the logarithmic parameter space:
\begin{eqnarray}
\avir \sim 10^{1.79\pm0.08}
\biggl(\frac{\Sigma}{100\ \Msol\,\pc^{-2}}\biggr)^{-0.52\pm0.08} \cdot
\nonumber \\ \biggl(\frac{\nHH}{10^4\ \rm
  cm^{-3}}\biggr)^{-0.61\pm0.12}, \label{eq:vtlaw2}
\end{eqnarray}
which includes additional dependence on surface mass density $\SigmaH\equiv M\,(\pi\,r^2)^{-1}$.  Factors depending on other parameters are omitted as their indices are similar to or less than theirerrors evaluated by JK method.

\begin{figure*}
\epsscale{1.}  \plotone{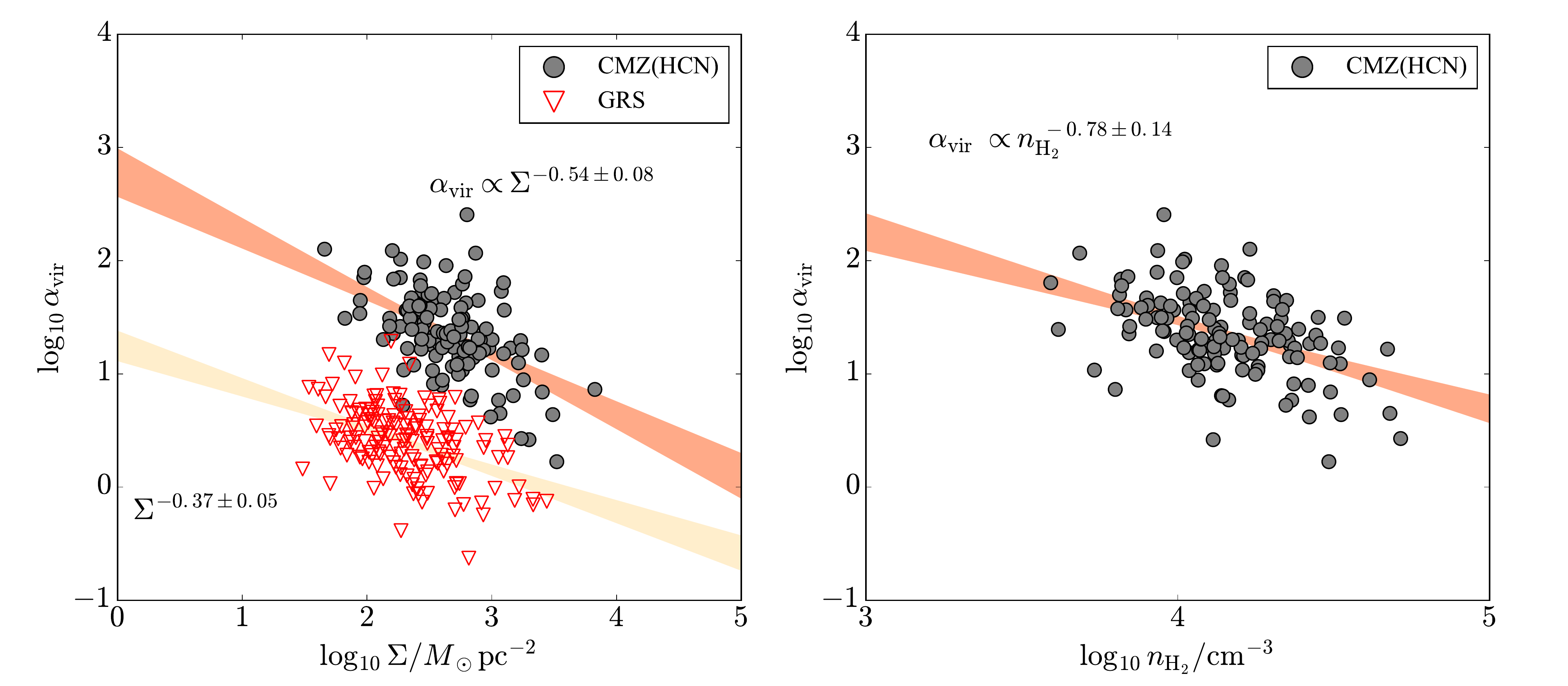}
\caption{ (a) Surface density (\SigmaH) -- virial parameter (\avir) and (b) Volume density (\nHH) -- \avir
  plots for the HCN clumps.  The same plots for the GRS clouds are overlaid on panel a.  The hatched regions
  indicate the result of linear regression with fitting uncertainty.\label{fig:virial}}
\end{figure*}

Figure \ref{fig:virial} plots \avir\ against \SigmaH\ and \nHH.
Note that the indices of the best-fit curves for the \SigmaH--\avir\ and \nH--\avir\ plots are steeper than the indices in Equation \ref{eq:vtlaw2} because of a weak positive correlation between \SigmaH\ an \nHH.
The \nHH- and $\SigmaH$-dependence of \avir\ is similar to the virial theorem law found for the Galactic disk clouds, in which \avir\ has negative dependence on the column density as $\avir\propto N^{-0.3\mbox{--}-0.6}$ \citep{Moore2015,Urquhart2014}.

Figure \ref{fig:virial} also shows that \avir\ has a large scatter around the best-fit \nHH--\avir\ and \SigmaH--\avir curves.
The component covariant with \nHH\ and \SigmaH\ occupies only 22\%\ of the total variance of \avir.
The variation in \avir\ of the HCN clumps is created by such intrinsic variation rather than by dependence on \nHH\ and \SigmaH.

\subsection{Summary of the PCA}

Thus PCA was able to find three independent parameters that describe the physical properties of the HCN clumps: scaling parameter (PC1), the strength of turbulent heating (PC2), and self-gravitational stability (PC5).
No straightforward physical interpretation appears to be present for PC3 and 4, except for that they relate $r$, \Dv, \nHH, and \Tkin\ in different manners from PC2.
They could be interpreted as a deviation from the turbulent heating model, possibly introduced by additional heating processes or by non-physical factors such as limited accuracy of the clump identification.

It is notable that PC1, PC2, and PC5 translate into relations between the clumps parameters that are consistent with the observational laws known from previous studies, such as the $r$--$M$, $r$--\Dv, and $M$--\Mvt\ relations. 
This ensures that the modification we added to the CLUMPFIND algorithm did not impair the accuracy of the clump parameters from previous cloud identifications.

\section{CONDITION FOR STAR FORMATION\label{section:logisticregression}}
\subsection{Logistic Regression}
The goal of this paper is to find the clump parameter or the combination of parameters that has the strongest correlation with SF activities, utilizing an objective statistical analysis.
PCA was successfully used in \cite{Komugi2018}\ to formulate SFR of GMCs in M33 with other observed physical quantities, but their approach is not applicable to our HCN clumps because SFRs is difficult to measure in spatial scales better than $\sim 1$ pc, i.e., size scale of HCN
clumps.
Therefore, we introduce a 0/1 binary parameter \fsf\ that indicates the absence/presence of SF signatures and apply logistic regression (LR).

\newcommand\LF{F}

Logistic regression predicts the value of a binary parameter \fsf\ of a given target from its known properties represented by vector \qv.
Probability distribution $\Prob{\fsf}$ is parameterized by a coefficient vector \av\ and a constant $\lrconst$ as
\begin{eqnarray}
\Prob{\fsf|\qv,\av,\lrconst} & =&
\LF\left(\qv,\av,\lrconst\right)^\fsf\cdot \nonumber \\ && \hspace{-1em}
\left\{1-\LF\left(\qv,\av,\lrconst\right)\right\}^{1-\fsf} \label{eqn:logisticRegression1},
\end{eqnarray}
where $\LF$ is a logistic function that represents the probability for the (\qv, \av, $\lrconst$) combination to give \fsf = 1,
\begin{eqnarray}
\LF\left(\qv, \av, \lrconst\right) & \equiv & \frac{1}{1 +
  \exp\left(-\av\cdot\qv +
  \lrconst\right)}. \label{eqn:logisticRegression2}
\end{eqnarray}
The function $\LF$ is $0.5$ when the logit $l \equiv \av\cdot\qv - \lrconst$ is 0 and approaches to 0 and 1 as $l\rightarrow -\infty$ and $+\infty$, respectively.
Individual elements of vector \av\ represent the weights of the respective parameters in predicting \fsf.
The optimal \av\ and $\lrconst$ are obtained by maximize the likelihood function :
\begin{eqnarray}
  L\left(\av,\lrconst\right) &= &\prod_{i}
  \Prob{\fsf_i|\qv_i,\av,\lrconst},
\end{eqnarray}
where subscript index $i$ specifies the data points (i.e., clumps).

The parameter that is best correlated with \fsf\ is found as the one with the largest positive or negative logit weights.
The performance of the best-fit logistic model is evaluated by the dispersion in the histogram of $l$.
If $l$ spans over a range sufficiently wider than the transition range of the logistic function, $|l|\lesssim 1$, the SF and non-SF clumps are well separated in the histogram, indicative of high accuracy of the model;
otherwise $l$ distribution confined in the transition range indicates that the SF and non-SF clumps are degenerate in the given parameter space.

We explore the best explanatory parameter or parameter combination to discriminate SF and non-SF clumps, by repeating LR analysis starting with the full parameter space and then with reduced parameter spaces chosen based on previous results.
Below are the parameter sets that we use:

\begin{description}
\item[HCN-A] The full parameter space, i.e., the five-dimensional parameter set of ($r$, \Dv, \nHH, \Tkin, \Mhnc).
\item[HCN-B] One-dimensional parameter space of (\avir), i.e., a reduced parameter space where $r$, \Mhnc, and \Dv, are replaced with \avir\ and \nHH\ and \Tkin\ are omitted.
\item[HCN-C] Two-dimensional parameter space of (\nHH,\SigmaH), i.e., a reduced parameter space where \avir\ in HCN-B is replaced with \nHH\ and \SigmaH\ combination, which is found to be correlated with \avir.
\end{description}

All parameters are in base-10 logarithm scale, whose normalization units are given in Table \ref{table:LRresult}.
Note that $l$ is a scalar value that is invariant with a regular linear transformation of the parameter space, and therefore any of these parameters can be replaced by linear combinations of them
as far as it does not reduce the degree of freedom;
e.g., \Mhnc, \Dv\, or $r$ is replaceable with \avir\ in HCN-A without changing the best-fit $l$ values.

\subsection{Identification of Star-forming Clumps\label{section:SFsignatures}}
In determining \fsf\ values we use catalogs of the 6.7 GHz class-II methanol masers \citep{Caswell2010,Ott2019}, $\mathrm{H_2O}$ masers \citep{Walsh2011,Chambers2014,Lu2015}, and the \rv{extended} green objects (\rv{EGOs})
detected with the {\it Spitzer} surveys \citep{Yusef-Zadeh2009}, as these sources are less affected by chance superposition of foreground or background sources than other infrared/radio sources.
Clumps are judged to have star formation signature (\fsf = 1) if either of these sources is located within the lowest-level contour in the \PPV\ three-dimensional space (for the case of the masers) or in the position--position (\PP) two-dimensional space (for the case of the \rv{EGOs}). 
We also include the 50-\kmps\ cloud (clump \#86) in star-forming clumps, which does not appear in either catalog but hosts a well-known cluster-forming region G$-0.02$$-0.07$.

\begin{revision}
  We have identified \rv{15} HCN clumps that have signatures of star formation, including four clumps with multiple sources (clumps \#7,\#116,\#135, and \#163), and two foreground clouds (\#37,\#94);  the latter are not used in the following analysis.
  The SF signatures associated with them are indicated in Table \ref{table:clumplist}.
  The above catalogs contain four sources (two $\rm H_2O$ masers and tow EGOs) that are not associated with any HCN clumps.
  Three out of them are located in a vicinity of the Sgr B2 core region, and their corresponding HCN \JJ{4}{3}\ peaks are likely missed due to the high degree of the crowdedness of the dense clumps in the Sgr B2 complex.
  The other is an $\rm H_2O$ maser source located in a blank sky region without significant HCN \JJ{4}{3}\ emission; possibly, this maser is not of SF origin.

  We note as a caveat that the above catalogs of masers and EGOs are not a uniform set of SF signatures. 
  The EGOs could be biased to less embedded phases of SF than those traced by masers, as mid-infrared emission is more susceptible to absorption.
  In addition, the lack of information on the line-of-sight velocities of the EGOs causes a higher possibility of chance superposition with background/foreground clumps than in the cases of the masers.
  These differences between the EGOs and masers are omitted in our analysis, in which SF activity is represented by a single binary flag. 
  We will examine the impact of this simplification on the analysis in a later subsection (\S\ref{subsection:EGOcheck}).
\end{revision}

\subsection{Control Data}
We may expect non-physical correlation between \fsf\ and clump parameters created by statistical effects and biases from the clump-identification process;
for example, clumps with larger \PPV\ volume are obviously more probable of having at least one SF signatures when their distribution is random.
Therefore, we generate control data from the observed HCN clumps and randomly created mock SF signatures, thereby intending to investigate specific
physical processes working in the CMZ clouds through comparison with the real observed data.

In the control data, SF clumps are assigned based on mock SF signatures that we randomly distribute in the \PPV\ space according to the probability density proportional to the HCN \JJ{4}{3}\ intensity.
The number of the SF clumps is controlled to be the same as that in the observed data.
We apply the LR analysis with the full parameter space and compare the results with the HCN-A run using the real data.

\subsection{Results with the HCN Clumps}
Table \ref{table:LRresult} tabulates the best-fit logit weights (elements of vector \av) and $\lrconst$ along with their standard deviation errors estimated using the JK analysis.
Figure \ref{fig:LRresult1} shows frequency histograms of the logit $l$ value of star-forming (SF; clumps with \fsf = 1) and non-star-forming (non-SF; clumps with \fsf = 0) groups, for all models.
The results for the control data are averages over 100 independent samples.
For the results with HCN-A and control data, the best-fit logit weights in the PC space are given in Table \ref{table:LRresultPC}.
Below we summarize and compare the results for the control data and the observed data with the four parameter models.

\paragraph{Control Data}
Significant positive logit weights are obtained for $r$ and \Dv, whose ratio is approximately 2:1; this translates into a statistical effect, since the chance for a clump to contain at least one mock SF signature is proportional to its \PPV\ volume $\propto r^2 \Dv$.
The dependence on the \PPV\ volume appears in the relatively large logit weight for PC1, as PC1 represents the scaling factor for $r$, $M$, and $\Dv$. 
The positive weights of \Tkin\ and \nHH\ are also readily understood as artifacts from the procedure of generating the control data;
we used the HCN \JJ{4}{3}\ map as the distribution function of the mock SF signature, which creates a correlation between \fsf\ and the excitation temperature of the HCN \JJ{4}{3}.
The $l$ histogram of the SF and non-SF clumps only differ by 1 in their peak values, indicative of low efficiency of the model.
Therefore, the above artificial effects are relatively weak compared with the random factor that dominates the SF probability in the control data.

\paragraph{HCN-A}
The results with the real data significantly differ from the control data in the weights of $r$, \Dv, and \Mhnc;
in particular, the large negative weights of $r$ and \Dv\ indicate that the statistical effect that was apparent in the control data is overwhelmed by another determinant factor.
The absence of the statistical effect is also directly indicated by the PC1 weight that is $\sim 0$.
The ratios of the logit weights of $r$, \Dv, and \Mhnc\ is approximately $1:2:-1$, which is the same ratio as that of the indices of respective parameters in $\avir \propto r\,\dv^2 M^{-1}$.
We show the logit weights expressed in the ($r$, $\Dv$, \nHH, \Tkin, \avir) coordinate system in Table \ref{table:LRresult}, which reveals that the significant explanatory parameters are reduced into one variable \avir\ without significant contribution from other parameters.
The histogram of $l$ spans in a wide range from $-15$ to $7$, in which the SF and non-SF groups are sharply divided.
Thus, \fsf\ is found to be primarily determined by \avir\ in the real observed data.

\paragraph{HCN-B}
This model uses a reduced parameter space where \avir\ is taken as the only explanatory parameter.
The separation of the SF and non-SF clumps in the $l$ histogram is not significantly worsened from HCN-A by the omission of the other four parameters;
this confirms that the fundamental role in predicting \fsf\ is played by \avir, while the other parameters have virtually no effect in the performance of the model.
The threshold $\avir$ value is obtained from the logit weight of \avir\ as $10^{\frac{\lrconst}{a_\avir}}\sim 6$.

\paragraph{HCN-C}
This model assumes the other extreme case from HCN-B; the parameter space does not include \avir, but instead \nHH\ and \SigmaH\ are taken as the explanatory parameters.
As \avir\ is found to be correlated with \nHH\ and \SigmaH\ (Section \ref{subsection:VT}), we examine whether their combination can replace \avir\ in the LR analysis using this model.
The best-fit logit weights of \nHH\ and \SigmaH\ are positive and significant, being consistent with their negative correlation with \avir.
The $l$ histogram shows that the predictability of HCN-C is obviously worse than HCN-B and no better than that of  the control data where the random factor dominates.

\bigskip
Thus we have found that the SF and non-SF clumps can be clearly  discriminated by the parameters measured with the HCN \JJ{4}{3}\ data.
The explanatory parameters are reduced into a single parameter, \avir;
clumps with \avir\ below $\sim 6$ are more likely to possess SF signatures than otherwise.
Addition of other parameters does not improve the predictability significantly, whereas the performance of the models without \avir\ is no better than the analysis with the control data where SF signatures are randomly distributed.
The stark difference between the results with the successful models (HCN-A, B) and the control data underlines that the \avir-dependence of \fsf\ is highly likely to have a physical background, not being a spurious relation created by statistical effects.

The HCN-B and HCN-C results demonstrate that \avir\ is the fundamental explanatory parameter.
As \avir\ decreases with increasing \nHH\ and \SigmaH\ (Section \ref{subsection:VT}), one might suspect they are the true explanatory parameters that create a spurious dependence on \avir;
however, the LR analysis has shown that \avir\ cannot be replaced with any other parameters without significantly impairing the predictability of \fsf.
This indicates that the decisive role in predicting \fsf\ is played by the intrinsic variation of \avir\ that is not covariant with \nHH\ and \SigmaH.
The same conclusion is drawn from the logit weights in the PC space;
the largest weight is obtained for PC5, which is understood as \avir\ from which dependence on $\nHH$ and \SigmaH\ is removed.

We note a possible bias in the LR analysis caused by the chemical characteristics of the HCN molecule, which is known to increase in shocked region from observations of proto-star jets \citep{Jorgensen2004,Tafalla2010} and broad-velocity clouds in the CMZ \citep{Tanaka2015,Tanaka2018a}.
As the shock chemistry is widespread across the CMZ clouds (e.g., \citealt{Requena-Torres2006};Paper I), the clump identification based on the HCN \JJ{4}{3}\ is potentially overestimating \Dv, and hence \avir.
We may identify the effect of this bias in the high threshold \avir\ of $\sim 6$, which exceeds the threshold for self-gravitating cloud, $\sim 2$. 
The significant positive logit weight for PC2 is also possibly a consequence of the bias;
since the larger PC2 score indicates the smaller contribution of shocks in heating and line-broadening (Section \ref{subsection:turbheat}), the positive PC2-dependence of \fsf\ could be understood as a correction for the overestimated \avir.
Nonetheless, we can consider that the effect on the LR analysis is limited, as the largest weight is obtained for PC5, which approximately represents the \avir\ variation that is independent of the increase in \Dv\ by shocks, which is represented by PC2.

We also note that the above results do not necessarily indicate the absence of effects other than \avir\ in the physical process, as our analysis is limited by the sample size, in particular that of the SF clumps.s
We may possibly find other parameters playing secondary roles in determining \fsf\ if we can reduce the uncertainties of the logit weights by analysis with an extended sample size.
In addition, our parameter set lacks a few important parameters such as magnetic field strength and cosmic-ray ionization rate, although the observed high values of them in the CMZ are frequently referred to as the cause of the inefficient SF \citep[e.g.][]{Morris1996,Yusef-Zadeh2007}.

\begin{figure*}[ttt]
\epsscale{0.8}
\plotone{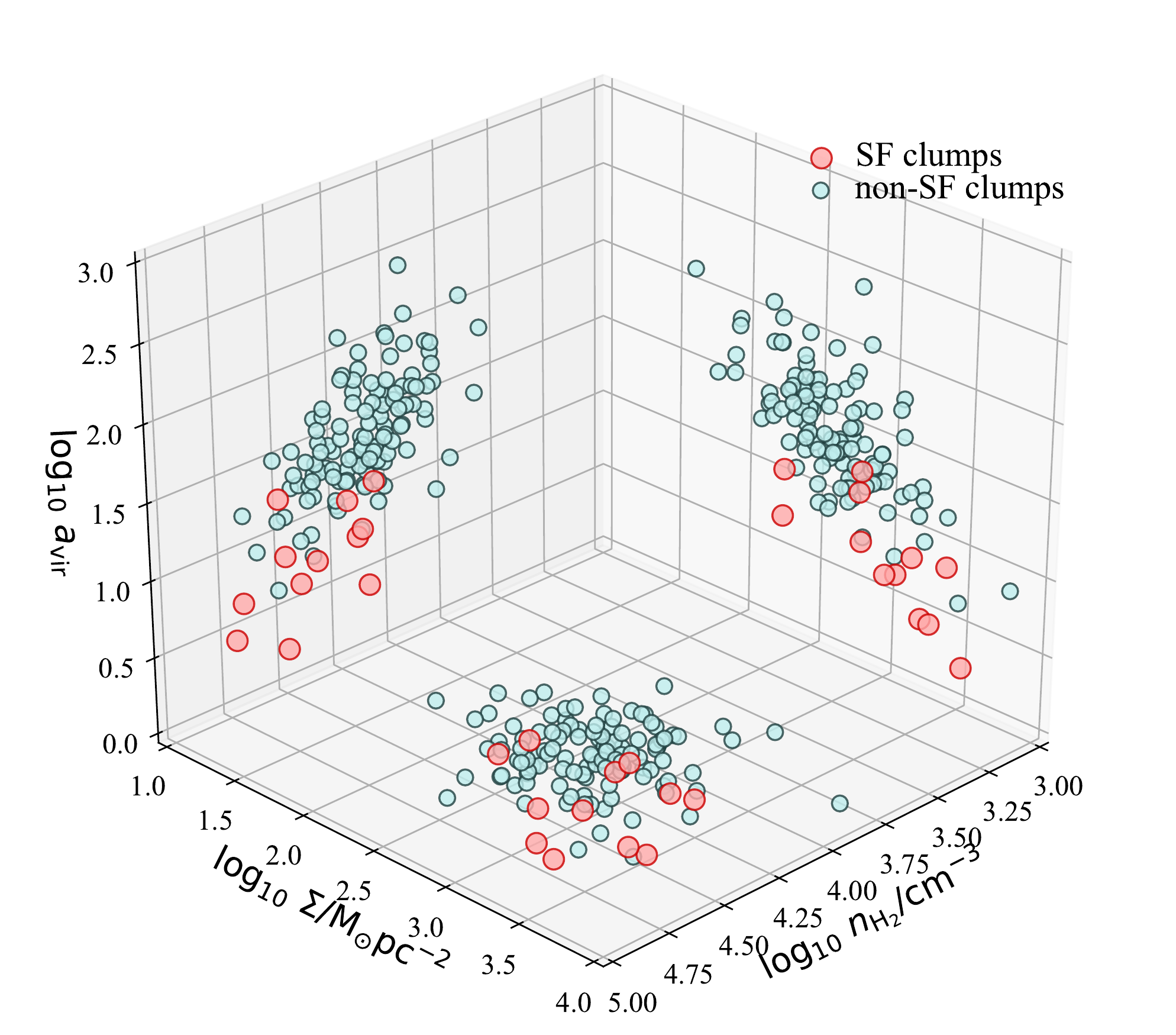}
\caption{HCN clumps in \nHH--\SigmaH--\avir\ 3D space, with colors according to \fsf. \label{fig:LRresult2}}
\end{figure*}

Figure \ref{fig:LRresult2} summarizes the result of the LR analysis, in which the SF and non-SF clumps in the three-dimensional space of \nHH--\SigmaH--\avir\ is presented.
The SF clumps preferably appear in the region with high-\nHH, high-\SigmaH, and low-\avir.
As the LR analysis has shown, \nHH\ and \SigmaH\ affect \Prob{\fsf}\ through their correlation with \avir.  We confirm this by the absence of a clear SF/non-SF border on the \nHH--\SigmaH\ plane.

\begin{revision}
  \subsection{Analysis without EGOs \label{subsection:EGOcheck}}
  Out of the 13 SF clumps used in the analysis, five are associated with EGOs but not with \methanol\ or \water\ masers.
  As noted in \S\ref{section:SFsignatures}, the EGO sources lacking maser emissions may be at a different phase of SF activity from those with maser sources, and possibly include misidentified clumps superposed with foreground/background sources by chance.
  We examine the impact of the inclusion of the EGO sources on the analysis by performing the LR analysis using only masers as SF signature sources.
  The same parameter as HCN-A is used except for the definition of \fsf.
  The results are shown in Figure \ref{fig:LRresult1} and Table \ref{table:LRresult}, labeled as HCN-A$^\prime$.

  We find that the exclusion of the EGOs does not change the results significantly.
  The difference between the best-fit logit weights of HCN-A and HCN-A$^\prime$ is within the statistical errors.
  The \avir\ parameter is the only significant explanatory parameter that distinguishes SF and non-SF clumps also in the HCN-A$^\prime$ result.
  The degree of separation between SF and non-SF clumps in the histogram of the $l$ value is slightly worse in the HCN-A$^\prime$, but is still greater than that in the control data.
  These comparisons indicate that the impact of the non-uniformity of the sample of the SF clumps due to the EGO sources, if any, is ignorable compared with the statistical errors.  
  \end{revision}

\subsection{Comparison with MT00 Clumps}
We apply the same LR analysis to the MT00 clumps identified on the CS \JJ{1}{0}\ map.
The \fsf\ values are determined based on a comparison between the \PPV\ distance of the SF signatures from the clump center positions and the clump sizes;
\rv{8} out of the 159 MT00 clumps are identified as SF clumps with \fsf\ of 1.
The parameter space is taken as $r$, \Dv, and \avir, as the MT00 catalog lacks \nHH\ and \Tkin\ values.
The best-fit logit weights and the distribution of $l$ value are presented in Table \ref{table:LRresult} and Figure \ref{fig:LRresult1}, respectively.

The result is qualitatively consistent with those with the HCN clumps;
significant logit weights are obtained only for \avir, and its sign is minus indicative of low-\avir\ clumps being more likely to possess SF signatures.
However, the narrow $l$ distribution confined in $-5 < l < 0$ and the coincidence of the frequency peak for the SF and non-SF groups in the $l$ histogram in Figure \ref{fig:LRresult1} indicate a poor prediction accuracy of the best-fit model.
\begin{revision}
This result indicates that SF and non-SF clumps cannot be discriminated by any multiplicative combination of  $r$, \Dv, and \avir\ (or $M$) of the MT00 clouds.
\end{revision}

\begin{deluxetable*}{lcccccccc}
  \tablecaption{Best-fit logit weights \label{table:LRresult}} \input{table_LR.etex}
\end{deluxetable*}

\begin{deluxetable*}{lcccccc}
  \tablecaption{Best-fit logit weights (PC) \label{table:LRresultPC}} \input{table_LR_pca.etex}
\end{deluxetable*}

\newcommand\plotInTab[1]{\hspace{-9pt}\includegraphics[clip,scale = 0.3]{#1}\hspace{-9pt}}
\begin{figure*}[ttt]
\epsscale{0.5}
\centering
\begin{tabular}{ccc}
  \plotInTab{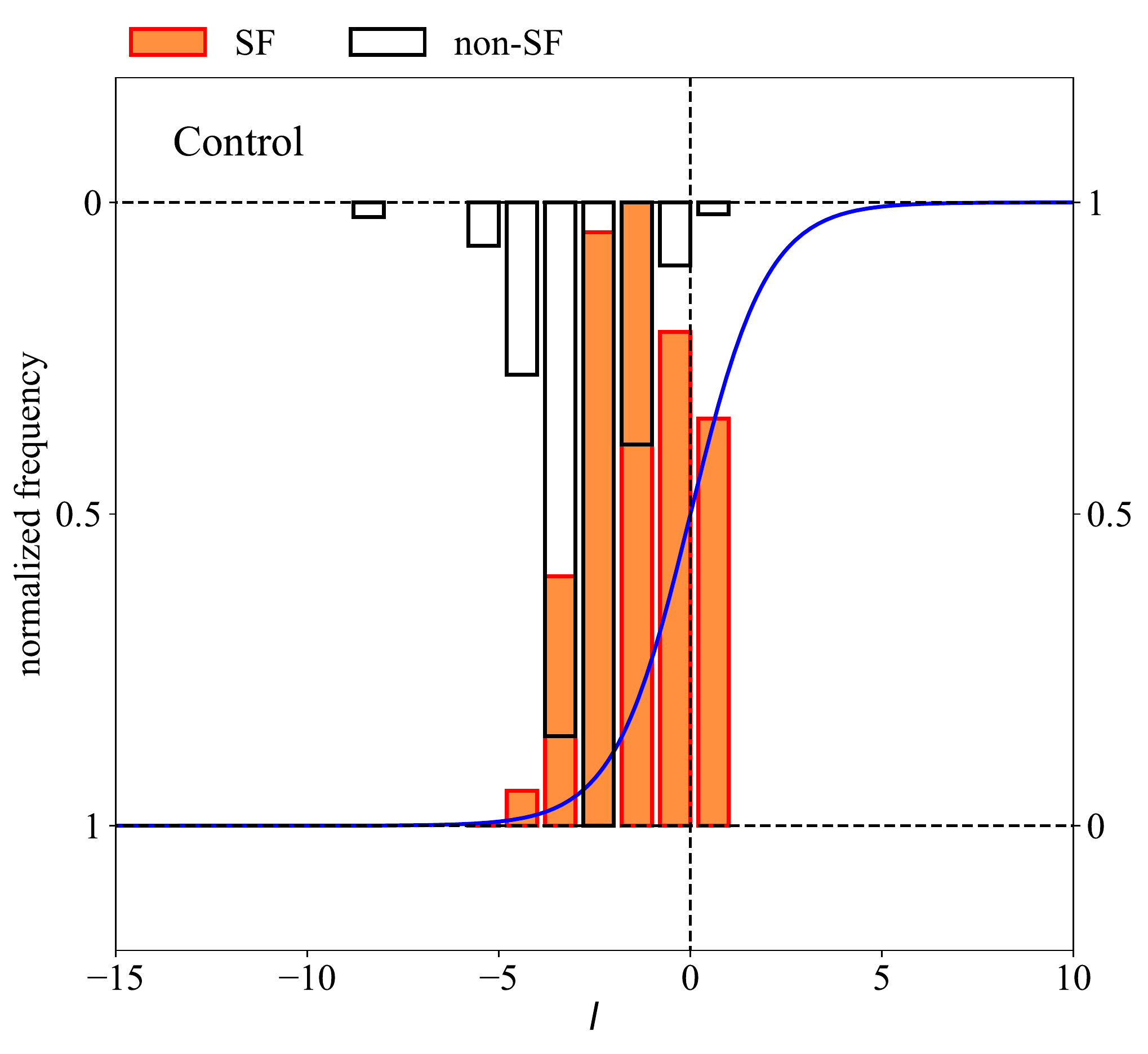} & \plotInTab{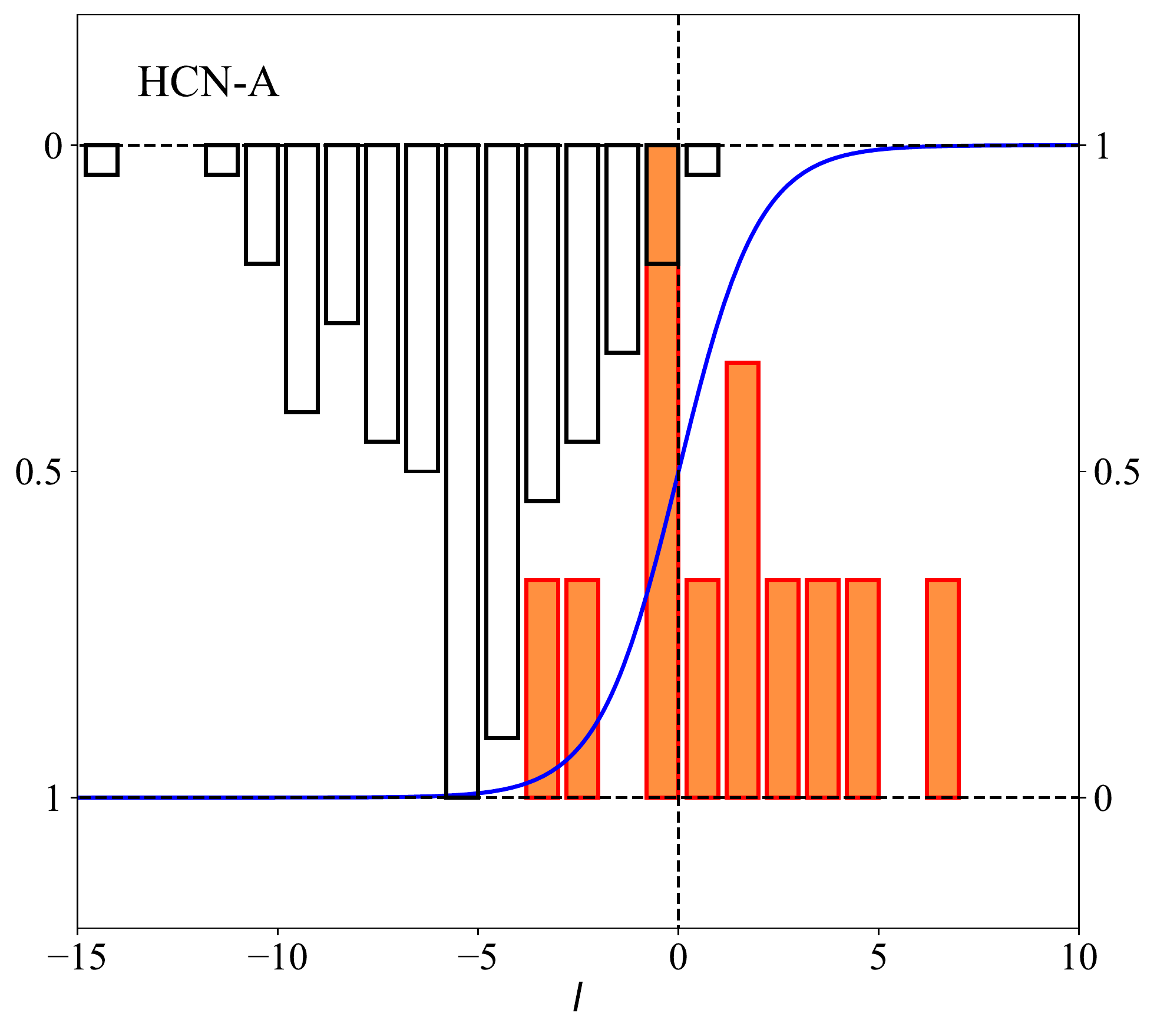} & \plotInTab{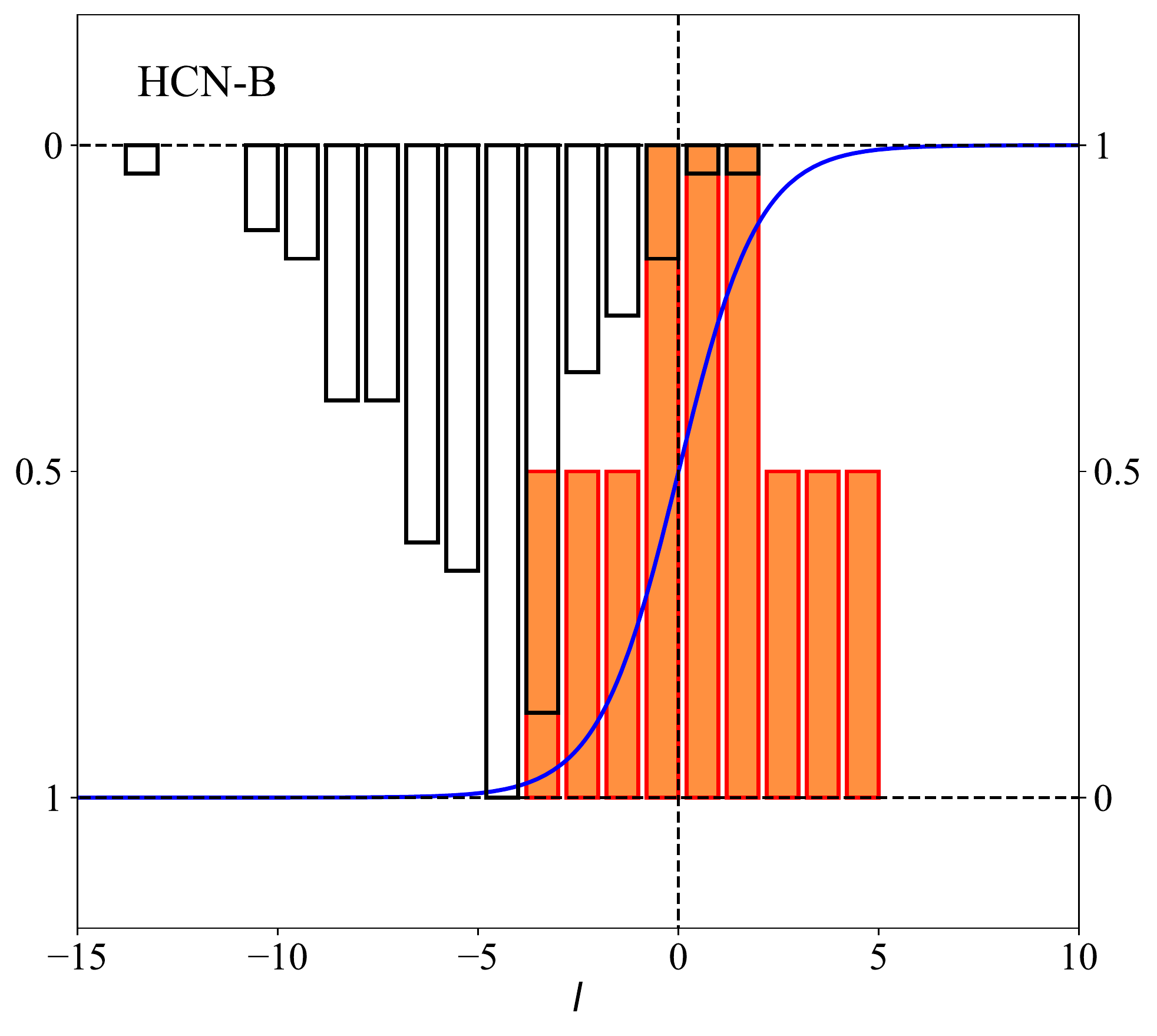} \\
  \plotInTab{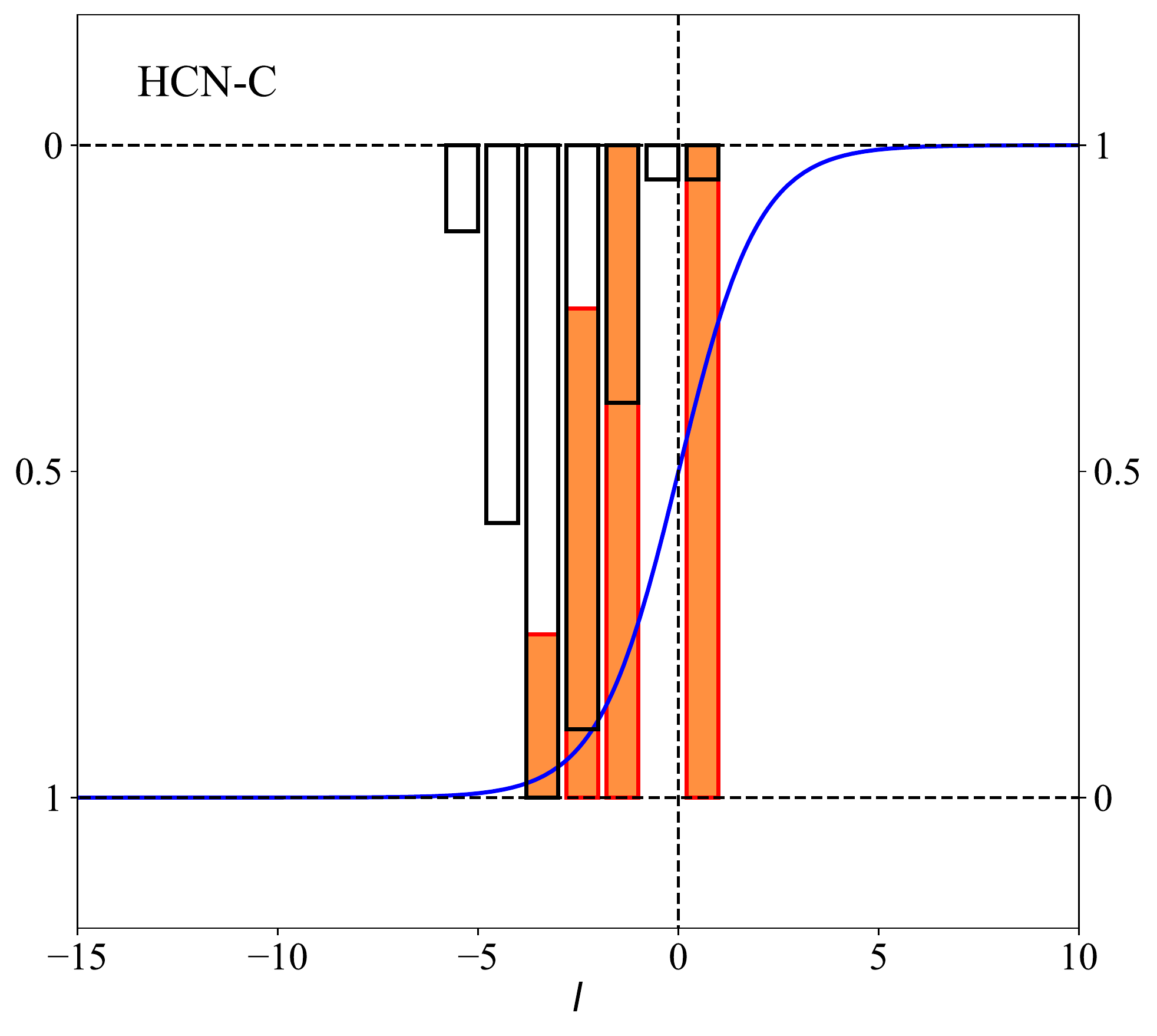} & \plotInTab{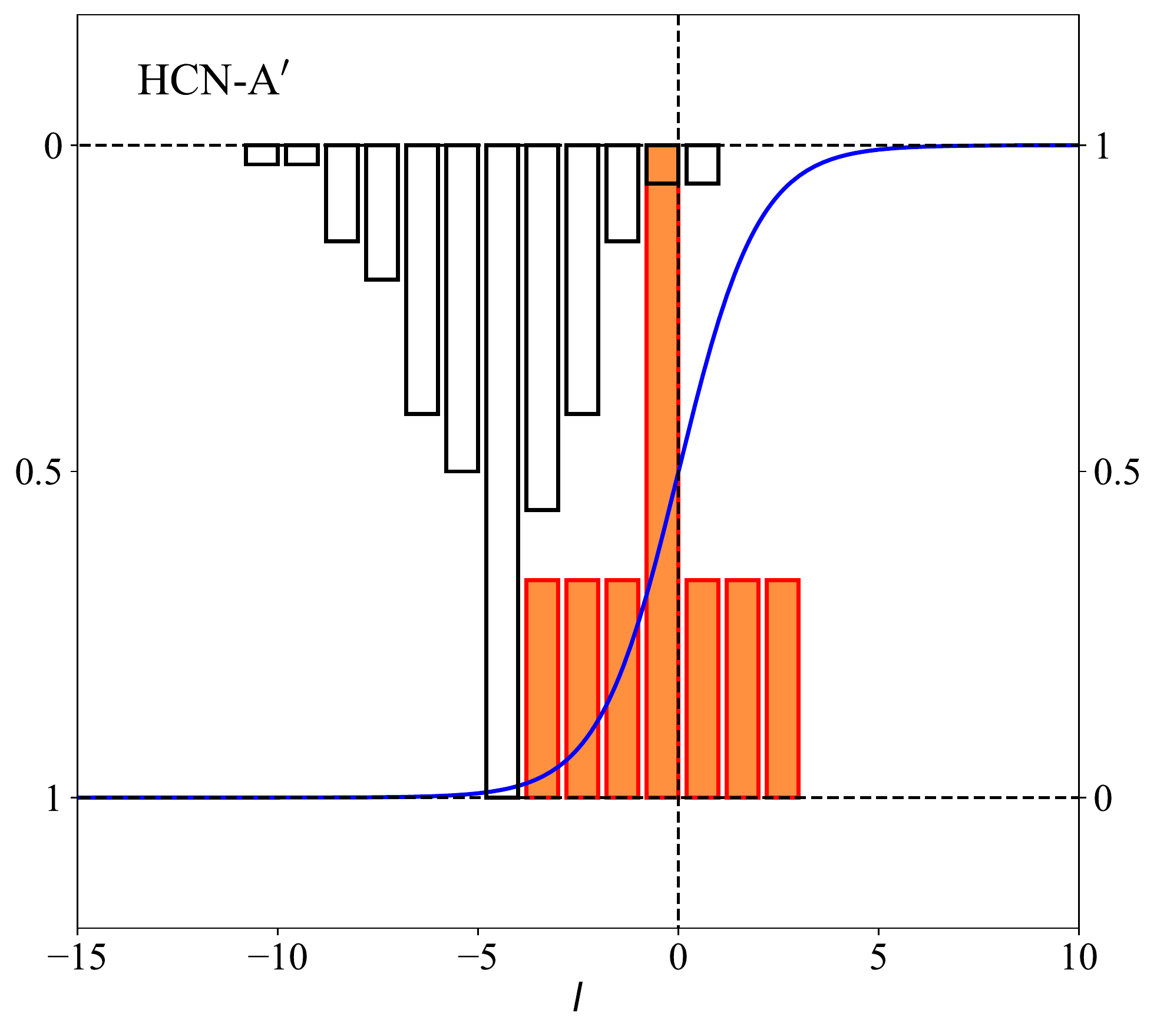} & \plotInTab{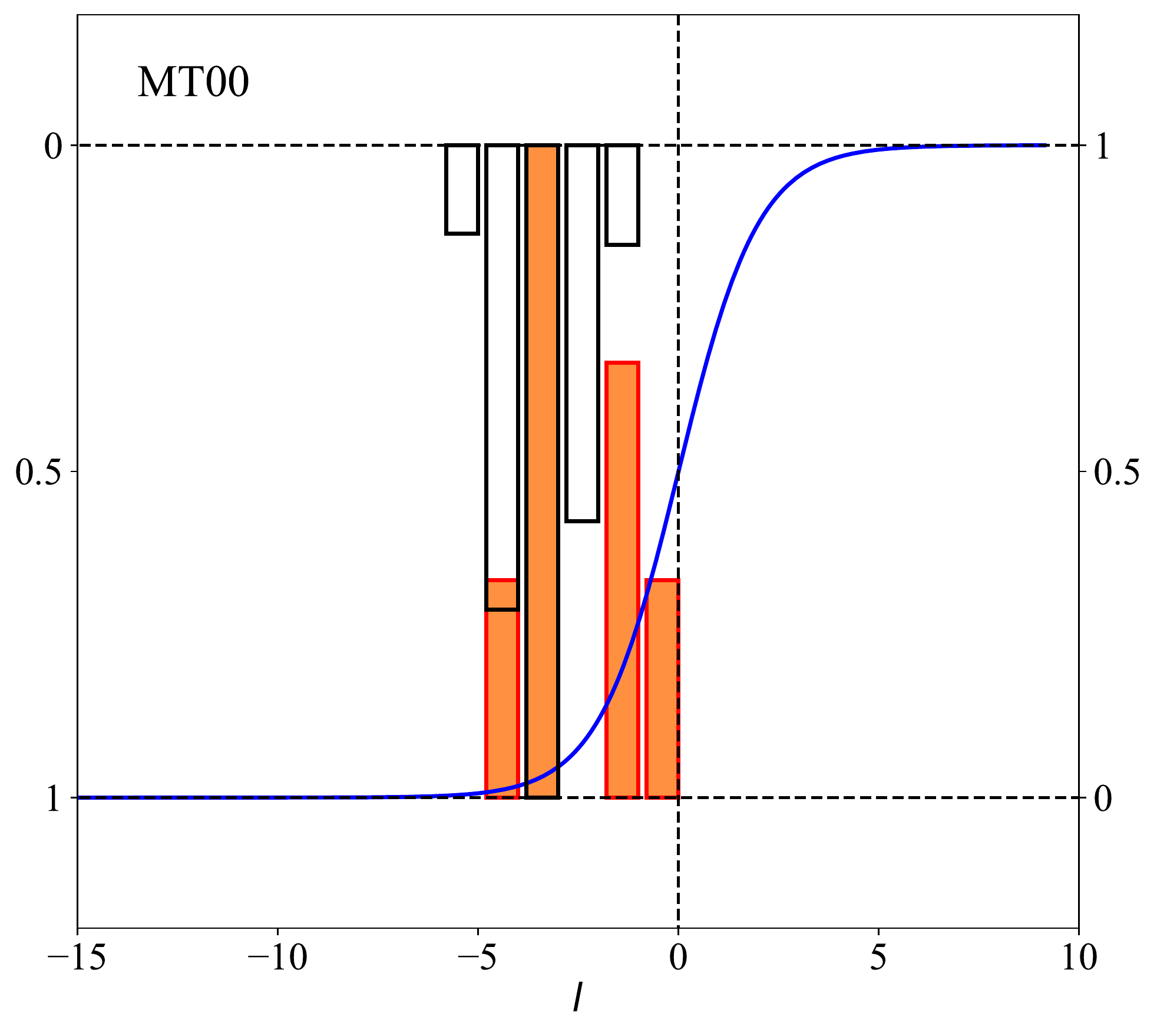} 
\end{tabular}
\caption{Frequency histograms of the best-fit logit function $l(\qv)\equiv \av\cdot\qv - \lrconst$ values of star-forming (SF) and non-star-forming (non-SF) groups, using the best-fit coefficients obtained for different parameter models, along with the logistic curve $\LF(l) = (1 + e^{-l})^{-1}$.
  Histograms are normalized so that the maximum frequency is 1 in each of SF and non-SF groups.
\label{fig:LRresult1}}
\end{figure*}

\section{DISCUSSION\label{section:discussion}}
\subsection{Origin of the Inefficient Star Formation in the CMZ}
The LR analysis has revealed that the probability of SF is primarily determined
by \avir\ of the HCN clumps;  SF and non-SF clumps are distinctly divided at the threshold \avir\ of $\sim 6$. 
If we may regard the clumps with $\avir < 6$ as SF or potential SF clumps,
the fraction of such low-\avir\ clumps to all identified HCN clumps (\etasf)
is approximately 0.1 both in mass and number count.
By considering the majority of the GRS clouds are in
the self-gravitating regime in the virial plot
(Figure \ref{fig:virial}), the scarcity of the low-\avir\ clumps can be
considered as one of the leading causes of the inefficient SF in the CMZ
compared with the typical Galactic disk environment.

If we assume the HCN clumps are units of SF activity, the degree of
suppression of the overall SFE from the typical Galactic
disk environment is expressed as $\etasf\cdot\etasfe\cdot\etadense$,
where \etasfe\ is the degree of suppression of the SFE in
individual SF clouds and \etadense\ is the fraction of the gas mass
associated with the HCN clumps to the total amount of the molecular
gas.
Although the accurate value \etadense\ is unknown,
\begin{revision}
  we may use the fraction of the HCN \JJ{4}{3}\ luminosity identified as the clump emission to the total HCN \JJ{4}{3}\ luminosity, which is $\sim 10\%$ (\S\ref{section:massEstimate}),
  as a measure of \etadense.
  This factor is further multiplied by the mass ratio of the high-density component of the multi-phase gas, which is 0.15--0.5 \citep{Mills2018,Tanaka2018b}, if the low-density component is not physically associated with the HCN clumps.
  This yields $\etasf\cdot\etadense\sim (0.15$--$0.5)\times 10^{-2}$, which does not greatly differ from the measured values of the degree of the suppression of the SFE
\citep{Longmore2012b,Kruijssen2014,Barnes2017}, even when $\etasfe\sim1$.
\end{revision}
Our analysis is not able to put significant constraints on the SFE in individual clumps,
but the above estimate indicates 
the efficiency may not
necessarily be suppressed from the standard value if we focus
on the star-forming HCN clumps.  This is consistent with the
measurements by \cite{Barnes2017}, who measured the SFE per
free-fall time, \epsff, to be 1--5\% for the dust-ridge clouds by
assuming the tidally-triggered SF
model \citep{Longmore2013,Kruijssen2015}; these values do not differ
from the standard \epsff\ value, $\sim
1\%$ \citep[e.g.][]{Krumholz2005}.

\begin{revision}
  Alternatively, we may assume the MT00 clouds as the unit of SF.
  The factor \etasf\ is $\sim 1$ for the MT00 clouds, as \fsf\ is only poorly dependent on \avir\ of the MT00 clouds.
  The masses of the MT00 clouds are systematically greater than those of HCN clumps, which decreases the factor \etasfe\ by 1--2 orders of magnitudes from the case with the HCN clumps. 
  In this regime, SF in the CMZ would appear to take place with low efficiency irrespective of the physical properties of the natal clouds with generally high \avir.
  The fundamental difference between the SF pictures in the HCN clumps and MT00 clouds is the difference in the spatial scales concerned.
  As we have seen in \S\ref{section:otherCats}, the cloud-identification based on the MT00 CS \JJ{1}{0}\ data is sensitive to a few to 10 pc scale structures, which are $\sim 0.8$ dex larger than the HCN clumps in median radius.   Such larger clouds tend to contain a larger fraction of the gas outside the SF regions, making the physical properties of the SF regions less visible in the cloud-averaged parameters.
  As a consequence of this dilution effect, SF probability appears more uniform in the MT00 clouds.

  Our results imply that the HCN \JJ{4}{3}\ emission does not predominantly originate from the SF gas with $\nHH > \ncritSF$.
  If the HCN \JJ{4}{3}\ luminosity scales the mass of the SF gas, we would expect a strong positive correlation between \fsf\ and the PC1 score (i.e., the scale factor for the clump size and mass), as we found with the control data; however, the LR analysis has shown that the PC5 score or \avir\ is the leading determinant factor of \fsf, while no significant PC1-dependence is found.
  This is consistent with the excitation analysis in Paper I, showing that typical \nHH\ of the HCN \JJ{4}{3}-emitting region is $10^{3.4\mbox{--}4.8}\ \pcc$, which is more than two orders of magnitude lower than the predicted \ncritSF\ in the CMZ.

  Our analysis also highlights that the HCN \JJ{4}{3}\ line is a good probe of the SF regions in the CMZ, in the sense that \fsf\ is tightly correlated with the physical properties measured with the HCN \JJ{4}{3}\ data.  %This is likely owing to the high \ncrit.
  %  This indicates that the contamination by the gas outside the SF gas is not severe enough to completely mask the physical properties of the SF regions, likely owing to the high \ncrit\ of HCN \JJ{4}{3}.
  This indicates that we successfully capture molecular cloud structures closely associated with SF by the HCN \JJ{4}{3}\ observations, likely owing to the high \ncrit\ of HCN \JJ{4}{3}.
  Although no clear threshold \nHH\ for SF clumps is found with the LR analysis, the threshold \avir\ of 6 approximately translates into a typical \nHH\ of SF clumps of $\gtrsim 10^{4.6}\ \pcc$ using the \nHH--\avir\ correlation found with the PCA (Equation \ref{eq:vtlaw1}).
  The \ncrit\ of HCN \JJ{4}{3}\ of $10^6\ \pcc$ is advantageous to distinguish those high-density structures from the gas with mean \nHH\ of the CMZ, $10^4\ \pcc$.

  On the other hand, \ncrit\ of the CS \JJ{1}{0}\ ($10^{4.8}\ \pcc$) differs from the mean \nHH\ by less than an order of magnitude.
  This indicates that the CS \JJ{1}{0}\ is more sensitive to the spatially extended gas than the HCN \JJ{4}{3}\ line, being consistent with the generally larger cloud sizes  and the consequent poor \avir--\fsf\ correlation in the MT00 data.
  We note that this does not indicate that the CS \JJ{1}{0}\ is always an inferior probe to HCN \JJ{4}{3}, as the clump/cloud identification is also dependent on the spatial resolutions and signal-to-noise ratios.
  %However, for the case of our HCN \JJ{4}{3}\ data and the MT00 CS \JJ{1}{0}\ data, the signal-to-noise ratio of the latter is similar to, or slightly higher than that of the latter (\S\ref{section:otherCats}).
  However, the signal-to-noise ratio of the MT00 CS \JJ{1}{0}\ data is similar to or slightly higher than that of the HCN \JJ{4}{3}\ data (\S\ref{section:otherCats}).
  The difference in the cloud/clump sizes is also sufficiently greater than the difference in their spatial resolutions.
  Hence, the difference in \ncrit\ is likely the primary cause of the difference in efficiency as an SF probe between the HCN \JJ{4}{3}\ and MT00 data.

\end{revision}

\subsection{Time Scale of Star Formation in the CMZ}
We have found a correlation between \avir, \nHH, \SigmaH\ thorough PCA.
The loci in the \nHH--\SigmaH--\avir\ plot
(Figure \ref{fig:LRresult1}) are understood to represent the evolutionary phase
of the clumps from the high-\avir, low-\nHH, and low-\SigmaH\ phase to
the low-\avir, high-\nHH, and high-\SigmaH\ phase.  If transitions across the threshold \avir\ with time exist,
the transition time scale $\ttr$ is estimated by scaling the free-fall
time \tff\ of the SF clumps by $\etasf^{-1}$ on the assumption that the number of the SF
clumps is in equilibrium.
The value of \tff\ is $\lesssim 0.16\ \Myr$ in the medium of $\nHH \gtrsim 10^{4.6}\ \pcc$,
yielding $\ttr \lesssim 2\ \Myr$; this is comparable to the time scale of the dissipation
of cloud via shear \cite{Jeffreson2018}, meaning that a significant
fraction of the dense gas in the CMZ dissipates before forming stars.

The high-\avir\ to low-\avir\ transition with timescale of $\lesssim 2\ \Myr$ is
unlikely driven by self-gravity of the HCN clumps, whose time scale is \tff\ of non-SF cloud, 0.3 Myr.
Therefore, we may assume an additional mechanism that regulates the evolution of the HCN clumps.
In the following we examine hypotheses of SF triggers in the CMZ, the tidal compression model,
gas accumulation associated with the X1/X2 orbit transition, and the CCC-triggered SF model.

The tidal compression model assumes that the SF is
initiated by the compressing tidal force of the stellar population near \sgras \citep{Longmore2013,Kruijssen2015,Barnes2017}.
This model places a series of GMCs from the brick cloud to the Sgr~B2 complex on a stream motion,
and translates their apparent difference in SFE into different durations of SF phase
after the pericenter passage.
The X1/X2 orbit transition model interprets the gas motion in the CMZ as the X1 and X2 orbit families created by
the bar potential of the inner Galaxy \citep{Binney1991}.  Along the course of the X1/X2 transition,
SF is considered to be promoted at the X2 apocenters where infalling gas is preferably accumulated \citep{Hasegawa1994}.
In both models, the time interval at which a cloud orbiting in the CMZ experiences the trigger 
is approximated by $2\pi \kappa^{-1} = 3\ \Myr$, where $\kappa = 2\ {\rm
rad}\ \Myr^{-1}$ is the epicyclic frequency at the Galacto-centric radius of 100
pc \citep{Sofue2013}.  As this timescale does not differ much from
the \ttr\ value estimated above, we can assume these events as the SF
trigger in the CMZ in terms of the timescale estimates.
We may expect a systematic spatial variation in \avir\ or \etasf\ along the gas stream/orbit if these triggering events dominate SF in the CMZ, as \cite{Barnes2017}\ found the variation in SFR in the dust-ridge clouds.
However, we did not detect significant evolution of \avir\ along the
stream/orbit in our data set; the null hypothesis of spatially
constant \etasf\ was not rejected, possibly limited by the small
sample size of the low-\avir\ clumps.
Therefore, we cannot
further pursue the spatial variation of \etasf\ in this paper.

The CCC frequency per cloud is estimated to be a few Myr from theories and
observations
\citep{Hasegawa1994,Tan2000,Tanaka2018a,Jeffreson2018}; this may allow us to
consider the CCC-triggered SF as a significant mode of SF in the CMZ
in terms of the comparison of the time scales.
On the other hand,
the effect of CCC is not substantiated by the LR analysis,
where \fsf\ is found slightly positively dependent on PC2;
as larger PC2 score indicates weaker turbulent heating, on-going CCC sites that are marked by
large negative PC2 score are equally or slightly less likely to possess SF signatures.
In addition, the effective timescale of CCC can be longer than the above value because of the low efficiency of CCC
to trigger active SF suggested by recent observations.
New CCC candidates found in the CMZ \citep{Tanaka2015,Tanaka2018a,Enokiya2019} often lacks
OB star formation, unlike previously reported CCC regions with cluster-formation such as
Sgr~B2 and the 50-\kmps\ cloud \citep{Hasegawa1994,Tsuboi2015,Uehara2019}.
\cite{Enokiya2019} argue that the typical CCC velocity in the CMZ, which is $\sim 10\ \kmps$, is too fast to collect dense gas
within interaction time, which results in cloud disruption rather than
promotes SF unless the cloud has huge column density $\gtrsim 10^{23}\ \psc$ that is rare except for a few massive GMC complexes in the CMZ. 
This inefficiency of the CCC suggests that the effective time scale
of CCC-triggered SF is further scaled by the inverse of the probability
of CCC events resulting in SF, which may increase the time scale by
an order of magnitude or more.
If this is the case, the CCC-triggered SF cannot contribute to the
evolution of \avir, unless the CCC frequency is substantially higher than our assumption.

\section{SUMMARY\label{section:summary}}
We have presented the clump identification using the HCN \JJ{4}{3} line reported in Paper I, which is the CMZ cloud catalog based on the molecular line that has the highest critical density for excitation (\ncrit) ever used in wide-field CMZ surveys.
We have applied a logistic regression (LR) analysis to the identified HCN clumps, randomly generated control data, and the CS \JJ{1}{0}\ clouds (MT00 clumps) from literature \citep{Miyazaki2000}, which traces larger and less dense structures than the HCN clumps.
Upon this analysis, we have explored the parameter that is the best correlated with the SF probability of the clumps.  Below summarizes the main results:

\begin{itemize}

\item We identified 195 clumps in the HCN \JJ{4}{3}\ data, using CLUMPFIND algorithm with a minor modification made to adjust to the CMZ data, where various scale structures overlap with a high volume-filling factor.
      Radius ($r$), mass ($M$), FWHM velocity width (\Dv), gas kinetic temperature (\Tkin), and hydrogen volume density (\nHH) were measured for the HCN clumps using the multi-line data and the excitation analysis presented in Paper I.
      The clump sizes range from a few 0.1 to 1 pc in $r$, which should be described as `clump-scale' structures in comparison to the `cloud-scale' structures of the MT00 clouds \citep{Miyazaki2000} identified using the CS \JJ{1}{0}\ cube.

\item We applied a principal component analysis (PCA) to the HCN clumps in the five-dimensional parameter space of ($r$, $M$, \Dv, \nHH, \Tkin).  PC1 and PC2 were found to represent the scaling relation among $r$, $M$, and \Dv, and the strength of turbulent heating, respectively.
      The obtained relations are consistent with previous studies, i.e., steeper $r$--$M$ and $r$--$\Dv$ relations than the Galactic disk clouds \citep{Miyazaki2000,Shetty2012,Kauffmann2017a,Kauffmann2017b}, and widespread turbulent heating \citep{Ao2013,Ginsburg2016}.
      PC5 indicates the correlation between the virial parameter (\avir), \nHH, and the surface density (\SigmaH), which is consistent with the results obtained for the Galactic disk clouds \citep{Urquhart2014,Moore2015}.

\item We conducted a LR analysis to find the function that best distinguishes the clumps with and without observed signatures of star formation (SF and non-SF clumps, respectively).
      The explanatory parameters in the best-fit model are reduced into the virial parameter, $\avir \propto r\cdot\Dv^2 \cdot M^{-1}$, without significant contributions from other parameters.
      On the other hand, the models where \avir\ is replaced by other parameters cannot predict the SF probability any better than the model using the control data where SF and non-SF are randomly assigned.

\item The LR analysis applied to the MT00 clumps showed that the SF probability is also dependent on \avir\ but in a much weaker manner than the results with the HCN clumps.
      The performance in discriminating SF and non-SF clumps is no better than the model using the control data.

\item \rv{The critical \avir\ for SF found with the LR analysis is $\sim 6$, which corresponds to \nHH\ of $10^{4.6}\ \pcc$ using the \avir--\nHH\ correlation found with the PCA.   The high \ncrit\ of the HCN \JJ{4}{3}\ ($10^6\ \pcc$) is advantageous in identifying those high-density clumps. }  

\item The fraction of the clumps with $\avir\ < 6$ is approximately 0.1 both in number and mass fractions.
      This scarcity of low-\avir\ (and high-\nHH) clumps is one of the immediate causes of the low overall SFE of the CMZ.

\item \rv{No significant correlation between the PC1 score and SF probability is found with the LR analysis, implying that the HCN $J$=4--3 does not predominantly originate from SF gas above a certain threshold density.}

\item By assuming that the transition from the non-SF regime ($\avir > 6$) to the SF regime ($\avir < 6$) with time exists, the time scale of the transition (\ttr) is estimated as $\lesssim 2$ Myr from the number fraction of the SF clumps ($\sim 0.1$) and the duration of the SF phase, i.e., free-fall time in the media with $\nHH \gtrsim 10^{4.6}\ \pcc$;
      this time scale is longer than the free-fall time of typical molecular clouds in the CMZ but is consistent with the predicted timescales of the events that may trigger SF along the gas orbit or stream.
\item  The above timescale \ttr\ is also consistent with the frequency of cloud--cloud collisions (CCCs) estimated by theories and observations.
       However, as typical CCC velocity in the CMZ is argued to be too fast to trigger SF efficiently by recent studies, the effective time scale of CCC-triggered SF can be substantially longer than \ttr.

\end{itemize}

\bibliographystyle{apj}
\bibliography{mendeley,local}

\end{document}